\documentclass[aps,pra,superscriptaddress,preprintnumbers,
showpacs,a4,twoside,twocolumn]{revtex4}

\usepackage{bm}
\usepackage{graphicx}
\usepackage{amsfonts}
\usepackage{amsmath}
\usepackage{amssymb}

\begin{document}

\title{Exploring the grand-canonical phase diagram of interacting bosons in optical lattices by trap squeezing}

\author{Tommaso Roscilde}
\affiliation{Laboratoire de Physique, \'Ecole Normale Sup\'erieure de Lyon,
46 All\'ee d'Italie, 69007 Lyon, France}


\begin{abstract}
In this paper we theoretically discuss how quantum simulators based on 
\emph{trapped} cold bosons in optical lattices can explore the grand-canonical phase diagram
of \emph{homogeneous} lattice boson models, via control of the trapping potential
independently of all other experimental parameters (trap squeezing). 
Based on quantum Monte Carlo, 
we establish the general scaling relation linking the global chemical 
potential to the Hamiltonian parameters for the Bose-Hubbard
model in a parabolic trap, describing cold bosons in optical lattices; we find that this 
scaling relation is well captured by a modified Thomas-Fermi
scaling behavior - corrected for quantum fluctuations - in the case of high enough
density and/or weak enough interactions, and by a mean-field Gutzwiller Ansatz 
over a much larger parameter range. The above scaling relation allows to 
control experimentally the chemical potential, independently of all other Hamiltonian
parameters, via trap squeezing; given that the global chemical potential 
coincides with the local chemical potential in the trap center, measurements
of the central density as a function of the chemical potential gives access
to the information on the \emph{bulk} compressibility of the Bose-Hubbard model. 
Supplemented with time-of-flight measurements of the coherence properties, 
the measurement of compressibility enables one to discern among the various
possible phases realized by bosons in an optical lattice with or without
external (periodic or random) potentials -- \emph{e.g.} superfluid, Mott insulator, band insulator, and Bose glass. 
We theoretically demonstrate the trap-squeezing investigation of 
the above phases in the case of bosons in a one-dimensional optical 
lattice, and in a one-dimensional incommensurate superlattice.

\end{abstract}
\maketitle

 \section{Introduction}
 
 The impressive recent advances in the engineering of
 interacting Hamiltonians for cold trapped atoms suggest
 the possibility of experimentally determining the equilibrium
 (and out-of-equilibrium) behavior
 of fundamental theoretical quantum many-body models.
  In particular experiments on cold atoms in optical lattices have demonstrated 
 the ability of implementing fundamental lattice models (of Hubbard type)
 with full tunability of all Hamiltonian parameters, as well
 as of the lattice geometry and dimensionality \cite{Blochetal08, Lewensteinetal07}. 
 Remarkably, the understanding of a large class of lattice many-body models (such 
 as interacting fermions, frustrated quantum magnets, etc.)
 keeps defying standard theoretical and computational 
 approaches. Hence, the perspective of realizing 
 analog quantum simulators \cite{Feynman82} based
 on cold atoms, literally implementing the physics of the
 challenging models in question, represents a most 
 promising and innovative route towards their understanding. 
 
   Nonetheless, several obstacles still separate the current 
 experiments from giving original answers to long-standing
 questions in quantum many-body physics on a lattice.
 One problematic feature of cold-atom systems in a lattice is thermometry
 \cite{Dieneretal07, HoZhou07, Capogrossoetal07, Polletetal08, 
 Weldetal09, McKayetal09 McKayetal09-2, ZhouHo09}: 
 given that trapped cold atoms are not coupled to a thermal reservoir, 
 their entropy can be controlled but not their temperature, and 
 the knowledge of the temperature as a function of entropy
 is a problem which requires the prior knowledge of the equation 
 of state of the many-body system under investigation
 \cite{Polletetal08, Trotzkyetal09, Jordensetal09}.  
 A second puzzling feature of current cold-atom 
 experiments - which will be the main concern of this paper - 
 is the intrinsic \emph{inhomogeneity} of these 
 systems, imposed by the presence of an overall parabolic 
 trapping potential. While the size of the lattices realized
 in experiments can easily beat the capability of current classical
 simulation methods (at least in dimensions $d>1$), the parabolic
 potential significantly limits the size of the lattice over
 which a uniform phase of the Hamiltonian model is realized. 
 Indeed, a parabolic trapping potential $V(i) = V_t (r_i-r_0)^2$ 
 (where $r_i$ is the position of the $i$-th lattice site and $r_0$ is 
 the center of the trap) imposes a site dependent chemical potential 
 and site-dependent energy jumps $\Delta E(i) = V(i+1) - V(i) = 
 V_t (r_{i+1}-r_i) (r_{i+1}+r_i-2r_0)$ between nearest-neighboring sites. If the local
 many-body phase at a given filling $n$, realized around $i$-th site at the 
 local chemical potential $\mu(i) = V(i)$, is not protected by 
 a particle/hole gap $\Delta_g \gg \Delta E(i)$, its existence is
 necessarily restricted to a narrow neighborhood of site $i$. 
 This aspect introduces significant limitations in the capability
 of realizing phases which are particularly sensitive to the filling, 
 and which display a small or no particle/hole gap over the 
 ground state (see \emph{e.g.} the supersolid phase of
 strongly correlated bosons \cite{Boninsegni03,Senguptaetal05,Scarolaetal06} or
 the Bose-glass \cite{Fisheretal89} for bosons in a random potential).  
 
  In principle one could consider the coexistence of several 
 different phases in the trap as an advantage, given that 
 one single experiment samples a significant portion of the 
 phase diagram of the Hamiltonian implemented in the system 
 via the chemical potential modulation imposed by the trap.
 \cite{Zhouetal09, HoZhou09}
 Unfortunately this form of ``parallelism" is generally hard to enjoy, 
 as most measurement protocols for trapped cold atoms
 employed to date give access to \emph{global} observables, 
 which collect a signal from all the different regions of the
 trap. This fact becomes particularly inconvenient when
 considering the measurement of the excitation spectrum,
 as sampled \emph{e.g.} via lattice-modulation and two-photon Bragg spectroscopy
 \cite{Stoeferleetal04, Fallanietal07, Clementetal09, Ernstetal09}.
  In fact it is quite difficult to associate 
 the different contributions to the excitation spectrum 
 with different spatial regions of the trap, a fact which makes it very hard
 to extract precise information on the structure of the 
 excitations associated with a particular phase of interest. 
 In particular, the presence of a trap imposes the existence of 
 a halo of dilute particles at the cloud boundary, which 
 in principle can always host low-energy excitations.
 Consequently, the fundamental question of the presence or absence 
 of an energy gap over the ground state of a particular phase, 
 realized locally in the trap, becomes a formidable task for current
 cold-atom simulators.
 
 It is important to mention that single-site addressability 
 in optical lattices is becoming reality in recent experiments on 
 quantum gas microscopy \cite{Nelsonetal07,Gerickeetal08, Gemelkeetal09, Bakretal09, Bloch09}.
 This achievement definitely allows one to enjoy the  above mentioned
 parallelism in the chemical potential exhibited by trapped experiments  
 \cite{Zhouetal09}, but only at the level of \emph{local} properties, defined on 
 spatial regions over which the variation of the local chemical potential can be 
 considered weak; and only as long as the so-called local-density approximation
 can be fully trusted. Yet \emph{non-local} properties of 
 homogeneous phases, including correlation functions and collective excitations,
 cannot be extracted from an inhomogeneous sample, except for
 those associated with the region of maximal homogeneity, namely
 the trap center.

 \begin{figure}[h]
\begin{center}
\includegraphics[
     width=75mm,angle=0]{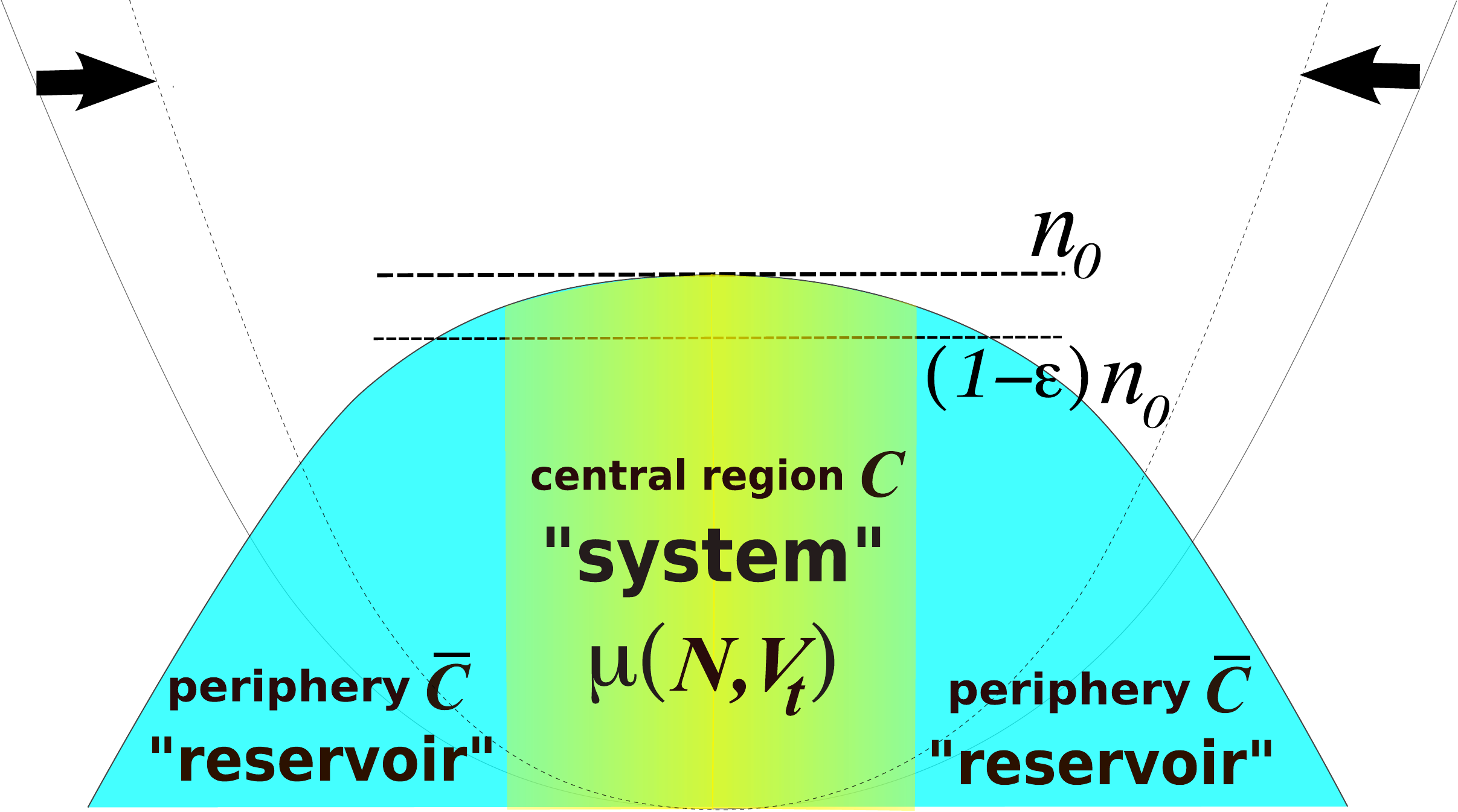} 
\caption{An atomic cloud in a smoothly varying trapping potential - such as a parabolic
trap - can be virtually divided into two regions: a central $C$ region over which the density
is essentially uniform (within a tolerance $\epsilon$), and whose local chemical 
potential $\mu(N,V_t)$ essentially corresponds to the global chemical potential of the
atomic cloud; and a periphery 
region $\bar{C}$, in which the density varies rapidly in space, due to the trapping 
potential. The nearly homogeneous central region can be regarded as the ``system"
in a grand-canonical ensemble, 
while the periphery can be regarded as a particle ``reservoir". Increasing the trapping 
potential (as indicated by the black arrows) increases the chemical potential 
of the reservoir $\bar{C}$, and, at equilibrium, that of the system $C$.  Hence the 
chemical potential of the system can be controlled continuously by the trapping
potential $V_t$.}
\label{f.sys+res}
\end{center}
\end{figure}

  In this paper we propose an experimental protocol which aims at 
 circumventing most of the difficulties listed above, while
 taking advantage of the trap as a low-energy probe for the 
 properties of the system (trap-squeezing spectroscopy \cite{Roscilde09}).
 The fundamental idea relies on the fact that trap effects are minimal
 in the center, both $V(i)$ and $\Delta E(i)$ vanish, so that 
 a local, homogeneous phase can be established over a significant 
 portion of the system, roughly of the order of $\sim ~10$ lattice 
 sites in each spatial direction. This sizable portion of the 
 lattice can be now regarded as the \emph{system} of interest, while 
 the rest of the lattice as the \emph{environment}, acting in 
 particular as a particle reservoir (see Fig.~\ref{f.sys+res} for a sketch). 
 In light of the above discussion, 
 we can conclude that the trap center 
 realizes the ``textbook" quantum simulation of the particular Hamiltonian
 implemented in the system in the grand canonical ensemble 
 (namely at a nearly \emph{uniform} local chemical potential). 
 What remains to be shown is how to control the crucial parameter
 of the quantum simulation at the trap center, namely its local 
 chemical potential, and how to retrieve selective information on the 
 local collective phase realized there. 
 
  We here specialize our discussion to the case of bosons in optical
 lattices, governed by the Bose-Hubbard Hamiltonian in an external
 potential, and leaving the case of fermions to future work. 
 In particular we quantitatively discuss how, tuning  
 the trap frequency independently
 of all others experimental parameters, one gains 
 direct access to the control of the chemical potential in the
 trap center. Remarkably, numerical simulations on the 
 Bose-Hubbard model on the hypercubic lattice show that there exists 
 a simple relation between the strength of the trapping potential
 and the chemical potential: this relation follows approximately the
 predictions which can be obtained both from the atomic limit
 in a lattice, and from 
 the Thomas-Fermi theory for weakly interacting
 gases in continuum space, with deviations due to quantum and lattice 
 corrections. Remarkably, such deviations are quantitatively captured at the 
 \emph{mean-field} level. This means that, unlike the case of the 
 temperature \cite{Trotzkyetal09, Jordensetal09}, the accurate knowledge of the chemical 
 potential of a strongly correlated bosonic system in an optical lattice
 might not require in general an extensive ab-initio calculation.
 This makes the trapping potential a fundamental experimental 
 knob for the quantum simulation, whose effect on the system's
 parameters can be readily assessed. 
 
 Nonetheless, the tuning of such a knob has to be done very carefully, 
 and in general it cannot be done after loading the atoms in the 
 optical lattices. 
 In fact, increasing the strength of the trapping potential 
 at equilibrium leads to the transfer of particles from the wings to 
 the center of the trap, but the tunneling amplitude associated
 with such a transfer can be extremely small in strongly
 interacting systems, requiring then exceedingly slow ramps
 of the trapping potential to enforce adiabaticity in the protocol.
 A simple way to circumvent this problem is proposed, based on
 the loading of atoms in the trap in the weakly interacting regime 
 (namely without optical lattice) followed by the ramp of the 
 optical lattice. 
 
  Once full control on the chemical potential at the trap center
 has been achieved, selective information can be retrieved
 on this region of space by microscopy of the atomic cloud. 
 We here propose a measurement scheme of the
 average central density based on two tightly focused crossed beams 
 resonant with different optical transitions of the atoms; 
 as already mentioned, other quantum gas microscopy schemes (with resolution
 as high as one lattice spacing) have been proposed
 or even become experimentally available in the recent past.   
 The knowledge of the central density as a function of the 
 chemical potential provides then the fundamental information
 on the \emph{bulk compressibility} of the Hamiltonian model
 implemented in the system, and hence on the particle/hole gap
 over the ground state. We present the application of trap-squeezing
 spectroscopy to the measurement of the bulk phase diagram of the 
 Bose-Hubbard model in $d=1$: in particular we discuss the case
 without any applied external potential, and the case in which
 an incommensurate potential with \emph{two} wavelength
 components is applied to the system, giving rise to an
 extended Bose-glass phase whose compressible nature is
 perfectly captured via trap-squeezing spectroscopy. 
    
  The structure of the paper is then as follows. Section \ref{s.LDA}
 introduces the model investigated, the local-density
 approximation and the fundamental role played by the 
 average central density in the trap; Section \ref{s.mu}
 discusses the relationship between the chemical potential in the
 trap center and the trap strength for the Bose-Hubbard model 
 in dimension $d=1$, 2, and 3, with or without an external potential;
 Section \ref{s.protocol} discusses the issue of equilibrium
 preparation of the system at a given trap strength, based
 on a two-step protocol for the ramp of the optical potentials;  
 Section \ref{s.imaging} discusses a proposal for the selective
 measurement of the average density in the trap center; 
 applications to the Bose-Hubbard model without and with 
 external superlattice potentials are presented in Section
 \ref{s.1DBH} and \ref{s.3color} respectively; Section \ref{s.centralvsglobal}
 compares the central compressibility 
 with the global compressibility of the atomic cloud, recently
 measured in experiments \cite{Schneideretal08};
 and finally
 Section \ref{s.conclusions} is devoted to conclusions.

 \section{Local density approximation, average central density
 and central chemical potential}
 \label{s.LDA}
 
  In this section we focus our attention to the general case
 of the Bose-Hubbard model on the $d$-dimensional hypercubic 
 lattice in an external
 potential, composed of a rapidly varying part, given by 
 a superlattice potential created by an additional standing wave
 applied to the system \cite{Fallanietal07,Foellingetal07,Anderlinietal07},
 and a slowly varying parabolic part, coming from the overall gaussian 
 profile of the lasers applied to the system. 
  
  \begin{equation}
{\cal H} (J,U,V_2,V_t) = {\cal H}_0(J,U,V_2) + 
V_t \sum_i (r_i-r_0)^2 ~n_i 
\label{e.Ham}
\end{equation}
\begin{eqnarray}
{\cal H}_0(J,U,V_2) = &-& J \sum_{\langle ij\rangle} 
\left(b_i^\dagger b_{j} + \rm{h.c.} \right) 
+ \frac{U}{2} \sum_i  n_i(n_i-1) \nonumber \\
&+& V_2 \sum_i 
g_i(\{\alpha_l\},\{\phi_l\})~ n_i
\label{e.Ham0}
\end{eqnarray}   
 where  
 \begin{equation}
 g_i(\{\alpha_l\},\{\phi_l\}) = \sum_{l=1}^d 
 \cos^2(2\pi \alpha_l~i_{l}+\phi_{l})-\frac{d}{2}
 \label{e.SL}
 \end{equation}
is a one-color superlattice potential. Here $i_l$ ($l=1,...,d$)
are the coordinates of the $i$-th site and $\langle ij\rangle$
are the pairs of nearest neighbors on the $d$-dimensional 
hypercubic lattice of size $L^d$.  
 
 Experiments on cold atoms in optical lattices are typically
 performed with a \emph{fixed} number of particles $N$. Nonetheless,
 given that the particle number is a good quantum number
 of the Hamiltonian, at $T=0$ the system will be in a 
 definite $N$-sector even in the grand-canonical ensemble.
 This allows us to regard the canonical system with $N$
 particles as equivalent to a system in the grand-canonical
 ensemble with Hamiltonian
 \begin{equation}
 {\cal H}_{\mu} = {\cal H} - \mu \sum_i n_i
 \label{e.Hmu}
 \end{equation}
 where $\mu=\mu(V_t,N,J,U,V_2)$ is the chemical potential which
 establishes $N$ particles in the ground state of the Hamiltonian
 ${\cal H} (J,U,V_2,V_t)$. Moreover we can rewrite Eq.~\eqref{e.Hmu}
 as 
 \begin{equation}
 {\cal H}_{\mu} = {\cal H}_0 - \sum_i \mu(i) n_i
 \end{equation}
 where we have introduced the site-dependent chemical
 potential $\mu(i) = \mu - V_t(r_i-r_0)^2$. If $\mu(i)$
 is a slow-varying function in space around the
 reference site $i^*$ over the typical
 length scale given by the correlation length of the
 Hamiltonian ${\cal H}_{\mu(i^*)} = {\cal H}_0 - \mu(i^*) \sum_i n_i$, 
 one can adopt the local-density approximation (LDA):
 this amounts to considering that the trapped system behaves around the
 site $i^*$ in the same way as its bulk conterpart at the homogeneous 
 chemical potential $\mu=\mu(i^*)$. In particular, in absence
 of a superlattice potential, the local
 density $\langle n_i \rangle$ for $i\approx i^*$ will be extremely 
 close to the homogeneous density in the ground state of
 ${\cal H}_{\mu(i^*)}$. In presence of a superlattice potential, 
 the local density $\langle n_i \rangle$ will be very close
 to the density of the bulk system at a point experiencing the 
 same superlattice potential; moreover the average density over
 a period of the superlattice (or quasi-period for incommensurate 
 superlattices) centered around the site $i^*$ will approximate
 very well the average density in the ground state of the
 bulk system. All these expectations for the behavior of the 
 density are fully verified by numerically exact calculations
 on the Bose-Hubbard model with or without a superlattice 
 \cite{Bergkvistetal04,Wesseletal04, Roscilde08}. 
 The LDA typically breaks down when the local chemical potential
 $\mu(i)$ approaches a critical value sitting at the boundary
 between two phases in the bulk system, so that the correlation
 length of the bulk system diverges. 
  
  The validity of the local-density approximation for parabolic traps implies that the 
  trapped system faithfully probes the density of the bulk system at many different 
 values of the chemical potential away from critical points.
 As mentioned in the introduction, this form of ``parallel" sampling 
 of the bulk phase diagram is only valid for local properties, 
 and it cannot be exploited experimentally unless single-site addressability 
 is achieved. On the contrary, 
 imaging of the atomic cloud over a length scale $R$ of $5-10 \mu$m  
 (corresponding to $\sim 10-20$ lattice sites of an optical 
  lattice with $\lambda\sim 800$ nm) can be achieved 
  more conventionally via
  large-aperture optics \cite{Sortaisetal07} (see Section \ref{s.imaging} 
  for a more detailed discussion). This means that 
  the information on the local density can be 
  retrieved if $\langle n_i \rangle$ does not change
  appreciably over the lengthscale $R$. For a weak enough
  trapping potential, such a condition can 
  be easily met at the \emph{trap center}, where the variation
  in the local chemical potential is the slowest. We hence introduce
  the \emph{average central density}
  \begin{equation}
  n_C =: \frac{1}{|C|}~\sum_{i\in C} \langle n_i \rangle
  \end{equation}
  where $|C|$ is the size of the $C$ region built around the
  trap center $i_0$. The $C$ region is then defined 
  as verifying a condition of quasi-homogeneity
  \begin{equation}
  \frac{|n_C  - \langle n_{i_0}\rangle|}{\langle n_{i_0}\rangle} \leq \epsilon~
  \label{e.quasih}
  \end{equation}    
  with $\epsilon \ll 1$.
  Another important observation singles out the trap center
 as the most interesting region of the trap. Indeed, even when
 it is possible to measure the density at the single-site level,
 one can exclusively achieve the knowledge of a local 
 observable $A_i$ which, via the LDA, can be associated with that ($A$) of 
 the bulk system as a function of the chemical potential, 
 $A(\mu) \simeq A_i(\mu_i = \mu)$. The complementary information on the 
 non-local correlation
 functions of the bulk system is instead completely missing, 
 and indeed the trapped system cannot faithfully reproduce
 the correlation properties of the bulk system in general
 (as testified by the poor performance of the LDA  
 at the level of correlations \cite{Roscilde08}). 
 Nonetheless, the trapped system can faithfully reproduce the 
 correlations around the trap center over a length scale
 corresponding to the extent of the quasi-homogeneous central
 region. This aspect will be discussed in details in Section 
 \ref{s.1DBH}, where it will be shown in addition that 
 the evolution of global correlation properties of the system 
 (as probed \emph{e.g.} by time-of-flight measurements) is dominated by the
 evolution of correlations in the trap center, so that
 correlations in the trap center can 
 effectively be accessed in the experiments. 

  The quasi-homogeneity condition Eq.~\eqref{e.quasih} 
 allows to identify $n_C$ as a close approximation to the
 ground-state density of the bulk Hamiltonian Eq.~\eqref{e.Hmu}
 (with $V_t=0$) at a chemical potential corresponding
 to the background chemical potential $\mu(V_t,N,J,U,V_2)$
 of the trapped system.  
 As further elaborated in Section \ref{s.imaging},
 $n_C$ is experimentally accessible with conventional 
 methods. In order to convert the information on 
 $n_C$ into information on the bulk phase diagram of the
 Hamiltonian ${\cal H}_0$ of Eq.~\eqref{e.Ham0}, we
 need to know at which chemical potential the experiment is operating
 once the experimental parameters $V_t$, $N$, $J$, $U$ and $V_2$ 
 are set. This is the goal of next Section.
 
 \section{Trap-squeezing control of the chemical potential}
 \label{s.mu}

   In this section we investigate 
  the dependence of the background chemical potential 
  $\mu(V_t,N,J,U,V_2)$, stabilizing a ground state with $N$
  particles, for the Bose-Hubbard Hamiltonian with parameters
  $J$, $U$ in a trapping potential of strength $V_t$ and, generally, 
  in a superlattice potential of strength $V_2$. We review the conventional
  Thomas-Fermi approximation and its generic prediction for
  the functional form of $\mu$ in the general case of
  a $d$-dimensional hypercubic lattice. Remarkably, a numerical
  investigation based on numerically exact quantum Monte Carlo
  shows that the Thomas-Fermi prediction, modified by
  quantum fluctuations, is rather accurate 
  at least in the regime of either weak enough interaction or 
  high enough density. As a result, the chemical potential 
  turns out to be simply controlled in the 
  experiments either via the control on the particle population 
  $N$ or the trap strength $V_t$, or on both. 

 \subsection{Atomic limit and Thomas-Fermi approximation}
 \label{s.ALTF}
 
   A simple theoretical approach giving the relationship between
  the chemical potential and the other Hamiltonian parameters
  for the Bose-Hubbard model is the atomic limit (AL), which 
  consists in discarding the 
  quantum kinetic term in the Hamiltonian and in solving for the 
  diagonal part. Such an approximation is reasonable in the
  strongly interacting and strongly trapped limit 
  $U, V_t R^2 \gg J$, where $R$ is the radius of the atomic
  cloud. Minimizing the potential energy part with respect
  to the density 
  \begin{equation}
  {\cal H}_{\rm pot} = \frac{U}{2} \sum_i n_i (n_i-1) 
  +V_2 \sum_i g_i n_i +  V_t \sum_i i^2~ n_i
  - \mu \sum_i n_i
  \end{equation}
  one finds 
  \begin{equation}
  n_i = \frac{\mu+U/2 - V_2 ~g_i - V_t~ (r_i-r_0)^2}{U}.
  \label{e.ni}
  \end{equation}
  Imposing the condition $N = \sum_i n_i$ and passing from the lattice 
  to the continuum formulation, one readily obtains the AL
  chemical potential for a $d$-dimensional system
\begin{eqnarray}
\mu_{\rm AL} + \frac{U}{2} &=& \left( \frac{d+2}{2}~ 
\frac{\Gamma(d/2+1)}{\pi^{d/2}} \right)^{\frac{2}{2+d}}~~
\left(U~N\right)^{\frac{2}{2+d}} ~~V_t^{\frac{d}{2+d}} \nonumber \\
&=& \begin{cases}
0.82548...~(U~N)^{2/3} ~~V_t^{1/3} ~~~~~~ (d=1) \\
0.79788...~(U~N~V_t)^{1/2} ~~~~~~~~~~ (d=2) \\
0.81346...~(U~N)^{2/5} ~~V_t^{3/5} ~~~~~~ (d=3)
\end{cases}
\label{e.AL}
\end{eqnarray}
 
  Notice that the superlattice term does not enter in 
  this formula because it has been chosen so as to take symmetric
  values around zero, and therefore it vanishes upon spatial
  integration. 

 
  On the other hand, a similar formula to Eq.~\eqref{e.qTF} can be 
 obtained in the weakly interacting case via a standard Gross-Pitaevskii (GP)
 approach plus a Thomas-Fermi (TF) approximation \cite{PethickSmith02}.
 Taking the mean-field approximation $a_i \approx \Psi_i$, 
 $a_i^{\dagger} \approx \Psi_i^*$ on the normal ordered Hamiltonian
 Eq.~\eqref{e.Ham} (where $\Psi_i$ is the condensate
 wave-function) one obtains the lattice GP energy functional
 \begin{equation}
 E_{\rm GP} = -J \sum_{\langle ij \rangle} 
 \left( \Psi^*_i \Psi_j + {\rm c.c.} \right)
  + \frac{U}{2} \sum_i |\Psi_i|^4  
  + \sum_i (V_i -\mu) |\Psi_i|^2
 \end{equation}
 where $V_i =  V_2~ g_i + V_t~ (r_i-r_0)^2 - \mu$. 
 
 The standard TF approximation
 consists in neglecting completely the kinetic term in the Gross-Pitaevskii
 functional. This is justified in the continuum case because the kinetic
 energy is suppressed when the wavefunction is slowly varying in space;  
 on the other hand, in the lattice case, a slowly varying wavefunction
 is such that $\Psi^*_i \Psi_j \approx |\Psi_i|^2 $, namely the kinetic
 term does not cancel, but it effectively adds up to the chemical potential 
 term, $\mu \to \mu - 2dJ$. Minimizing the GP functional leads then to the 
 lattice TF equation for the density:
  \begin{equation}
  |\Psi_i|^2 = \frac{\mu + 2dJ - V_2 ~g_i - V_t~ (r_i-r_0)^2}{U}
  \label{e.TFlattice}
  \end{equation}
   Integrating over space in the continuum 
 limit leads to the result $\mu_{\rm TF} = - 2dJ + \mu_{\rm AL} + U/2$.
 
 The central prediction is the linear dependence of $\mu$ on the 
 combination $(N^2 V_t^d)^{1/(2+d)}$, with a slope dependent on 
 $U^{2/(2+d)}$: both predictions will be verified by a numerically 
 exact calculation in a large parameter 
 regime. What is completely missing in the above
 simple approach are: 1) lattice commensuration effects, relevant in the
 case of the appearance of band/Mott insulator states 
 (given that the analytical expression Eq.~\eqref{e.AL} is obtained in the continuum limit);
 2) a proper treatment of the quantum kinetic $J$-term in the Hamiltonian. 
 The lattice effects  
 can be readily restored by numerically performing
 the sum $N = \sum_i n_i$ instead of analytically integrating the continuum
 generalization of Eq.~\ref{e.ni}, at the expense of losing the 
 closed-form prediction of Eq.~\eqref{e.AL}.
 On the other hand, a full account of the quantum corrections 
 requires a more extensive numerical treatment.
 This will be provided in the following, where we will see that such effects 
 do not alter too drastically the TF/AL predictions, and they can be captured 
 already at the mean-field level.

 \subsection{Quantum Monte Carlo results}
 
  Here we present the results of quantum Monte Carlo 
 simulations of the $d$-dimensional Bose-Hubbard model 
 on $L^d$ hypercubic lattices, based on the Stochastic
 Series Expansion method with directed-loop updates
 \cite{SSE}. We have performed 
 simulations at low temperatures $T \sim J/L$ in order to remove significant 
 thermal effects. We perform the simulation in the grand-canonical
 ensemble, and we fine-tune the chemical potential $\mu$
 which stabilizes a given particle number $N$ for the Hamiltonian
 Eq.~\eqref{e.Ham}; this allows us to numerically
 sample the function $\mu(V_t,N,J,U,V_2)$. In the following
 we take $J$ as the energy scale, and express all other
 quantities in units of $J$. We have performed the simulations
 for two values of the $U/J$ ratio for each dimensionality $d$
 and both for zero superlattice potential ($V_2=0$) and
 for an intense superlattice potential ($V_2=U$). We have 
 considered both incommensurate superlattices
 ($\alpha=830/1076$ as in the experiment of Ref.~\onlinecite{Fallanietal07})
 and commensurate ones ($\alpha=3/4$ and $1/2$).
 Results for one spatial dimension have already been partially
 reported in Ref.~\onlinecite{Roscilde09}.

  \subsubsection{d=1 without superlattice}
  
\begin{figure}[h]
\begin{center}
\includegraphics[
     width=65mm,angle=0]{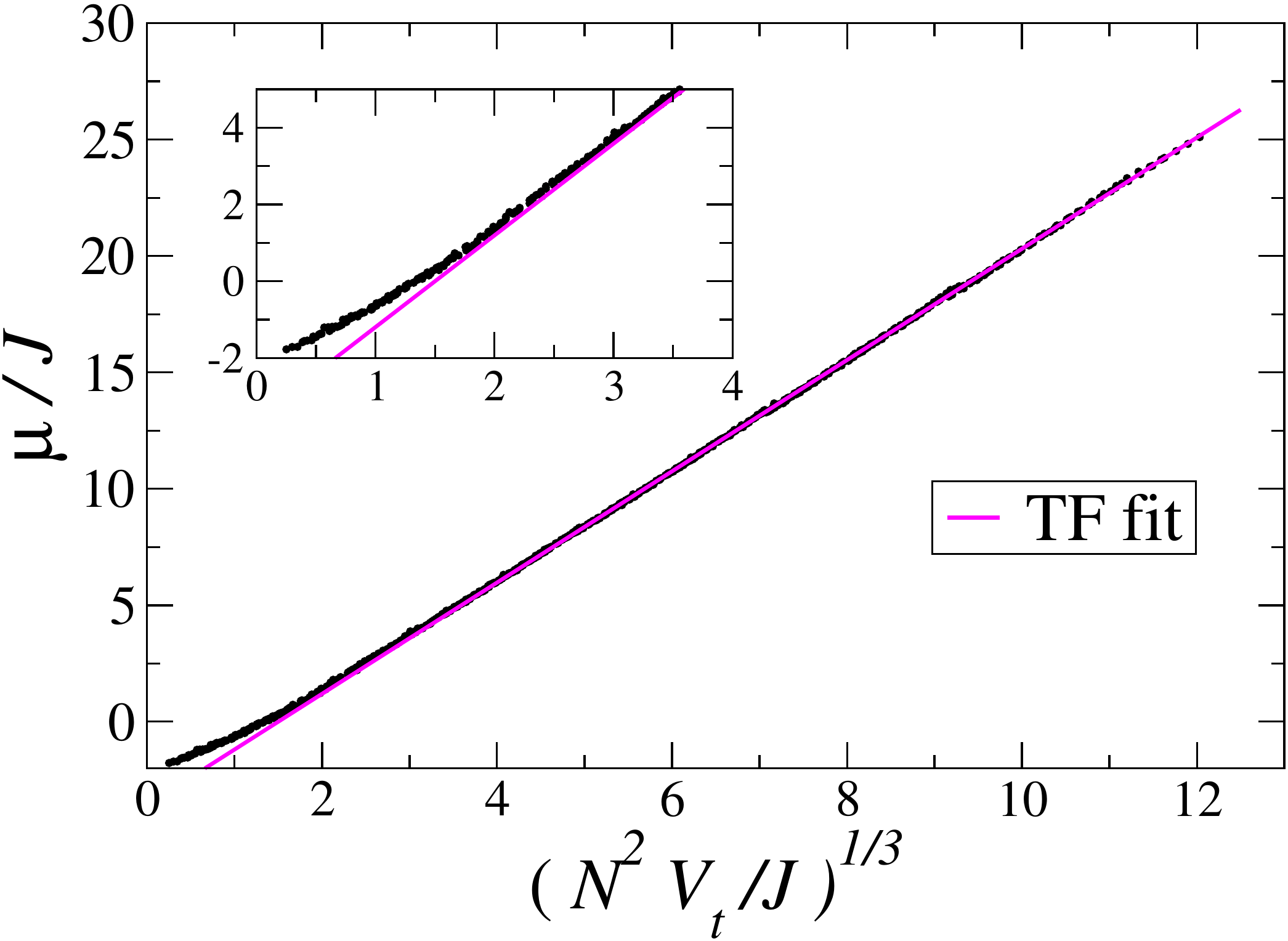} 
\caption{Chemical potential for the 1$d$ Bose-Hubbard model with 
$U =5~J$. The universal curve $\mu = \mu(N^2 V_t/J)$ has
been obtained via the collapse of many data sets, with
$N=20$, ...,$120$ and $V_t=0.004$, ..., $0.140~J$. The solid
curve is a linear fit to the Ansatz Eq.~\eqref{e.qTF}.
Inset: zoom on the crossover from the mTF behavior to the 
low-$N$ and low-$V_t$ behavior.}
\label{f.1dmuU5}
\end{center}
\end{figure}

\begin{figure}[h]
\begin{center}
\includegraphics[
     width=105mm,angle=0]{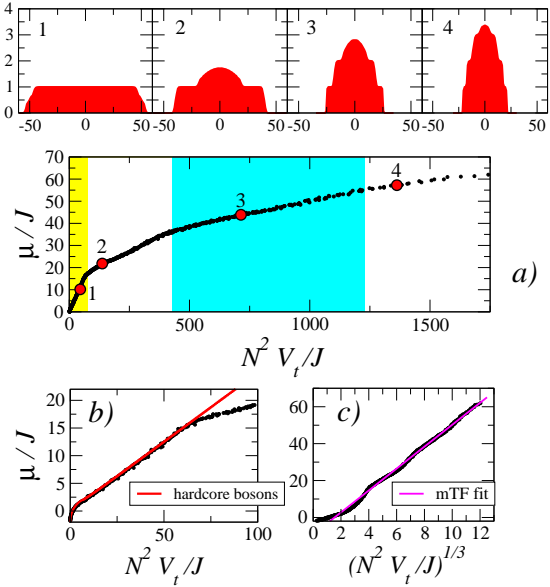} 
\caption{Chemical potential for the 1$d$ Bose-Hubbard model with 
$U =20~J$. \emph{a)} Main panel: Universal curve $\mu = \mu(N^2 V_t/J)$.
The curve is obtained via the collapse of many data sets
(same parameters as in Fig.~\ref{f.1dmuU5}). The differently colored
regions correspond to the number of shells populated in the
``wedding cake" structure, from 1 up to 4. Upper panels: four representative
density profiles showing four different shell occupations 
($N=100$ and $V_t=0.004, 0.014, 0.070$ and $0.135~J$). \emph{b)}
Zoom on the low-$N$/low-$V_t$ region (first shell occupied only), 
showing agreement with the corresponding data for a hardcore
boson system. \emph{c)} Universal curve presented as a function 
of $N^{2/3} (V_t/J)^{1/3}$: the solid curve represents a linear 
fit to the Ansatz Eq.~\eqref{e.qTF}.}
\label{f.1dmuU20}
\end{center}
\end{figure}

  We begin our discussion with the case of absence of a superlattice
  ($V_2=0$).
  Figs.~\ref{f.1dmuU5} and  Figs.~\ref{f.1dmuU20} show results for 
  the two complementary regimes of $U/J=5$ and $U/J=20$. Judging
  from the phase diagram of the 1$d$ Bose-Hubbard model 
  \cite{Rizzietal05}, for $U/J=5$ the kinetic and the 
  potential part of the bulk Hubbard Hamiltonian are in strong 
  competition for most filling values $n\lesssim 3$, so that
  a variation of the chemical potential brings the system through
  an alternation of correlated superfluid phases and 
  Mott insulating phases separated by quantum critical 
  points. Despite the important quantum effects taking place
  in the system, it is remarkable to observe in Fig.~\ref{f.1dmuU5}
  that all $\mu$ values obtained by various combinations
  of $V_t$ and $N$ collapse onto the same universal curve, 
  which is a homogeneous function of $N^2 V_t/J$, in agreement
  with the AL/TF prediction Eq.~\eqref{e.AL}. Refs.~\cite{Rigoletal03, 
  Rigoletal04} have introduced the the so-called ``characteristic density", 
  $\tilde \rho = N(V_t/J)^{d/2}$, as the relevant parameter to characterize
  the density of a trapped system in $d$-dimensions. Our data suggests 
  that the chemical potential is a homogeneous function of the characteristic density. 
  This result will be further confirmed in the case $d=2, 3$; see also 
  Sec.~\ref{s.discussion} for a general discussion.

  Even more remarkably,
  for large enough $N$ and/or $V_t/J$ values, the universal curve
  obeys a modified TF (mTF) scaling of the kind
  \begin{eqnarray}
  \mu_{\rm mTF}(V_t,N,J,U,V_2)/J =~~~~~~~~~~~~~~~~~~~~~ && \nonumber \\
  C_d \left(U/J\right)^{\frac{2}{2+d}}  
  + \gamma_{d}
  \left(U/J~N\right)^{\frac{2}{2+d}} (V_t/J)^{\frac{d}{2+d}}&.&
   \label{e.qTF}
  \end{eqnarray}  
  where $C_{d} = C_{d}(U/J,V_2/J)$. 
  Here $\gamma_d$ is a $d$-dependent slope: numerical evidence
  shows that $\gamma_d$ does \emph{not} seem to depend on the other 
  Hamiltonian parameters (see Ref.~\onlinecite{Roscilde09},
  and the discussion below). A fit to the $d=1$ data
  for $U/J=5$ gives us $\gamma_{d=1}=0.817(2)$, which compares
  surprisingly well with the AL/TF prediction of Eq.~\eqref{e.AL}.
  Moreover $C_{d}$ is a $d$-dependent offset term 
  which is also found to depend on $U/J$ and on $V_2/J$, 
  and which contains the most relevant quantum corrections to 
  the TF/AL result. It is important to stress that in general
  $C_d~ (U/J)^{\frac{2}{2+d}}$ is \emph{not} the correct value of $\mu/J$ for 
  $V_t/J, N \to 0$. In fact in this limit the system
  becomes extremely dilute and the dependence on  
  $U/J$ should drop out. Indeed, in the limit of low $N$ and $V_t$,
  the $\mu$ curve crosses over to the correct dilute limit
  (see inset of Fig.~\ref{f.1dmuU5}) which gives the well known
  result $\mu(N=0,V_t=0) = -2dJ$ as obtained from the solution
  of the tight-binding model. As seen in Sec.~\ref{s.ALTF}
  this result coincides with the $N=0, V_t = 0$ value predicted by the 
  lattice TF theory (Eqs.~\eqref{e.AL} and \eqref{e.TFlattice}) --
  as it should be expected, given that, in the dilute limit, the
  bose gas becomes ideal. This also means that the quantum 
  offset term ($-2dJ$) of the chemical potential from lattice TF theory  
  does not correspond at all to the quantum correction appearing
  in the mTF scaling, Eq.~\eqref{e.qTF}. Hence the form of Eq.~\eqref{e.qTF}
  is far from being trivial.
  
 The results for $U=20 J$ are instead more elaborate.
 In this case the inspection of the 1$d$ Bose-Hubbard phase 
 diagram reveals that the system with a filling $n\lesssim 3$ 
 is a Mott insulator for most of the chemical potential values. 
 Hence the lattice stabilizes commensurate insulating regions
 in the trap which are completely missed in the continuum
 limit leading to the AL/TF formula Eq.~\eqref{e.AL}.
 Nonetheless it is remarkable to observe that all the 
 data obtained for different $N$ and $V_t$ values
 collapse onto the same universal curve which is again
 a homogeneous function of $N^2 V_t$, consistently with
 the AL/TF prediction. Yet this curve shows a succession
 of kinks separating various regimes of filling and trapping:
 an inspection in the microscopic structure of the states
 corresponding to the various regions shows that each kink
 marks the appearance of particles in a new ``shell" of the
 trapped cake structure of the density profile 
 (see Fig.~\ref{f.1dmuU20}(a)). When the system has filling
 $\langle n_i \rangle < 1$ throughout the trap, the large
 $U/J$ value suppresses multiple occupation and the system
 behaves effectively as a hardcore-boson system. This is 
 shown by the excellent agreement between the low-$N$/low-$V_t$ 
 data for the softcore boson system and the same data for a
 hardcore-boson system which can be obtained exactly by 
 standard Jordan-Wigner diagonalization \cite{Liebetal61}.
 Increasing the filling and/or the trapping, the condition
 of single occupancy is eventually violated and the second shell
 of the cake starts to be filled. In the AL, the
 condition for the presence of a site with occupation $n=2$
 in the center of the trap in a $1d$ system is $V_t ~(N/2)^2 \geq U$
 which gives the critical value $(N^2 V_t/J)_{n=2} = 4~U/J$.
 Assuming that the population of further shells appears upon
 depletion of the shell with single occupation, we find that
 the $n=3$ shell starts being occupied when 
 \begin{equation}
 V_t\left(N_1/2\right)^2 = V_t\left(N_2/2\right)^2 + U = 2U
 \label{e.N1N2}
 \end{equation}
 which means that the $n=2$ shell stops growing upon
 increasing $V_t$ and the $n=3$ shell starts building up.
 Here $N_i$ indicates the number of particles in the $i$-th
 shell. With the condition $N_1+N_2=N$ Eq.~\eqref{e.N1N2}
 gives
 \begin{equation}
 (N^2 V_t/J)_{n=3} = 4(1+\sqrt{2})^2 ~U/J
 \end{equation}
 A similar reasoning leads to the critical value for the 
 population of the $n=4$ shell
 \begin{equation}
 (N^2 V_t/J)_{n=4} = 2(1+\sqrt{2})^2 (1+\sqrt{3})^2~U/J
 \end{equation}
 and so on. Hence for $U = 20~J$ we obtain 
 $(N^2 V_t/J)_{n} = 80$, 466.27, and 1740.16 for $n=2,3,$ and 4;
 the corresponding numerical estimates from the QMC simulations 
 are $(N^2 V_t/J)_{n} \approx 66$, 400 and 1200 which, 
 not surprisingly, is lower than the AL prediction due to quantum
 effects allowing the particles to tunnel from the 
 wings of the cloud into the center at a lower trapping strength/
 particle number.
 
  It is evident from Fig.~\ref{f.1dmuU20} that lattice commensuration
 effects become weaker and weaker on the $\mu$ curve for increasing 
 filling. In particular, when plotting the curve as a function
 of the natural TF parameter $(N^ 2 V_t/J)^{1/3}$, as done in 
 Fig.~\ref{f.1dmuU20}(c), we observe that the mTF scaling of Eq.~\ref{e.qTF} sets in approximately 
 once the $n=3$ shell starts to be filled. A fit of the large-$N$/large-$V_t$
 data with Eq.~\ref{e.qTF} delivers a value of $\gamma_{d=1}$ consistent
 with the value obtained from the $U/J=5$ data, confirming that
 $\gamma_d$ only depends on dimensionality.

 \subsubsection{d=1 with superlattice}
 
\begin{figure}[h]
\begin{center}
\includegraphics[
     width=90mm,angle=0]{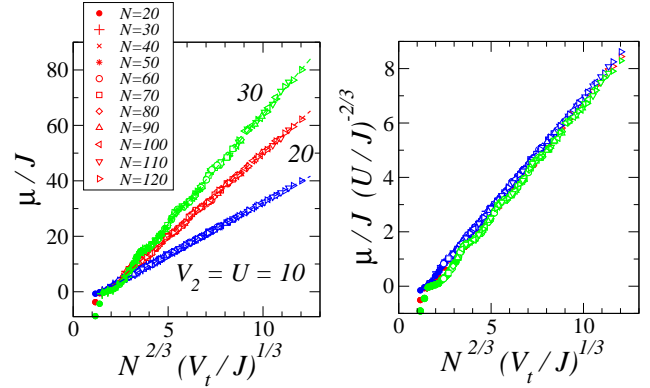} 
\caption{Chemical potential for the trapped 1$d$ Bose-Hubbard model 
with a commensurate superlattice potential ($V_2=U$,$\alpha=3/4$).
The right panel shows that all curves have a universal scaling
of the the slope with the ratio $U/J$. Dashed lines
are fits to the Ansatz Eq.~\eqref{e.qTF}.}
\label{f.1dmuSLcomm}
\end{center}
\end{figure}

 Fig.~\ref{f.1dmuSLcomm} shows the $\mu$ curve for the 1$d$ Bose-Hubbard
 model in a strong superlattice with $\alpha=3/4$, $\phi=0$ and $V_2=U$, and 
 for three values of  $U/J= 10, 20$ and 30. Again the chemical potential
 is clearly a homogeneous function of $N^2 V_t/J$, as shown by the collapse
 of all the $\mu$ values onto the same universal curve dependent
 on $U/J$ and $V_2/J$. For large $U/J$, $V_2/J$ and/or for small
 $N^2 V_t/J$ we observe kinks in the $\mu$ curve characteristic
 of lattice commensuration effects: in the case of a commensurate
 superlattice as the one considered here these effects are clearly
 related to the fractional filling shells \cite{Roscilde09} appearing in the 
 AL of the system at fillings $(2n+1)/4$ with $n=0,1,2,...$. Yet the overall 
 trend of the curve for high enough filling follows the mTF 
 behavior Eq.~\ref{e.qTF}, and a fit to that equation gives again a $\gamma_d$
 consistent with what found in the previous section in absence
 of the superlattice. This confirms the TF/AL prediction 
 that the presence of a superlattice is not influencing the overall slope
 of the $\mu$ curve. Similar curves for the incommensurate superlattice
 case $\alpha = 1076/830$ have already appeared in Ref.~\onlinecite{Roscilde09}.

\begin{figure}[h]
\begin{center}
\includegraphics[
     width=80mm,angle=0]{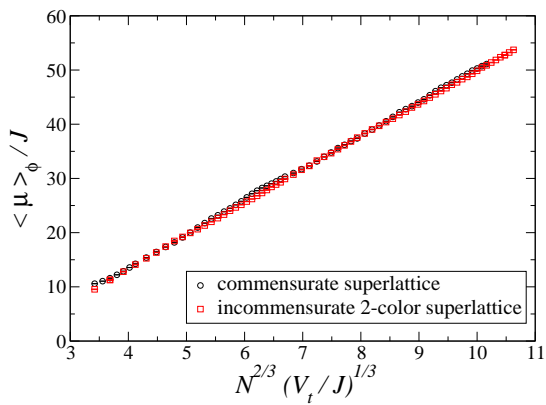} 
\caption{Phase-averaged chemical potential for the trapped 1$d$ 
Bose-Hubbard model with a commensurate superlattice potential 
and with a 2-color incommensurate potential (see text). For
both cases $V_2=U=20 J$ and $N=100$.}
\label{f.1DmuSLav}
\end{center}
\end{figure}

  It is important to stress that lattice commensuration effects
 in the presence of a superlattice are strongly related to the possibility
 of maintaining the spatial phase $\phi$ of the superlattice fixed in the
 experiments. If this phase is allowed to fluctuate, (as it generally
 happens from shot to shot of the same experiment \cite{Fallaniprivate}, 
 unless phase locking is explicitly enforced \cite{Foellingetal07}),
 the chemical potential in the experiment is going to change
 accordingly, assuming that all other experimental parameters remain
 unchanged. In this case it is then more convenient to introduce
 a \emph{phase-averaged} chemical potential $\langle \mu \rangle_{\phi}$ 
 - namely averaged over random fluctuations of the phase $\phi$. 
 Fig.~\ref{f.1DmuSLav} shows $\langle \mu \rangle_{\phi}$ for two
 superlattices: for the commensurate one discussed above, and for
 an incommensurate one which has one additional color component,
 namely 
 \begin{equation}
 g_i(\alpha,\alpha',\phi,\phi') =  
 \cos^2(2\pi \alpha~i+\phi) + \cos^2(2\pi \alpha'~i+\phi')-1
\label{e.3SL}
 \end{equation}
 with $\alpha=1076/830$ and $\alpha'=1473/830$. This particular superlattice
 will be further discussed in Section \ref{s.3color}.
 The chemical potential values are typically
 averaged over a sample of $\sim 100-200$ phase values. 
 Comparing the average results with typical ones we notice that
 fluctuations are small around the average, so that,
 for each different realization of the superlattice, the 
 average chemical potential $\langle \mu \rangle_{\phi}$ stabilizes 
 in the system a number of particles $N$ which is close to the 
 desired one.

 As a summary of the study of one dimensional systems,  Table \ref{t.1dmu} shows the result 
 of the fit of the various cases considered (no superlattice, 
 commensurate and incommensurate superlattice) to the Ansatz 
 of Eq.~\eqref{e.qTF}. We observe that the coefficients 
 $\gamma_{d=1}$ are essentially all consistent within error bars, 
 as the TF/AL theory would simply predict; more quantitatively,
 they are all close to the TF/AL prediction 
 $\gamma_{d=1}=0.82548..$, which is a remarkable fact given
 the simplicity of that theory. Moreover the constant terms
 $C_{d=1}(U/J,V_2/J)$ appear to be only weakly dependent
 on the Hamiltonian parameters; in particular, within error
 bars they appear not to depend on the parameter $\alpha$
 of the superlattice but only on the superlattice strength.
 
  \begin{table}
  \begin{tabular}{|l|c|c|c|r|}
  \hline
  $U/J$ & $V_2$ & $\alpha$ & $\gamma_{d=1}$ & $C_{d=1}$ \\
  \hline
  \hline
    5  & 0 & - & 0.818(1) & -1.23(1) \\
    10  &  & - & 0.818(2) & -1.21(2) \\
    20  &  & - & 0.815(5) & -1.32(4) \\
   \hline
    10  & $U$ & 3/4 & 0.817(2) &  -1.23(2) \\
    20  &  & 3/4 & 0.820(2) & -1.40(1) \\
    30  &  & 3/4 & 0.822(3) & -1.58(1) \\
  \hline
    10  & $U$ & 830/1076 & 0.816(3)  & -1.22(3) \\
    20  &  & 830/1076 & 0.822(4)  & -1.42(4) \\
    30  &  & 830/1076 & 0.819(5) & -1.54(5) \\
    \hline
 \end{tabular}
 \caption{Results of the fit of the chemical potential 
 data to the scaling Ansatz, Eq.~\eqref{e.qTF}, for the 
 1$d$ trapped Bose-Hubbard model without any superlattice, 
 and with a commensurate/incommensurate superlattice.}
 \label{t.1dmu}
 \end{table}

\begin{figure}[h]
\begin{center}
\includegraphics[
     width=90mm,angle=0]{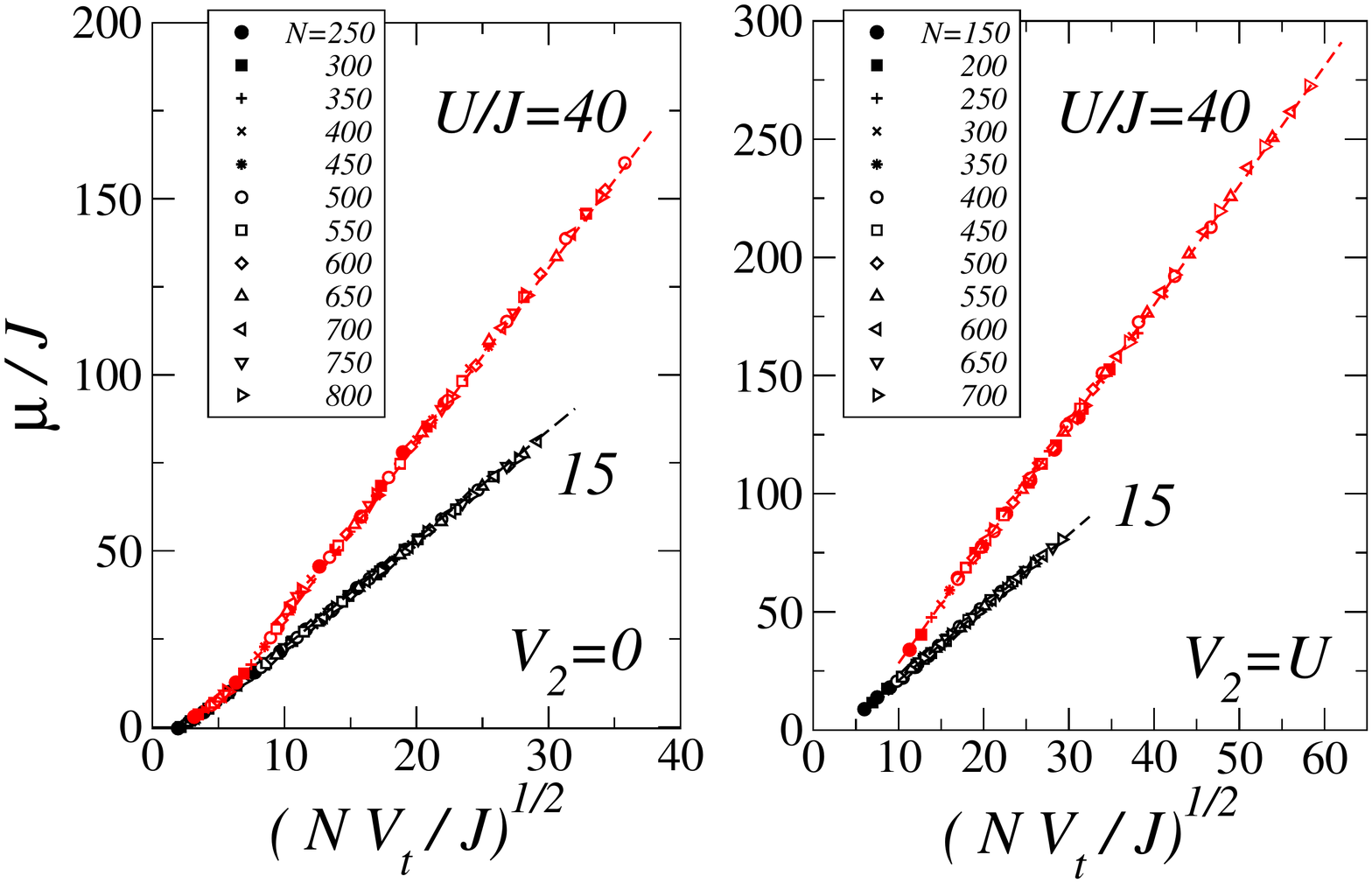} 
\caption{Chemical potential for the trapped 2$d$ 
Bose-Hubbard model without (left panel) and with (right panel)  
a superlattice potential ($\alpha=830/1076$ in all
spatial dimensions). Dashed lines are fits to the 
Ansatz Eq.~\eqref{e.qTF}.}
\label{f.2Dmu}
\end{center}
\end{figure}

 \subsubsection{d=2, 3}
 
  Fig.~\ref{f.2Dmu} shows the chemical potential for the $d=2$ Bose-Hubbard
  model as numerically determined via quantum Monte Carlo. 
  We have considered the two values $U/J=15$ and $40$ for the interaction: 
  according to the phase diagram of the homogeneous 2$d$ Bose-Hubbard
  model \cite{Capogrossoetal08}, for the first value the system is always in a 
  superfluid state at all fillings, while for the second value the 
  system experiences Mott insulating phases at fillings $n=1,2,..$
  upon changing the chemical potential. 
  From Fig.~\ref{f.2Dmu} we see that, no matter the phase the system is 
  in, data sets for different particle numbers and varying trapping potentials 
  all collapse onto the same curve when plotted as a function of 
  $(N V_t)^{1/2}$, in agreement with the TF/AL prediction, both
  in absence and in presence of a superlattice potential. For sufficiently
  high $N$ and/or $V_t$ the dependence of $\mu$ on $(N V_t)^{1/2}$
  is linear and can be well fitted to the Ansatz Eq.~\eqref{e.qTF}
  giving the results summarized in Table \ref{t.2dmu}. It is remarkable
  that the numerical results for the slope $\gamma_{d=2}$ are
  very close to being all numerically consistent with each other, and 
  with the TF/AL prediction $\gamma_{d=2}=0.79788$. 
  On the opposite end, for small $N$ and/or low $V_t$, the local filling
  can drop to values $\langle n_i \rangle \leq 1$, which, in absence of 
  a superlattice, amounts to the onset of a $n=1$ Mott plateau for 
  $U/J=40$: this strong lattice commensurability
  effect, completely neglected in the TF/AL approach in continuum space, is
  responsible for the deviation of the $U/J=40$ results from the linear
  behavior. As already observed for the $d=1$ case \cite{Roscilde09} 
  the $\mu$ curve exhibits a crossover to a low-density regime which loses
  the dependence on the $U/J$ value, as shown by the merging of the $U/J=40$ 
  curve with the $U/J=15$ one. In general one does expect all $\mu$ curves at 
  arbitrary $U/J$ values to merge onto the same universal curve for a diluted
  system when $N, V_t \to 0$. Yet a large enough $U/J$ suppresses double
  occupancy in a system with filling smaller than unity, so that the
  the actual value of $U/J$ becomes irrelevant and the system enters
  the so-called Tonks or hardcore regime: this is the effect responsible for 
  the merging of the $\mu$ curves at $U/J=40$ and $U/J=15$ in Fig.~\ref{f.2Dmu},
  well before the regime of extreme dilution is attained (the two curves
  come close to each other for $(N V_t)^{1/2}$ values which correspond
  to a filling $n\sim1$ in the trap center). On the contrary, for the $N V_t$ values we considered,
  the hardcore regime is not observed  in presence of a 
  strong superlattice $V_2=U$, which competes with the repulsion and maintains 
  doubly occupied sites down to the lowest trap fillings we have explored.
  
\begin{figure}[h]
\begin{center}
\includegraphics[
     width=90mm,angle=0]{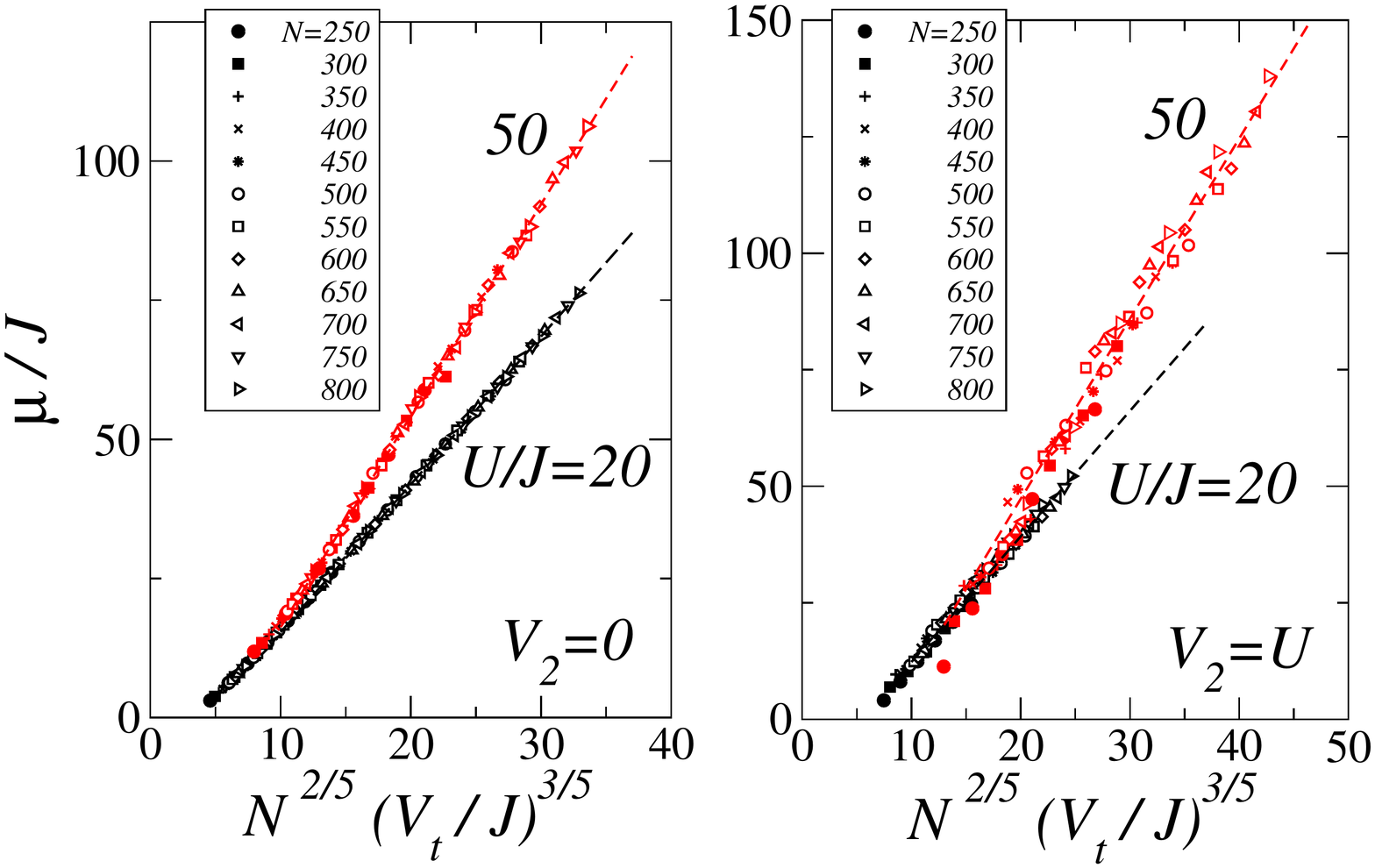} 
\caption{Chemical potential for the trapped 3$d$ 
Bose-Hubbard model without (left panel) and with (right panel)  
a superlattice potential ($\alpha=830/1076$ in all
spatial dimensions). Dashed lines are fits to the 
Ansatz Eq.~\eqref{e.qTF}.}
\label{f.3Dmu}
\end{center}
\end{figure}

  To conclude, Fig.~\ref{f.3Dmu} shows analogous results for the $d=3$ case.
  We again choose two values of repulsion: $U/J=20$, for which the homogeneous
  system is superfluid at all fillings \cite{Capogrossoetal07-2}; and $U/J=50$, for which
  the system has Mott insulating regions at fillings $n=1,2...$ .
  In agreement with the TF/AL prediction, the chemical potential is a homogeneous
  function of the product $N^{2/5}V_t^{3/5}$, and it becomes linear for 
  large enough filling in the trap center ($n\gtrsim 1$); the results of
  fits to Eq.~\eqref{e.qTF} are summarized in Table~\ref{t.3dmu}, and
  show numerical consistency between the estimated slopes $\gamma_{d=3}$ 
  and the proximity to the TF/AL prediction $\gamma_{d=3} = 0.81346$,
  both in absence and in presence of a superlattice \cite{note.on.3DSL}.
  In the opposite limit of low filling $n\lesssim 1$, and in absence
  of the superlattice, we observe again that the $\mu$ curves
  at different $U/J$ values tend to merge together, which corresponds
  to the onset of the hardcore regime with suppression of double
  occupancy.

 \begin{table}
  \begin{tabular}{|l|c|c|c|r|}
  \hline
  $U/J$ & $V_2$ & $\alpha$ & $\gamma_{d=2}$ & $C_{d=2}$ \\
  \hline
  \hline
    15  & 0 & - & 0.795(3) & -2.1(1) \\
    40  &  & - & 0.790(4) &  -3.1(1) \\
   \hline
    15  & $U$ & 830/1076 & 0.803(6) & -2.6(2) \\
    40  &  & 830/1076 & 0.799(3) & -3.53(8) \\
    \hline
\end{tabular}
\caption{Results of the fit of the chemical potential 
 data to the scaling Ansatz Eq.~\eqref{e.qTF} for the 
 2$d$ trapped Bose-Hubbard model without any superlattice, 
 and with an incommensurate superlattice.}
\label{t.2dmu}
\end{table}

 \begin{table}
  \begin{tabular}{|l|c|c|c|r|}
  \hline
  $U/J$ & $V_2$ & $\alpha$ & $\gamma_{d=3}$ & $C_{d=3}$ \\
  \hline
  \hline
    20  & 0 & - & 0.801(3) & -3.38(8) \\
    50  &  & -  & 0.795(5) & -4.55(11) \\
   \hline
    20  & $U$ & 830/1076 & 0.812(14) &  -4.4(2)  \\
    50  &     & 830/1076 & 0.81(5)   &  -6.3(1.7) \\
    \hline
\end{tabular}
\caption{Same as in Table \ref{t.2dmu} bur for the 
 3$d$ trapped Bose-Hubbard model.}
\label{t.3dmu}
\end{table}

\begin{figure}[h]
\begin{center}
\includegraphics[
width=90mm]{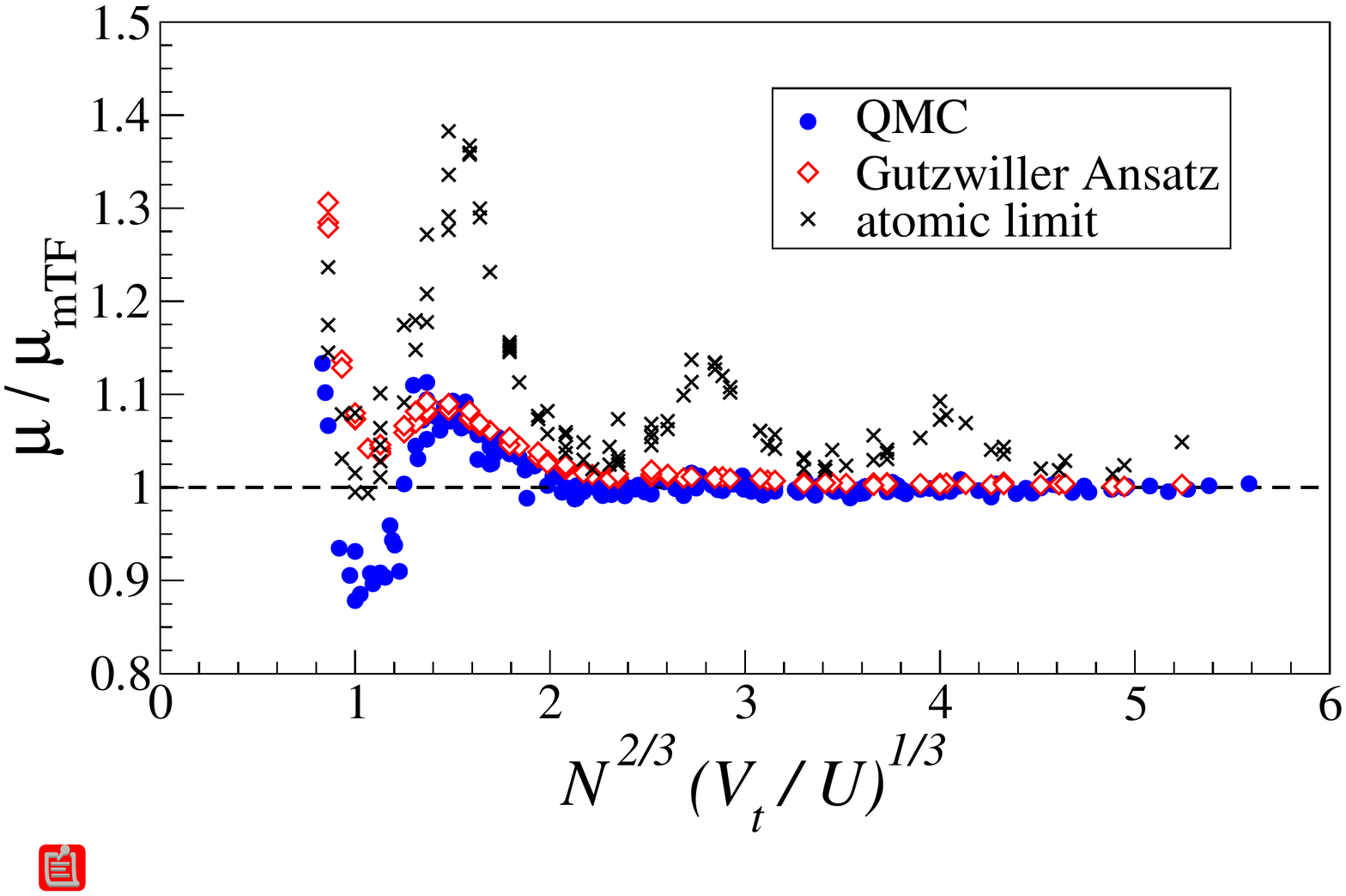}
\null\vspace*{-2cm} 
\includegraphics[
width=90mm]{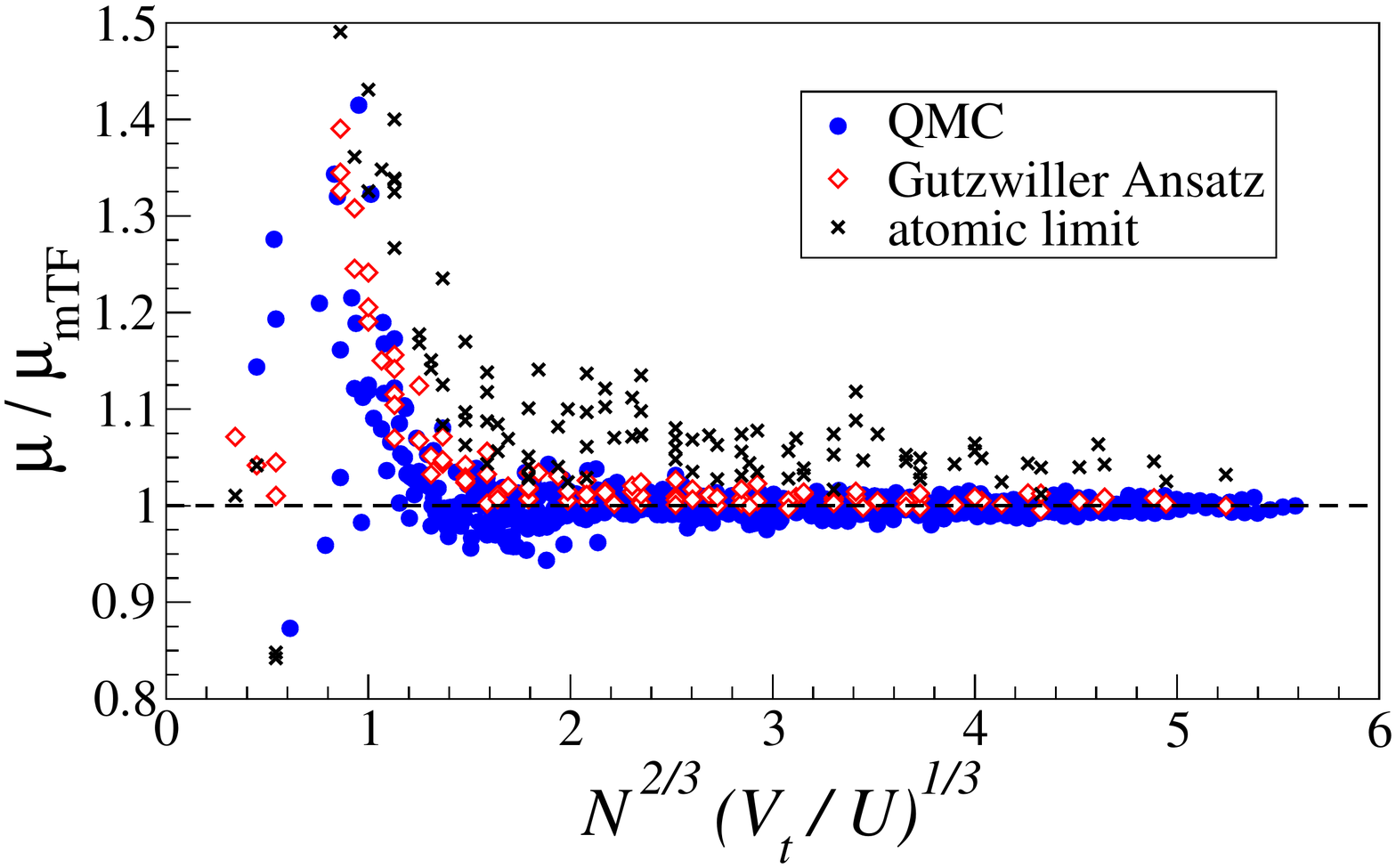}
\null\vspace*{-1.5cm} 
\caption{Chemical potential for the trapped 1$d$ Bose Hubbard model
with $U = 10 J$, as obtained via quantum Monte Carlo, 
Gutzwiller Ansatz and atomic limit calculations. $\mu_{\rm mTF}$  
is the modified Thomas-Fermi prediction of Eq.~\eqref{e.qTF} with
fitting parameters from Table \ref{t.1dmu}.  \emph{Upper panel}: $V_2=0$. 
\emph{Lower panel}: $V_2=U$, $\alpha=830/1076$.}
\label{f.Gutz1d}
\end{center}
\end{figure}

\subsection{Quantum Monte Carlo vs. Gutzwiller mean-field results} 

  In the previous section we have seen that the chemical potential 
 of strongly correlated bosonic systems can be described via
 a modified Thomas-Fermi scaling for sufficiently high densities
 or sufficiently strong trapping potentials, at which lattice commensuration
 effects become very weak. On the other hand, lattice effects can be 
 easily taken into account at the classical level via a numerical calculation 
 in the atomic limit, neglecting intersite hopping. Yet such a drastic 
 approximation is going to be unreliable when the hopping energy becomes
 comparable with the repulsion one, typically when $2dnJ \sim U n^2$. 
 Nonetheless, quantum effects can be restored at an approximate
 level within Gutzwiller mean-field theory, which keeps the cost
 of numerical calculations at a minimum, while producing predictions
 which turn out to be in a suprisingly good agreement with those of quantum 
 Monte Carlo. 
 
  Gutzwiller mean-field theory assumes a factorized Ansatz for the
  wavefunction of the many-body system
  \begin{equation}
  |\Psi \rangle = \otimes_{i=1}^N \left( \sum_{n=0}^{n_{\rm max}} f_n^{(i)} |n\rangle_i \right)
  \label{e.Gutz}
  \end{equation}    
where $|n\rangle_i$ is a Fock state with $n$ particles at site $i$, and
$n_{\rm max}$ is a suitable truncation imposed on the local Hilbert space
for numerical purposes.  Assuming all $f_n^{(i)}$ coefficients to be real, 
the expectation value of the Hamiltonian,
Eq.~\eqref{e.Ham}, on the Gutzwiller Ansatz (GA) wavefunction takes the form
\begin{equation}
 \langle {\cal H} \rangle  = \sum_i \varepsilon_i
\end{equation}
where
\begin{eqnarray}
  \varepsilon_i = 
 &-& J \gamma_i  \sum_n \sqrt{n+1} ~f_n^{(i)}  f_{n+1}^{(i)}  \nonumber \\
&+& \sum_n |f_n^{(i)}|^2 \left[ \frac{U}{2} n(n-1) +  V_i n\right]
\end{eqnarray}
and 
\begin{eqnarray}
\gamma_i &=&  \sum_l \sum_n  \sqrt{n+1}~ f_n^{(i+l)}  f_{n+1}^{(i+l)}  
 \nonumber \\
V_i &=& V_2 ~g_i +V_t(r_i-r_0)^2 - \mu~.
\end{eqnarray}
Here $\sum_l$ runs on the $z = 2d$ nearest neighbors in a 
hypercubic lattice. 

The ground state corresponds then to the minimization of the 
Hamiltonian expectation value with respect to the $N n_{\rm max}$
coefficients. Instead of proceeding with a full minimization, a usual
procedure is to minimize with respect to the local variables $f_n^{(i)}$
at a site $i$ while holding fixed the variables at all the other sites.
We perform random sweeps over the lattice sites, touching
each site once per sweep on average, until convergence is reached -- 
typically less than a hundred sweeps are necessary for convergence.  For the 
Hubbard model under investigation, this leads to a very efficient
location of the absolute energy minimum, which can be verified
by repeating the minimization starting from a different initial 
condition. 

\begin{figure}[h]
\begin{center}
\includegraphics[
width=90mm]{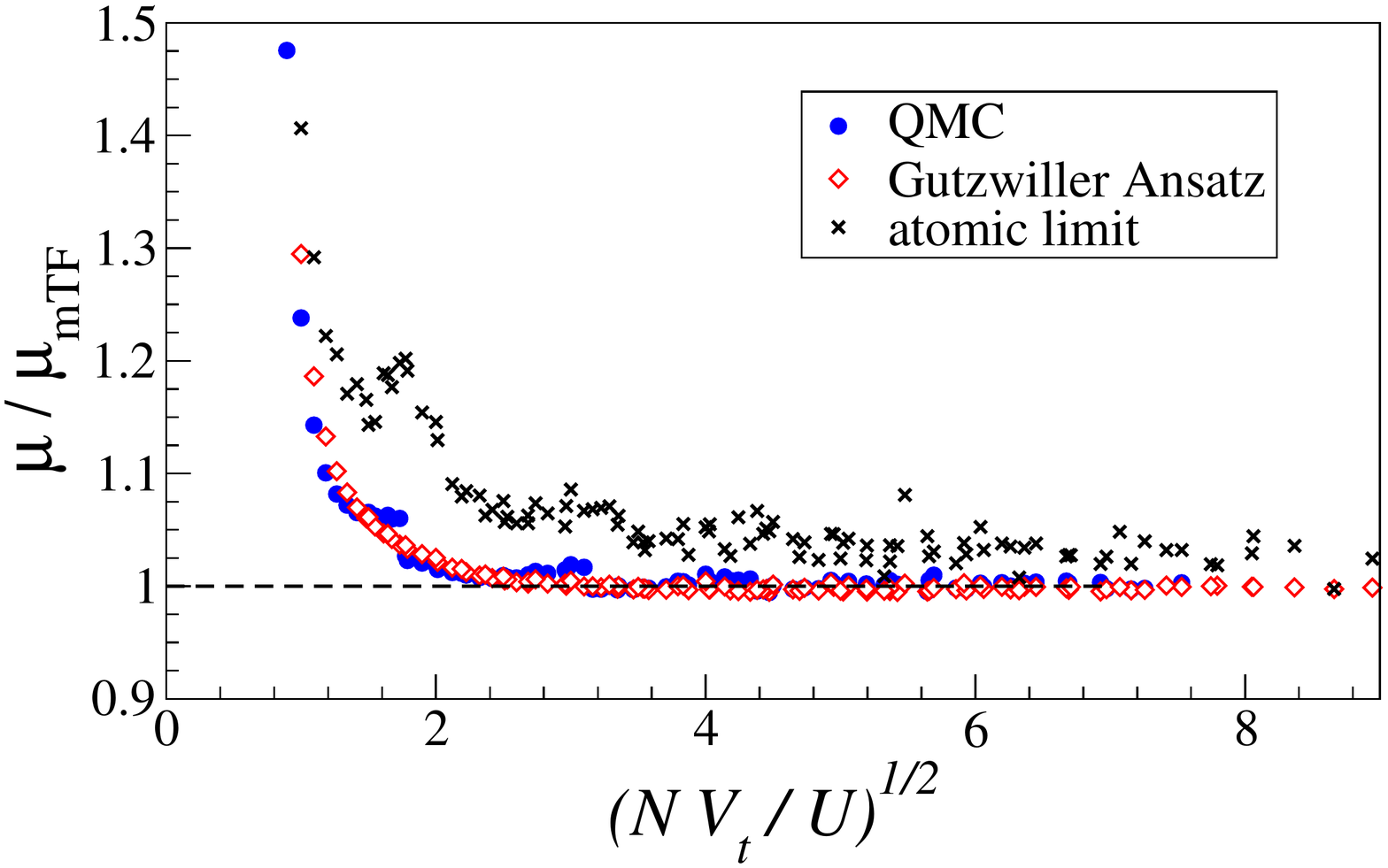}
\null\vspace*{-2cm} 
\includegraphics[
width=90mm]{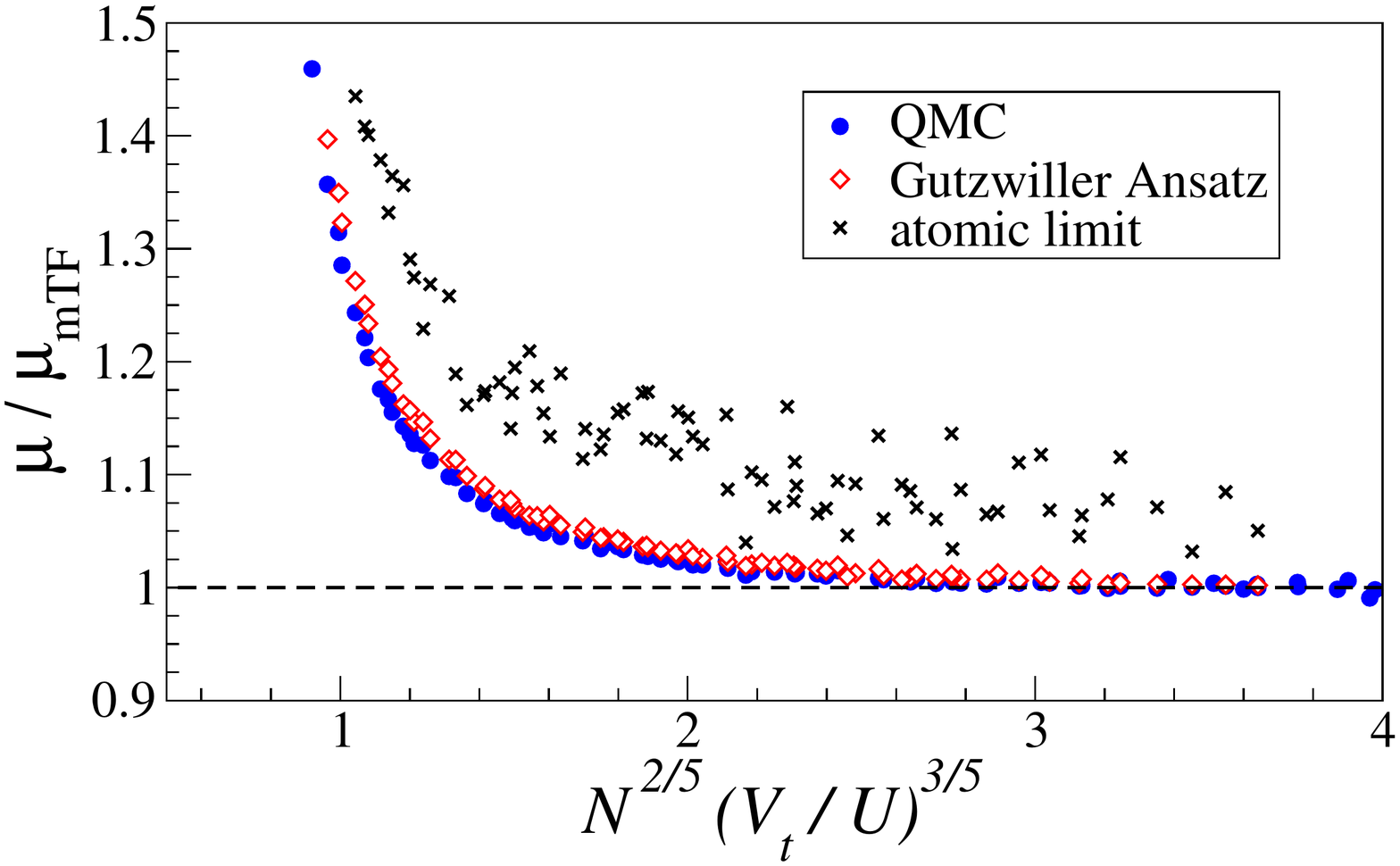}
\null\vspace*{-1.5cm} 
\caption{Chemical potential for the trapped 2$d$
and 3$d$  Bose Hubbard model. $\mu_{\rm mTF}$  
is the modified Thomas-Fermi prediction of Eq.~\eqref{e.qTF} with
fitting parameters from Tables \ref{t.2dmu} and \ref{t.3dmu}.  
\emph{Upper panel}: $d=2$, $U= 15 J$, 
$V_2=0$. \emph{Lower panel}: $d=3$, $U= 20 J$, 
$V_2=0$.}
\label{f.Gutz2d3d}
\end{center}
\end{figure} 

 Eq.~\eqref{e.Gutz} generally describes a state which does
 not have a well defined particle number. Given that the minimization
 of the energy is done in an unconstrained way at each site, 
 we need to adjust the chemical potential $\mu$ \emph{a posteriori}
 in order to achieve a desired average particle number $N = \sum_{i,n}  
 |f_n^{(i)}|^2 n$, with an analogous procedure to that used with
 quantum Monte Carlo. Hence this procedure samples the 
 function $\mu=\mu(V_t,N,J,U,V_2)$ of interest. We use different 
 particle numbers  $N$ and trapping potentials, ranging in the
 same intervals as those explored in quantum Monte Carlo 
 calculations.

 Figs.~\ref{f.Gutz1d} and \ref{f.Gutz2d3d} compare the 
 results of the GA with those of quantum Monte Carlo  
calculations for the trapped Bose-Hubbard model in dimensions
$d$=1,2 and 3. Data obtained in the atomic limit $J=0$ are also
shown as a reference. All data are normalized to the 
mTF behavior, Eq.~\eqref{e.qTF}, attained in the 
large density / strong confinement limit. 
The results in the atomic
limit are seen to generally overestimate the chemical potential --
this is not surprising, given that there is no kinetic energy gain
in adding particles to the system in the atomic limit, and hence
all the energy gain to balance repulsion has to be provided by
the chemical potential. Quite surprisingly, instead, we observe that
the mean-field GA accounts very well 
for the quantum corrections to the atomic limit for \emph{all} 
system dimensions, and that the 
chemical potential predicted via the GA is 
typically less than $4\%$ off with respect to the quantum Monte
Carlo prediction. Indeed the GA, while successfully
describing the main features of the phase diagram of the Bose-Hubbard
model, is not particularly accurate in determining the phase boundaries;
given its mean-field nature, it is not quantitatively trustworthy in 
low dimensions, and its predictions on the location of phase boundaries
can be up to $100\%$ off in $d=1$.  
We find that the GA works equally well even in presence
of an incommensurate superlattice potential, despite its inability to 
describe Anderson localization and Bose-glass physics (data for
the case $d=1$ are shown in Fig.~\ref{f.Gutz1d}). 

 A semi-quantitative justification of the success of the GA
in predicting the chemical potential of the trapped Bose-Hubbard model
can be provided based on LDA. LDA predicts that the total density
of the system can be written as 
\begin{equation}
N = \sum_i n(\mu_i)
\end{equation}
where $\mu_i = \mu - V_t (r_i - r_0)^2$ is the local chemical potential, 
and $n(\mu)$ is the density as a function of the chemical potential
for the bulk system. The GA might provide a very poor
prediction $n_{\rm GA}(\mu)$ for the $n(\mu)$ function of the bulk system, especially in $d=1$, 
and this might suggest that the global chemical potential $\mu$ which 
stabilizes $N$ particles in the system should be also poorly estimated. 
Yet what matters in the determination of $N$ is not the whole 
$n(\mu)$ curve, but only its integral. If the difference 
 $\Delta n(\mu) = n(\mu) - n_{\rm GA}(\mu)$ oscillates from positive to negative
 and back, it will be averaged to zero upon integration. This is most
 likely the case for the Hamiltonian parameters we have considered. 
 As observed \emph{e.g.} in Fig.~\ref{f.Gutz1d}, the most serious problems arise 
 for low $N$, which, in a trapped system, implies a chemical potential excursion 
 (from the tails to the center) over which the $\Delta n$ function has  
 completed only a few oscillations (if any).

\subsection{Discussion}
\label{s.discussion} 
 
 From the above results we can conclude that the chemical potential 
 for the Bose-Hubbard model in a trap and in a superlattice potential
 has a simple scaling form as a function of the experimental parameters. 
 On the one hand an accurate determination relies in general on 
 numerics due to the strong interactions in the system; yet we find
 that the chemical potential
 appears as a homogeneous function of $N^{2/(2+d)}~V_t^{d/(2+d)}$,
 and in particular a linear function thereof for sufficiently high
 filling in the trap center (typically $n\gtrsim 1$ 
 in absence of a superlattice, and even lower in presence of 
 a superlattice), and/or weak enough interaction, verifying this way 
 a quantum-modified version of the TF/AL prediction, Eq.~\eqref{e.qTF}.
 Ref.~\cite{Batrounietal08} has shown that, under the assumption that
 LDA holds, one can prove the relationship $\tilde \rho = I_d(\mu/J;U/J)$,
 where $\tilde \rho$ is the characteristic density, already introduced above,
 and $I_d$ is an unknown function. By inverting the previous relation, 
 this amounts to say that $\mu$ is a homogeneous function of $\tilde\rho$,
 in agreement with our findings. Therefore our results allow to explicitly reconstruct
 the relation between $\mu$ and $\tilde\rho$. This correspondence of 
 our results with the LDA prediction suggests that $\mu$ is \emph{exactly} 
 a homogeneous function of $\tilde\rho$ if (and possibly only if) 
 LDA is exact. Indeed deviations from perfect homogeneity are seen 
 in our numerical data, for instance in Fig.~\ref{f.Gutz1d}, and they
 can be attributed in part to numerical uncertainty; in part to the fact
 that in \emph{any} finite system different $\mu$ values correspond to 
 the same $N$ value (which is discretized), so that there is a natural
 uncertainty on $\mu$; and in part to possible systematic deviations, enhanced
 in the presence of a rapidly oscillating superlattice. Yet we observe 
 that the homogeneity property is an important simplifying assumption, 
 and it is verified with an accuracy of a few percent in the worst case.

 Moreover, in the case of the Bose-Hubbard model without superlattices,
 Ref.~\cite{Rigoletal09} has numerically shown that density profiles (and profiles
 of other local observables) for trapped
 bosons having the same characteristic density $\tilde\rho$, and the same 
 Hamiltonian parameters $U$ and $J$, appear to be invariant
 up to a rescaling of the space coordinates with the characteristic length $\eta = \sqrt{J/V_t}$.
 This implies that, under rescaling of the coordinates with $\eta$, the density at site $i$, 
 $\langle n_i \rangle = \langle \tilde n  ({\bm r_i}/\eta) \rangle$, is a unique function of $\tilde \rho$.  
 At the same time, the local chemical potential, $\mu_i$, expressed 
 in terms of rescaled variables, turns out as well to be a unique function of $\tilde\rho$ 
 \begin{eqnarray}
 \mu_i &=& \tilde \mu_{{\bm r}_i/\eta} = \mu(\tilde\rho;U,J) + V_t (r_i-r_0)^2  \nonumber \\
&=& \mu(\tilde\rho;U,J) + J [(r_i-r_0)/\eta]^2
 \end{eqnarray}
thanks to the result that $\mu = \mu(\tilde\rho;U,J)$.  As a consequence, 
the results of Ref.~\cite{Rigoletal09} suggest that 
$\langle \tilde n ({\bm r_i}/\eta) \rangle $ is a unique function of  $\tilde \mu_{{\bm r}_i/\eta}$. 
In the limit of an infinitely shallow trap, $V_t\to 0$ and $N\to \infty$ at fixed $\tilde\rho$,
the dependence of $\langle \tilde n  ({\bm r_i}/\eta)\rangle$ on $\tilde \mu_{{\bm r}_i/\eta}$
must reproduce the dependence of the bulk density on the chemical potential, 
$n(\mu)$. Hence the unique function relating $\langle \tilde n ({\bm r_i}/\eta)  \rangle$ to
$\tilde \mu_{{\bm r}_i/\eta}$ must be identical with that of the bulk limit. As a consequence, 
the invariance of the density profiles (and, with the same reasoning, of any other 
local observable in the trap) under rescaling with $\eta$ implies that the LDA is exact. 
It is trivial to show that the the exactness of the LDA implies the invariance of 
local quantities under rescaling with $\eta$. Hence we can conclude that 
the scale invariance of trapped systems at fixed characteristic density is 
exact \emph{if and only if} LDA is exact. Yet, even when LDA is not exact, 
the system might display scale invariance with a better accuracy than that
of LDA, as shown numerically in Ref.~\cite{Rigoletal09} in the case of density fluctuations
and local compressibility.

 At the experimental level, the above results reveal the possibility of controlling the chemical 
 potential in the trapped system by: 1) controlling the trapping frequency at 
 fixed particle number, or viceversa 2) controlling the particle number
 at fixed trap strength, or 3) controlling both simultaneously. Hence the present 
 experimental setups have direct access to the measurement of phase 
 diagrams of strongly correlated bosons in the \emph{grand-canonical} 
 ensemble at variable chemical potential. 
  Most noticeably, an accurate knowledge (within a few percent)
  of the zero-temperature chemical potential as a function of the Hubbard Hamiltonian
  parameters can be obtained with little numerical effort via the mean-field
  Gutzwiller Ansatz, namely without fully solving numerically the many-body
  problem. This aspect is to be contrasted with the case of accurate 
  thermometry of strongly correlated bosons, which at the moment can
  be reliably assessed only via exhaustive ab-initio simulations \cite{Trotzkyetal09, Jordensetal09}.


 \section{Trap-squeezing protocol}
 \label{s.protocol} 

\begin{figure}[h]
\begin{center}
\includegraphics[width=50mm]{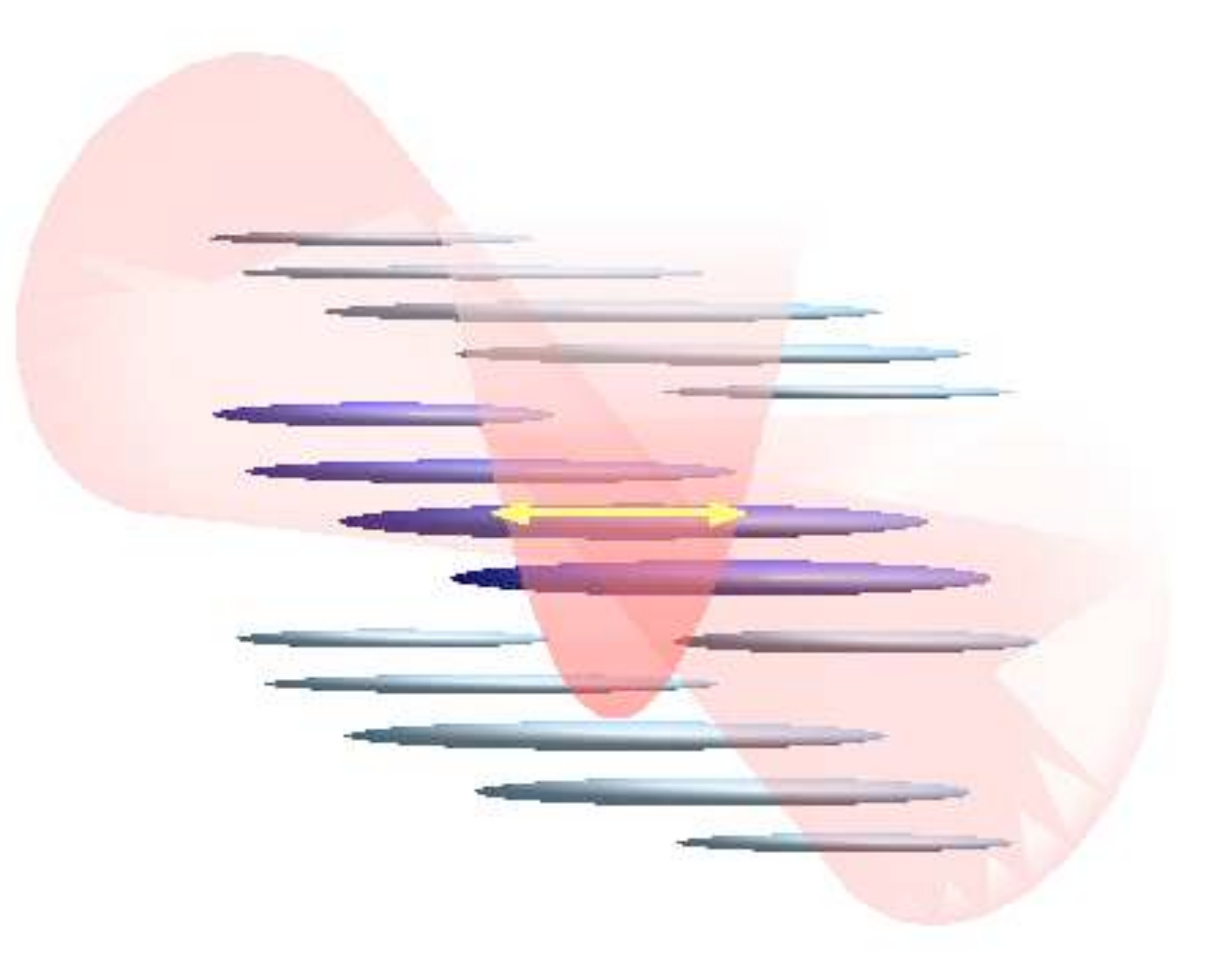} 
\caption{Schematic view of the confinement control on the 
longitudinal direction in a set of one-dimensional tubes.}
\label{f.squeeze}
\end{center}
\end{figure}

   The previous section has shown how the chemical potential
  of a trapped system is simply controlled by the number of 
  particles $N$ and by the trapping potential $V_t$. Indeed recent experiments
  have probed the compressibility of fermions loaded in optical lattices 
  by varying the trapping potential \cite{Schneideretal08} or the
  atom number \cite{Jordensetal08, Scarolaetal09}, although 
  the explicit relation between the experimental parameter and the chemical 
  potential was not known for that system.  
  In the experiments, the atom number is typically subject to significant 
  fluctuations due to shot noise of order $\sqrt{N}$, as well as to other
  systematic effects. On the contrary the trapping potential 
  $V_t$ can be controlled on a much finer scale via the application
  of a dipolar trap along the spatial direction of interest,
  namely by shining $d$ red-detuned running laser waves onto the 
  sample to control the confinement in $d$ spatial dimensions
  (this is sketched in Fig.~\ref{f.squeeze} for the case of 
  variable one-dimensional confinement). 
   The dipolar trap adds up to the confining (anti-confining) 
  potential given by the overall gaussian intensity profile of 
  the red-detuned (blue-detuned) optical lattice, and it gives
  the possibility of varying the trap strength $V_t$ 
  \emph{independently} of the Hamiltonian 
  parameters $U$, $J$, $V_2$, offering in this way the possibility
  of simulating the Bose-Hubbard Hamiltonian in a variable trapping
  potential \cite{Kinoshitaetal06, Schneideretal08}.

  In the remainder of the paper we hence focus on the proposal
 of extracting the phase diagram of the Bose-Hubbard model in 
 the grand-canonical ensemble by controlling the chemical potential 
 of the system via trap-squeezing and at fixed particle number. 
 Two fundamental issues are addressed in the next subsections
 concerning the experimental feasibility of this proposal:
 1) we discuss the fundamental difficulty in achieving adiabatic 
 trap-squeezing once the particles are loaded in the optical lattice;
 2) hence we propose a simple loading sequence which overcomes this 
 difficulty, and which allows to control the particle number in the 
 central tubes/layers in the case of 1$d$/2$d$ confinement.

 \subsection{The adiabaticity issue}
 \label{s.adiabaticity}  
 
  Optical lattice experiments have revealed the unique possibility 
 of tuning the Hamiltonian parameters in real time, \emph{e.g.} via
 changing the intensity of the standing wave which controls the 
 $U/J$ parameter \cite{Greineretal02}. A fundamental issue raised
 by such real-time control is the possible generation of excitations in the system 
 by a non-adiabatic change
 of the Hamiltonian parameters. This becomes particularly dramatic 
 when the parameter change implies the crossing of a quantum critical 
 point, at which the gap over the ground state vanishes, leading to 
 efficient Landau-Zener tunneling \cite{Dziarmaga05}. Similarly
 to the change in the optical lattice strength, a change in the
 dipolar trapping strength can also lead to quantum phase transitions,
 driven this time by the chemical potential, and hence it is subject
 to adiabaticity issues. In fact adiabaticity is even more problematic
 in the case of trap squeezing: regarding the center of the trap
 as the system, and the trap periphery as the particle reservoir
 (as mentioned in the Introduction), the process
 of transfer of particles between system and reservoir under trap squeezing 
 might be pathologically slow if the particles are strongly localized in one part or 
 the other, due to strong interactions in presence of an optical lattice
 or to (quasi-)disorder potentials. 
 
 \begin{figure}[h]
\begin{center}
\includegraphics[
bbllx=40pt,bblly=160pt,bburx=590pt,bbury=600pt,%
width=90mm]{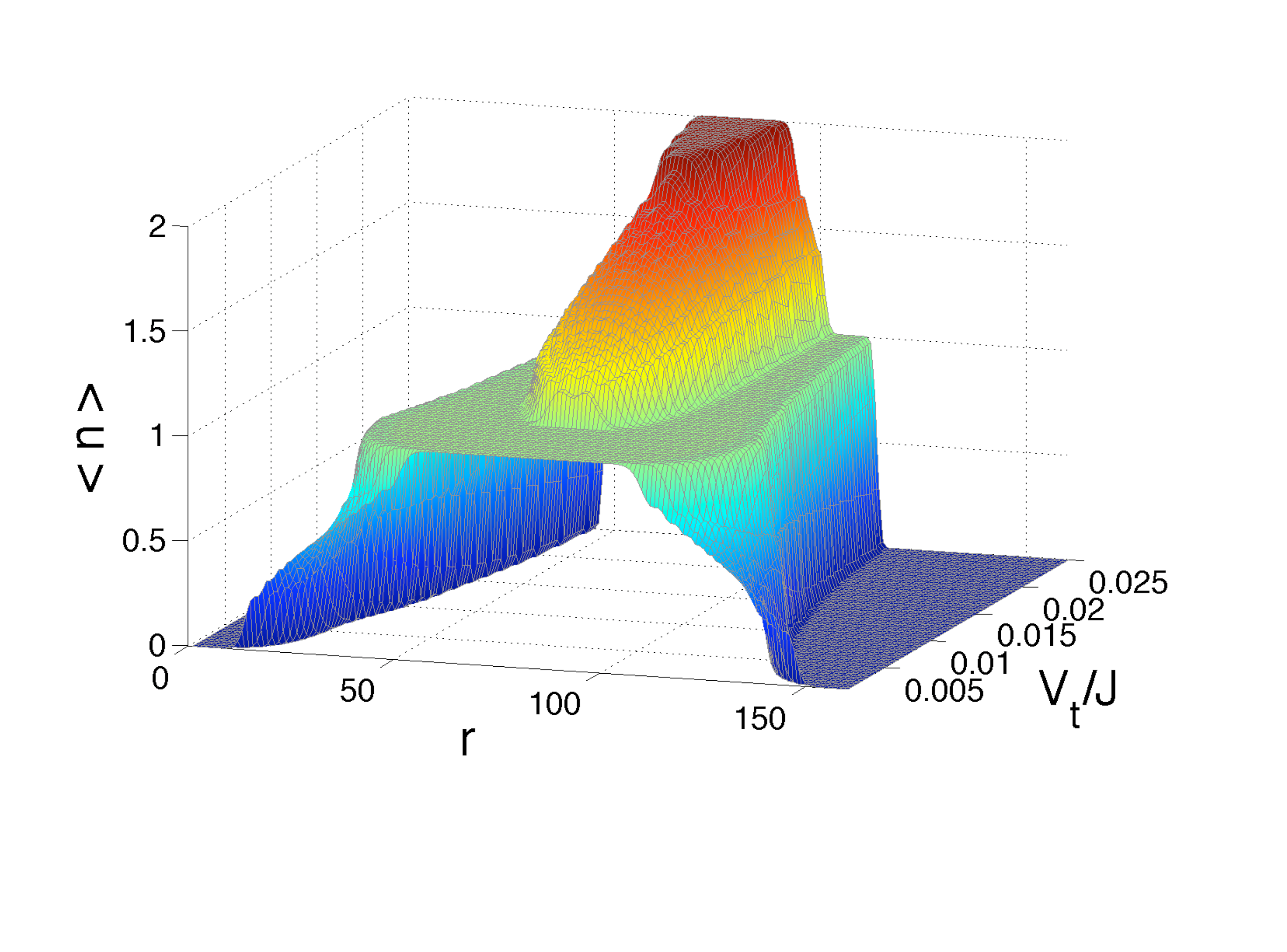} 
\caption{Density profile as a function of trap strength
for the Bose-Hubbard model with $N=100$ particles and with 
$U=20/J$, as resulting from an extended fermionization
calculation (see text and Appendix~\ref{a.fermionization}).}
\label{f.EFdensity}
\end{center}
\end{figure} 

\begin{figure}[h]
\begin{center}
\includegraphics[
width=80mm]{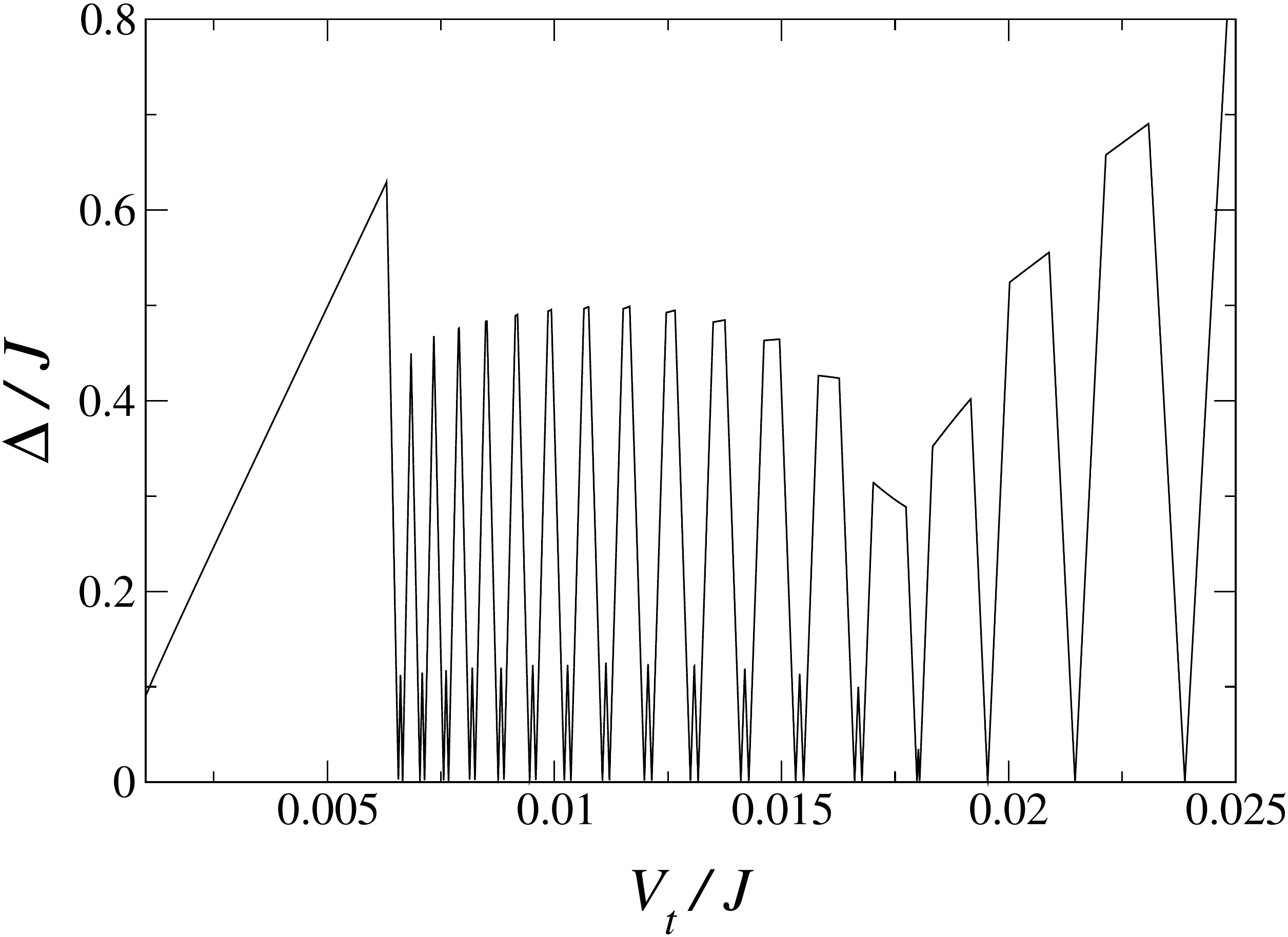} 
\caption{Lowest energy gap $\Delta$
as a function of trap strength
for the Bose-Hubbard model. Parameters as in Fig.~\ref{f.EFdensity}
\label{f.EFgapBH}}
\end{center}
\end{figure} 

 We exemplify the above discussion in the case of lattice 
 \emph{hardcore} bosons, for which the exact many-body spectrum can be 
 calculated exactly via Jordan-Wigner diagonalization 
 \cite{Liebetal61}, in the case of density $n\leq 1$ on 
 each site. The fermionization approach can be extended to 
 the case $n > 1$ \cite{Pupilloetal06} under the condition of having
 a well developed layer-cake structure of the density profile
 (see Appendix \ref{a.fermionization} for a detailed discussion).  
Fig.~\ref{f.EFdensity} shows the evolution of the density profile 
upon trap squeezing for the trapped Bose-Hubbard model with $U=20 J$ and 
with $N=100$ particles, as obtained via extended fermionization. For the 
chosen range of trap strengths, we observe that the system goes from
a Mott-insulating phase with $n=1$ particles in the trap center to 
a superfluid phase with $1<n<2$ and finally to a Mott-insulating phase
with $n=2$ particles. In particular, the ``local transition"
from Mott insulator to superfluid in the trap center at fixed particle number
is exclusively due to the redistribution of particles from states in the 
wings of the trap to states in the
center. The states in the trap periphery are localized by the joint effect 
of the strong repulsion exerted by the particles in the trap center and of
the confining potential. On the other hand the states in the trap center are localized by 
the confining potential over a length scale which is well below the width
of the atomic cloud. Hence the spatial overlap between such states is 
negligible, and they are separated by a high potential barrier - which
is represented by the atoms forming the $n=1$ Mott plateaus. This implies
that the tunnel splitting between the states in the center and the
states in the wings can be extremely small, and hence the ground
state can become nearly degenerate with the first excited state. This is 
indeed revealed by the direct investigation of the lowest energy gap, 
which, within the extended fermionization approach, is estimated as
the lowest particle-hole excitation energy for the spinless fermions, 
and it is shown in Fig.~\ref{f.EFgapBH}. We observe that such a gap 
goes to a value very close to zero \cite{noteongap} when the first particles
move from the trap wings to the trap center, corresponding to the 
occurrence of an insulator-to-superfluid transition in the trap core. 
The gap shows as many dips as 
particles transfered to the center, revealing the tightly avoided
level crossings corresponding to the particle redistribution -- 
more precisely, each deep minimum of the $\Delta$ \emph{vs.} $V_t$ curve 
corresponds to the successive migration of two particles from the wings to 
the center, because such particles occupy states which are nearly degenerate 
symmetric/antisymmetric superpositions of localized states on the two opposite
wings of the trap).  
 Hence this succession of tightly avoided level crossings makes adiabatic
trap squeezing essentially impossible in presence of Mott-insulating
regions, as squeezing times which are several orders of magnitude 
bigger than the typical tunneling time ($\sim$ms) would be required. 

\begin{figure}[h]
\begin{center}
\includegraphics[
width=80mm]{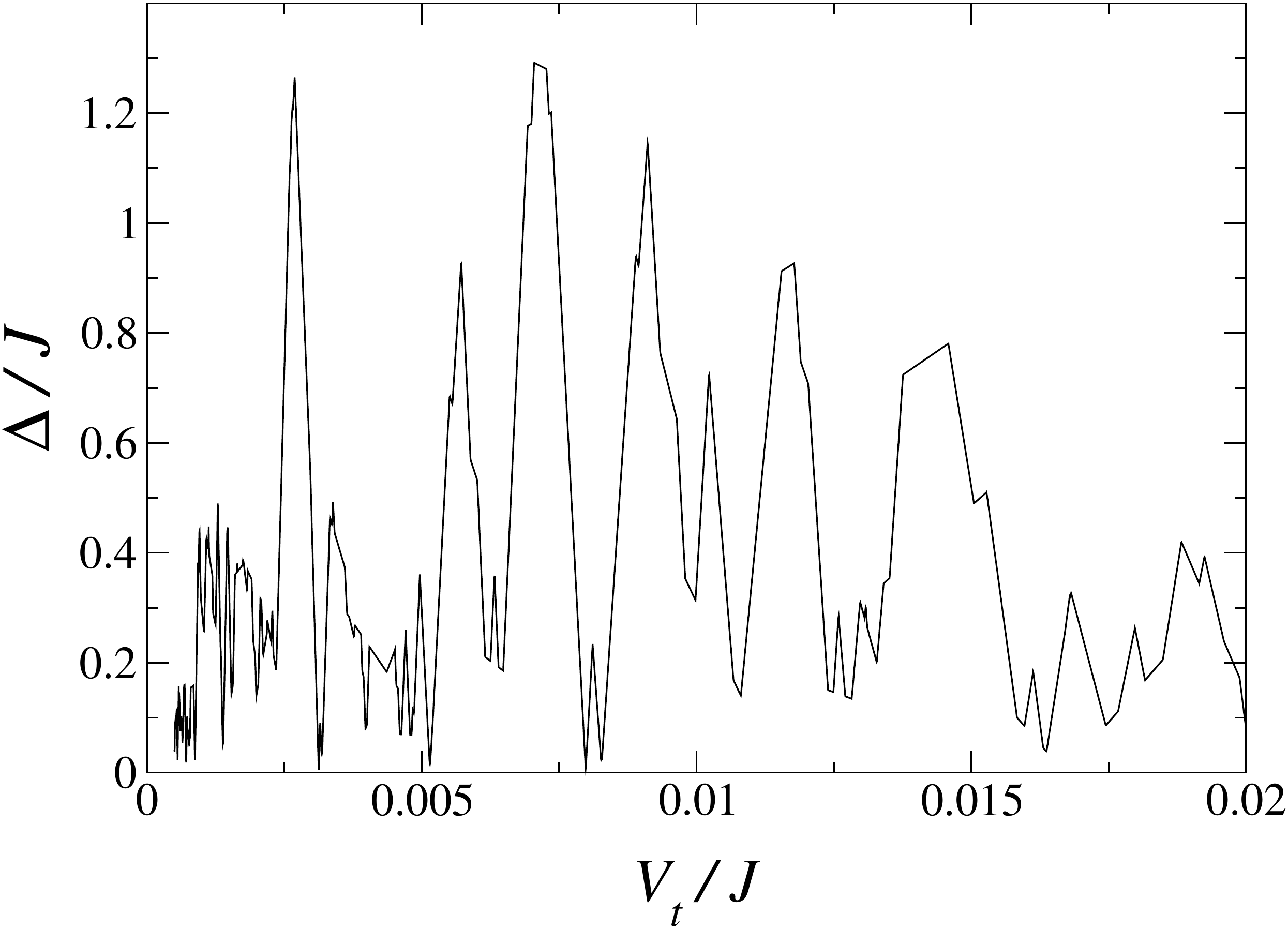} 
\caption{Lowest energy gap $\Delta$
as a function of trap strength
for $N=50$ hardcore bosons in an incommensurate
superlattice ($\alpha=830/1076$) of strength $V_2=10 J$.}
\label{f.SLgap}
\end{center}
\end{figure} 

 As a second example we consider the case of $N=50$ hardcore bosons 
in an incommensurate superlattice potential Eq.~\eqref{e.SL}
with $\alpha=830/1076$ (as experimentally realized in  
Ref.~\onlinecite{Fallanietal07}) and $V_2=20 J$. 
Such a potential is known to lead to Anderson localization of all 
single-particle states for $V_2 > 4J$ \cite{AubryA79, Sokoloff85}.
In this case standard fermionization provides the exact spectrum, 
and in particular the lowest energy gap as a function of the 
trap strength, as shown in Fig.~\ref{f.SLgap}. We observe that 
the gap can become extremely small, this time due to 
the quasi-degeneracy of Anderson localized states which 
are spatially well separated, so that the tunnel splitting 
between them is very small. 

 Hence from these two examples we can conclude that either
strong repulsion induced by the optical lattice, or a (pseudo-)
disorder potential, give rise to localized single-particle states, 
which become quasi-degenerate upon changing the trap frequency. 
This leads to a very efficient Landau-Zener tunneling which essentially makes
adiabatic trap squeezing impossible. Although the results shown here are obtained 
for a one-dimensional system, their underlying mechanism is quite
general and applies to higher dimensions as well. 
Hence we can generally conclude that adiabatic trap-squeezing
for a system of particles loaded in a strong optical (super)lattice
is virtually impossible. Luckily the statement
of the problem already contains in itself the solution: 
trap squeezing has to be performed \emph{before} loading the
particles in the optical (super)lattice, namely when
the particles still enjoy their full mobility, so that they
can adiabatically follow the variation in the trapping 
potential. In the case of spatially anisotropic optical lattices,
discussed in the next subsection, this consideration applies
to the loading of the particles in the optical lattice along
the spatial dimensions in which the particles are more
weakly confined.

 \begin{figure}[h]
\begin{center}
\includegraphics[
width=80mm]{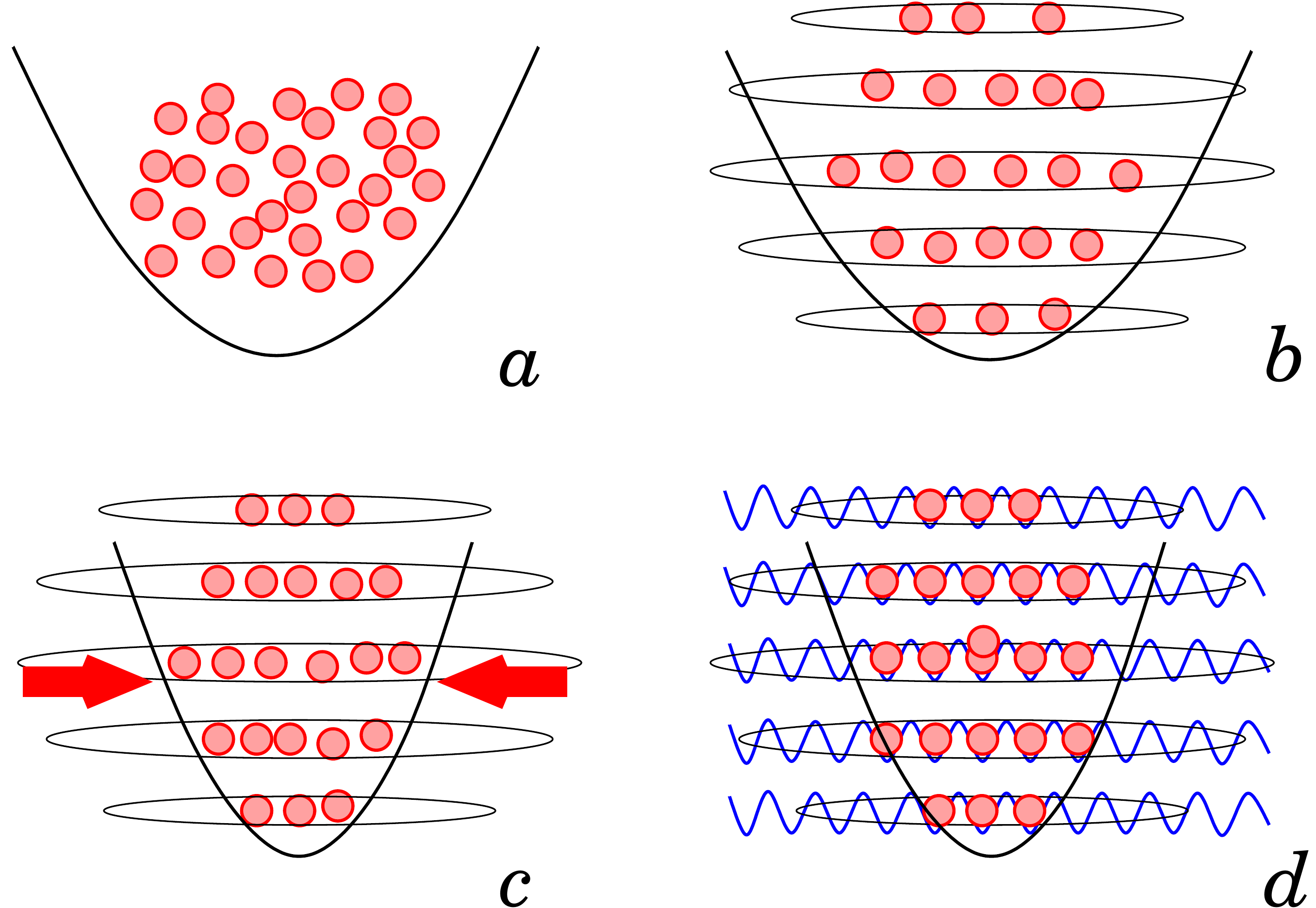} 
\caption{Sketch of the proposed loading and squeezing protocol
of atoms in a spatially anisotropic optical lattice. 
The description of the various
stages is presented in Sec.\ref{s.TF}.}
\label{f.load-squeeze}
\end{center}
\end{figure}

 \subsection{Trap squeezing protocol  for low-dimensional systems}
 \label{s.TF}

 In the case of spatially anisotropic optical lattices, allowing to
simulate the 1$d$ or 2$d$ Bose-Hubbard model \cite{Stoeferleetal04,
Spielmanetal07}, the control of the chemical potential can be 
achieved by trap squeezing along the spatial dimensions of 
interest, namely along the tubes in 1$d$-anisotropic
optical lattices and along the layers in 2$d$-anisotropic
ones. Yet, if the strength of the trapping potential
is changed before confining the atoms in tubes or in layers,
the final atomic population present in the tubes/layers will be 
significantly affected, so that one violates the condition of working 
at fixed particle number in the low-dimensional elements
(tube or layer) of the system, and in particular in the
central ones. The problem of guaranteeing at the same time
the control on the particle number in the tubes/layer and 
adiabatic trap squeezing can be easily solved by considering
the following loading protocol, sketched in Fig.~\ref{f.load-squeeze}: 

$a)$ Particles are initially trapped in a dipolar trap or a magneto-optical 
trap (MOT) at fixed strength; 

$b)$ A deep one-dimensional/two-dimensional optical lattice is then 
ramped up, defining layers/tubes in which the atoms are still
moving in continuum space. For a given initial atom number $N$ 
the overall trapping potential (dipolar trap or MOT plus the optical 
lattice beam profile) uniquely defines the populations in each tube/layer.
These populations remain essentially fixed for the rest of the experiment, 
thanks to the deep optical lattice. 
Given that the atoms are still in the weakly interacting regime,
we can determine these populations via the Thomas-Fermi approximation --
the results are given by Eqs.~\eqref{e.N1d} and \eqref{e.N2d} in  
Appendix \ref{a.TF}); 

$c)$ Varying the dipolar trapping along the tube/layer
dimensions gives rise to trap squeezing, which can be
easily kept adiabatic for weakly interacting particles
\cite{Hanetal01}; 

$d)$ At this stage the optical lattice can be adiabatically
ramped up along the tubes/layers, finally realizing a 1$d$/2$d$
Bose-Hubbard Hamiltonian at a given particle number and 
in a given longitudinal trapping potential.

 \section{Imaging techniques}
 \label{s.imaging}
 
  The two previous sections have discussed how to achieve full
 control on the chemical potential of the Bose-Hubbard model 
 realized in optical lattice experiments via the control on the 
 trapping potential at fixed particle number. 
 As discussed in Section \ref{s.LDA}, the overall chemical potential 
 of the trapped system corresponds in particular to the local
 chemical potential at the trap center: hence, getting access
 to the average central density in the trap as a function of
 the chemical potential gives the possibility of extracting 
 experimentally the density curve of the \emph{bulk} 
 Bose-Hubbard model in the grand-canonical ensemble and hence
 its compressibility.

\begin{figure}[h]
\begin{center}
\includegraphics[
width=80mm]{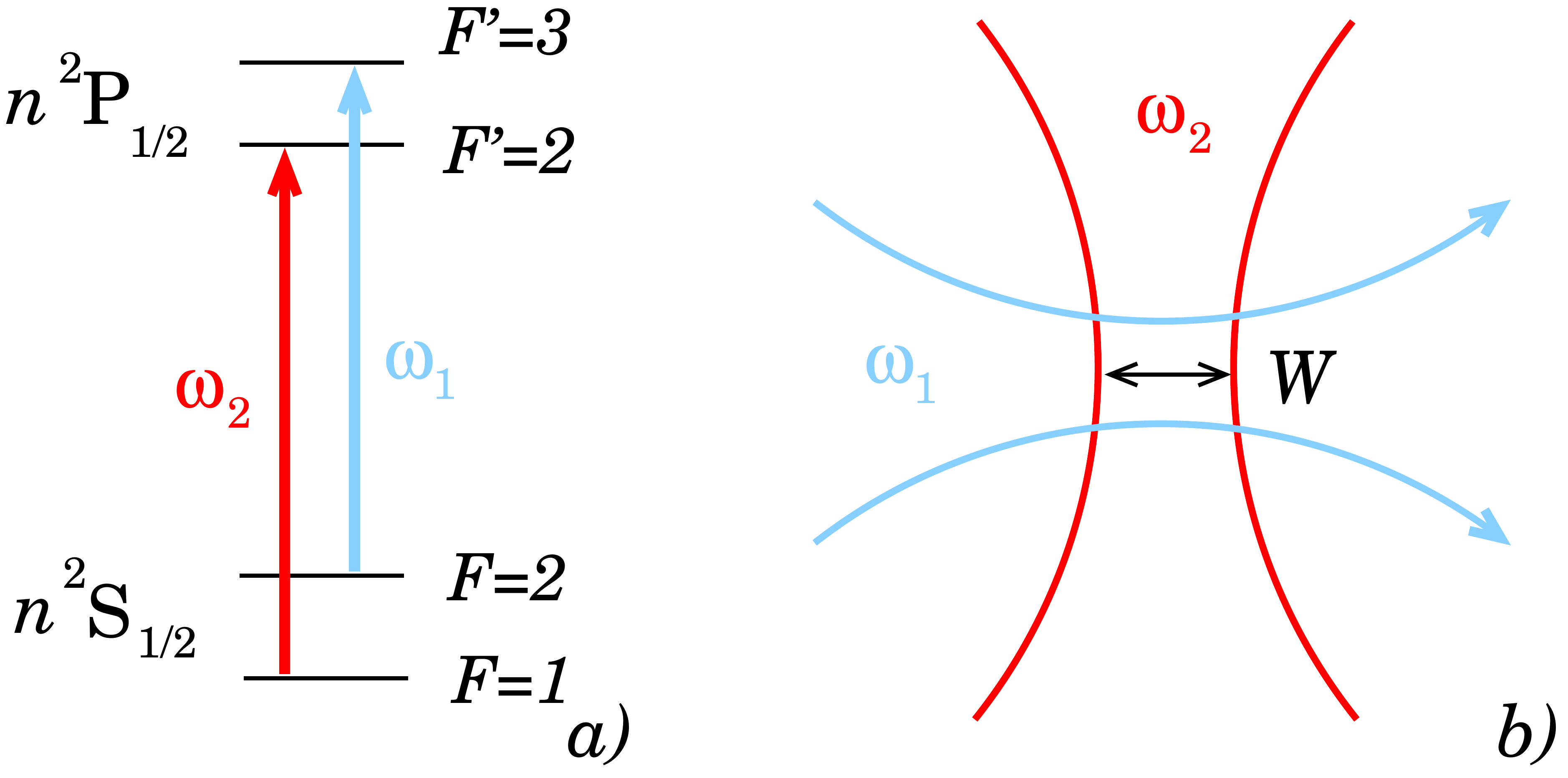} 
\caption{Sketch of the proposed imaging protocol for alkali atoms. A repumping 
($\omega_2$) and a cooling ($\omega_1$) beam are tightly 
focused on the center of the trap, and they are crossed so as
to overlap in a region of width $\approx W$ (minimal beam waist)
in all spatial directions. This setup realizes a spatially-resolved optical 
cycling transition.}
\label{f.imaging}
\end{center}
\end{figure} 
  
  As it will be further seen in the next Sections,  
  measuring an average central density which mimics closely the one 
 of the bulk system requires imaging the atomic cloud over a
 typical length scale of 10-20 lattice sites in each spatial direction
 for typical
 atom numbers and trapping frequencies used in current experiments,
 corresponding to a spatial resolution of $\sim 5-10~\mu$m.
 To achieve this resolution, a possible technique is given by
 spatially resolved fluorescence imaging for atoms 
 (see Fig.~\ref{f.imaging}). In the case of alkali 
 atoms initially prepared
 in an $F=1$ hyperfine state of the $n ^{2}$S$_{1/2}$ level, 
 a repumping beam, resonant with the transition to an $F'=2$ state 
 of the $n ^{2}$P$_{3/2}$ level, optically pumps the atoms
 into an $F=2$ state \cite{Ketterleetal93}; a second beam resonant 
 with the $F=2 \to F'=3$
 transition (cooling beam) is then used to image the atoms.
 If the two beams are tightly focused to a beam waist of 
 $\sim 5-10~\mu$m and are perpendicular to each other such 
 that they meet at the focal point, they selectively
 image atoms in a region of space of the desired size. 
 Typical optical lattice experiments have $n>1$ particles
 per lattice site in the trap center, so that imaging
 a central region with $10^3-20^3$ lattice sites
 involves imaging at least as many atoms, which is possible 
 over a time $\lesssim$ 100 ms (see \emph{e.g.} 
 Ref.~\onlinecite{Chuuetal05}). Before imaging, the optical 
 lattice could be rapidly increased to the maximum height 
 in order to freeze the atomic cloud profile during the
 successive imaging time.  
  
   The proposed imaging technique has the advantage of being 
   generally compatible with common atom trapping setups
   based on magneto-optical or dipolar traps, and to 
   require only an intermediate imaging resolution. On the 
   other hand, ultra-high resolution optics has given 
   access to few-site/few-atom or even single-site/single-atom imaging, 
   in a variety of very recent experiments 
   \cite{Nelsonetal07, Gerickeetal08, Gemelkeetal09, Bakretal09, Bloch09}. 
   This level of resolution far exceeds the one required by our
   present proposal.  An alternative technique for the indirect extraction of the local
   density in the trap consists in the high resolution \emph{in-situ} 
   imaging of the atomic cloud density integrated along the line
   of sight, from which the full three-dimensional atomic 
   distribution is reconstructed via inverse Abel transformation
   \cite{Shinetal08, Liaoetal09}. Given the diversity of techniques
   mentioned above,  we are confident that the extraction of the local density 
   properties in selected regions of the trap will become an experimental
 routine in the near future. 
 
  After the \emph{in situ} imaging stage, turning off all trapping
 potentials and imaging the expanded cloud gives
 access to the total number of atoms. This piece of information is
 fundamental to \emph{post-select} the measurements with a 
 given total atom number $N$, which corresponds to the desired
 value of the chemical potential to be realized in the experiment. 
 Strictly speaking, once the calibration curve relating $\mu$ to 
 $N$ is known, the final measurement of the particle number
 has only the role of assigning the measurement of the central
 density to the proper place in the grand-canonical phase diagram
 of the bulk system. Hence one can regard the fluctuations
 in the total particle number $N$ as a source of random sampling 
 of the $\mu$ axis in that phase diagram. In this perspective,  
 all measurements give useful information, provided that 
 particle number fluctuations are not bringing the chemical 
 potential $\mu$ too far from the region of interest in the phase diagram. 
   
\begin{figure*}[ht!]
\begin{center}
\mbox{ \includegraphics[
width=0.5\textwidth]{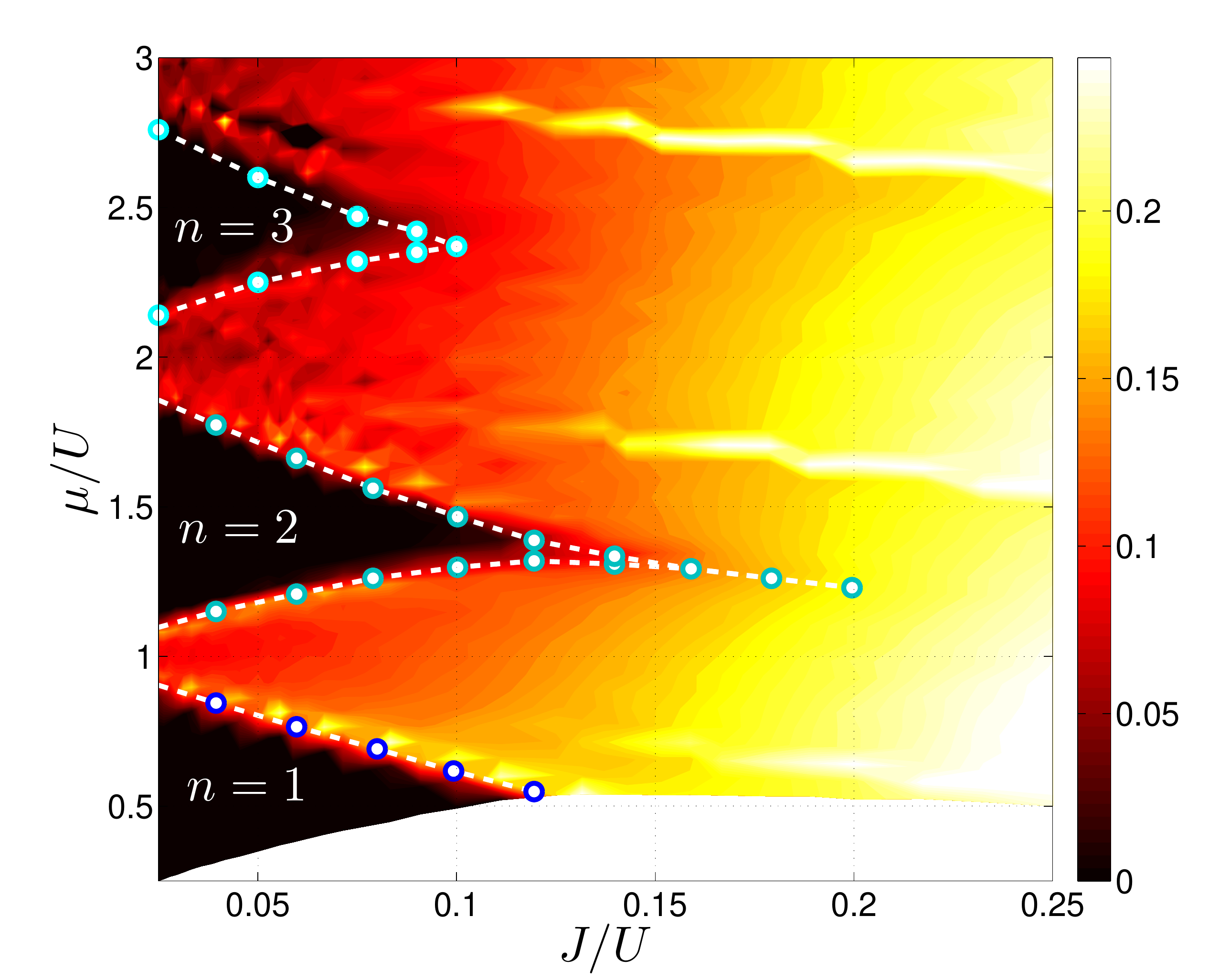} \hspace{-2.7cm} {\large $\kappa_{_C} U$, $10$ sites} 
 \includegraphics[
width=0.5\textwidth]{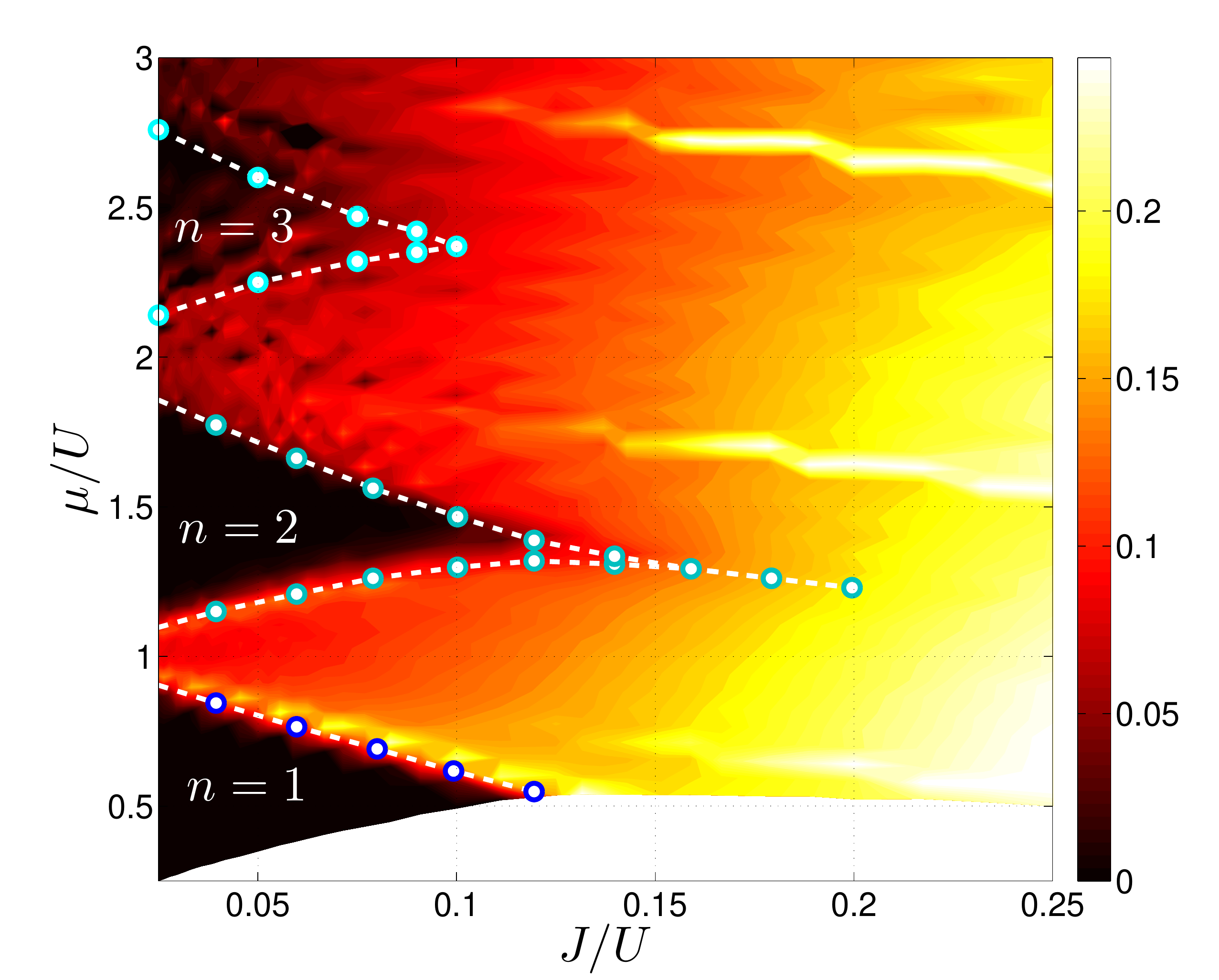} \hspace{-2.7cm} {\large $ \kappa_{_C} U$, $20$ sites}} 
 \caption{Central compressibility $\kappa_{_C} = \partial n_{_C} / \partial \mu$ (in units of $U^{-1}$)
 of $N=100$ bosons under variable trapping frequency. The central density $n_{_C}$ is obtained
 by imaging over 10 sites (left panel) and 20 sites (right panel). The white curves report the boundary
 lines between superfluid and Mott-insulating phases as determined in Ref. \onlinecite{Rizzietal05}
 (for the $n=1$ and $n=2$ lobe) via density-matrix renormalization group, and 
 independently in this work (for the $n=3$ lobe) via quantum Monte Carlo.  } 
\label{f.kappa1DBH}
\end{center}
\end{figure*}

 \section{Applications: phase diagram of the 1$d$ Bose-Hubbard model}
 \label{s.1DBH}
 
As a first application of trap squeezing, we investigate the ground-state
phase diagram of the one-dimensional Bose-Hubbard model. 
We consider the situation of $N=100$ bosons trapped in the
central tube of a strongly anisotropic optical lattice \cite{Stoeferleetal04}, 
and addressed with spatially resolved imaging described in the
previous section. We present quantum Monte Carlo results for the 
central properties in the trap, as well as for the global time-of-flight
properties. The chemical potential $\mu$ is adjusted so that
the ground state of the grand-Hamiltonian, Eq.~\ref{e.Hmu},  
has the desired number of bosons. 

\begin{figure}[h]
\begin{center}
\includegraphics[
width=84mm]{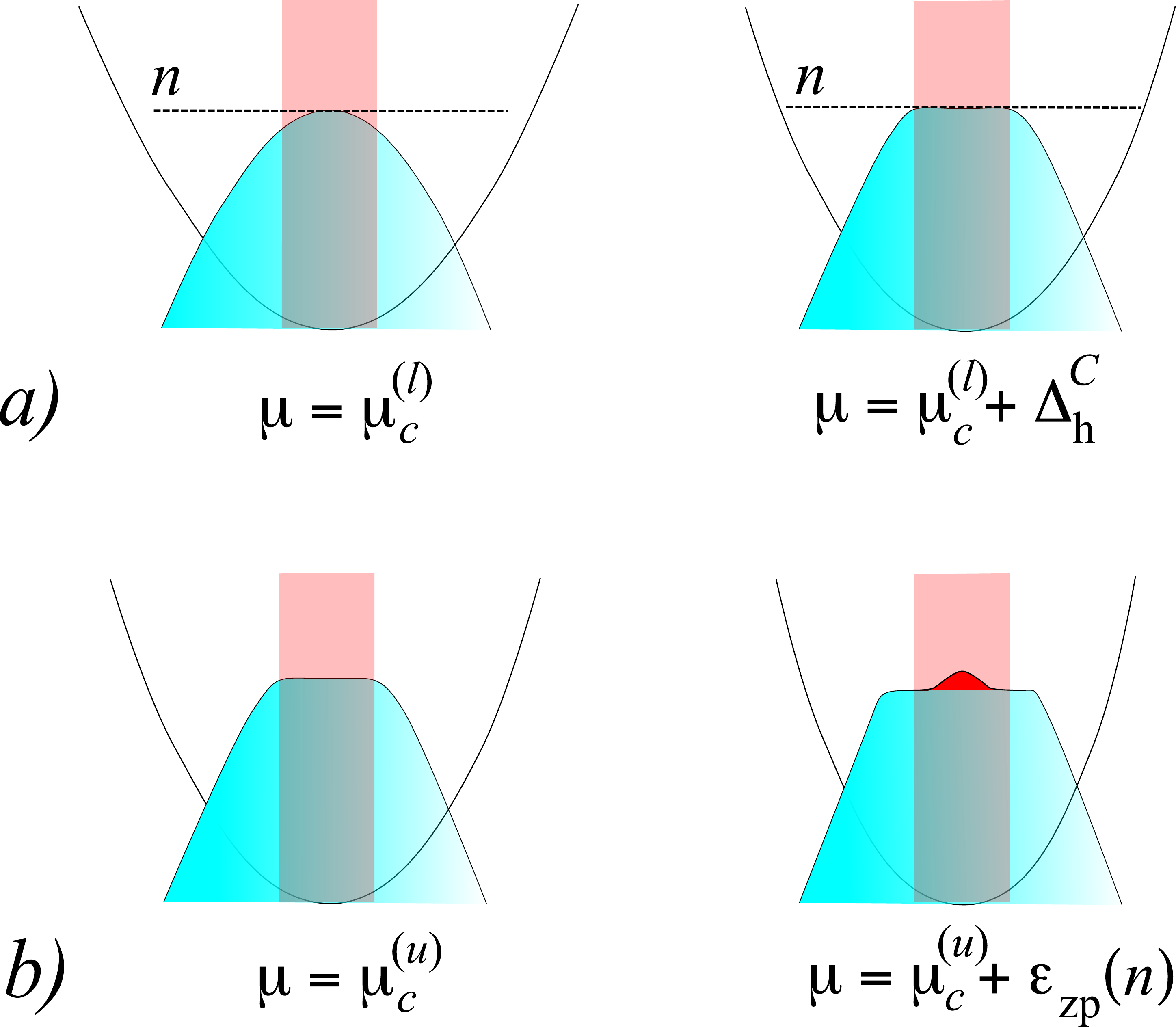}
\caption{Mechanisms of blue shift of the Mott lobes obtained from the 
central compressibility. $a)$ At the lower critical chemical potential
$\mu_c^{(l)}$ a Mott region only forms on the central site, and 
one needs an increase of the chemical potential by a term
 $\Delta_{\rm h}^C$ to establish a Mott phase throughout the
 imaging region (pink-shaded).
 $ b)$ At the upper critical potential $\mu_c^{(u)}$ of the 
 bulk system, the trapped system  has not yet developed 
 a superfluid core, because of the additional 
 zero-point energy $\varepsilon_{\rm zp}(n)$ due
 to trapping in the harmonic oscillator potential. 
 Only when the chemical potential reaches
 the blue-shifted value $\mu_c^{(u)} + \varepsilon_{\rm zp}(n)$,
 the first particle of the superfluid core appears (marked in red).}
\label{f.blueshift}
\end{center}
\end{figure}

\subsection{Central compressibility}

 Investigating the phase diagram of the Hubbard model via trap 
 squeezing implies that the superfluid-insulator transitions 
 that one can probe are accessed via a variation of the chemical potential, 
 namely they are of the commensurate-incommensurate kind in a
 system without disorder. Such transitions have the advantage
 of being very sharp, as they are characterized roughly
 speaking by the appearance (or disappearance) of a
 superfluid component in the system, induced by doping
 particles or holes into a Mott insulating state at integer
 filling. The superfluid-insulator
 transition can be completely characterized by the compressibility 
 $\kappa = \partial n/ \partial \mu$, which, similarly to the 
superfluid density, is identically zero in the Mott-insulating phase
and finite in the superfluid phase, for all spatial dimensions. 
 In particular the compressibility has quite a sharp variation
 at the transition in low-dimensional systems. Indeed its critical scaling 
 with system size $L$ 
 obeys the law $L^{d-z}$ \cite{Fisheretal89} where $z=2$ is the dynamical
 critical exponent; hence, for $d=1$, it jumps from a value diverging
 linearly with system size to zero at the transition point (while in
 $d=2$ the jump takes place from a value diverging logarithmically 
 with the system size). This very sharp feature makes it possible to 
 detect the transition unambiguously even on a small system size, 
 which is effectively the case here. 
In the following we deliberately limit our attention to the $n>1$ region of
the phase diagram, as this is a typical situation in the experiments.

 We focus our attention on the behavior of the \emph{central compressibility},
\begin{equation}
\kappa_{_C} = \frac{\partial n_{_C}}{\partial \mu}~
\end{equation}
corresponding to the response of the density in the 
central region $C$ to a variation of the chemical potential, 
driven exclusively by a change in the trapping potential
(which leaves the parameters $J$ and $U$ unchanged).
Upon choosing a sufficiently narrow central region  $C$ satisfying 
the condition of Eq.~\ref{e.quasih}, the central density is supposed to 
reproduce closely the behavior of the compressibility in
 the bulk system. Fig.~\ref{f.kappa1DBH} shows indeed that this
 is the case. In fact we observe that the central compressibility 
 jumps from from a finite value to zero for critical values of the chemical 
 potential which reproduce closely the shape of the Mott-insulating
 lobes of the \emph{bulk} phase diagram of the 1$d$ Bose-Hubbard
 model. (The Mott insulating boundaries for the bulk system have been determined 
 in Ref.~\cite{Rizzietal05} via density-matrix renormalization group
 for the $n=1$ and $n=2$ lobes, while we have determined
 the $n=3$ lobe via quantum Monte Carlo). 
 Remarkably, upon doubling the central region 
$C$ from 10 to 20 sites (namely halving the resolution of the 
imaging in the experiment), the main features of the reconstructed
phase diagram are only slightly altered, at least for what 
concerns the lobes $n=1$, $2$. This shows the robustness of 
the proposed approach under realistic experimental conditions. 
An interesting phenomenon to be observed in Fig.~\ref{f.kappa1DBH} 
is the "blue shift" (towards higher chemical potentials) of the 
$n=2$ and $n=3$ Mott lobes 
predicted by the vanishing of $\kappa_{_C}$ with respect to what is observed 
in a bulk system. This shift can be easily understood as a result of the
finite width of the imaging region and as an effect of the finiteness of 
the central region exhibiting bulk physics, and, as we will discuss, it can 
be corrected for systematically. 

\emph{Blue shift of the lower lobe boundaries}. The lower lobe boundaries are
associated with a SF-MI transition induced by an increase of the chemical 
potential above a lower critical value $\mu^{(l)}_c(J/U)$. A chemical potential 
$\mu > \mu^{(l)}_c$ forces the atom density
to an integer value, for which a Mott gap opens if $J/U < (J/U)_c$
corresponding to the lobe tip. When this phenomenon occurs in the center
of the trap, the MI region which appears at first is extremely 
narrow in space (see Fig.~\ref{f.blueshift}(a) for a sketch). 
Indeed it extends over a region of width $\Delta r$ 
such that the chemical potential modulation due to the trap over this 
region is lower than the MI hole gap $\Delta_{\rm h}$, namely
$V_t (\Delta r/2)^2 \approx \Delta_{\rm h}$. 
The central compressibility detects the appearance of an incompressible 
MI phase in the trap center only when $\Delta r$ reaches the size $|C|$ 
of the imaging region, namely for a critical bulk gap 
$\Delta^{C}_{\rm h} \approx V_{t,C} (|C|/2)^2$. 
Here $V_{t,C}$ is the trapping
 potential corresponding to the effective chemical potential
 $\mu(V_{t,C}, N, J, U)$ at which the hole
 gap takes the value $\Delta^{C}_{\rm h}$. 
 This chemical potential is simply given by $\mu^{(l)}_c + \Delta^{C}_{\rm h}$.
 This implies that, for a strong confinement $V_t$, as that required to 
obtain $n=2$ or $n=3$ MI lobes in the system, one observes the onset of a MI phase 
for a blue-shifted critical chemical potential 
\begin{equation}
(\mu^{(l)}_c)'  =  \mu^{(l)}_c + \Delta^{C}_{\rm h}~.
\label{e.mulshift}
\end{equation}
Remarkably this shift depends exclusively
on experimentally known parameters $V_t$ and $|C|$, so that one can 
correct \emph{a posteriori} for it when determining the phase 
diagram experimentally. 

\emph{Blue shift of the upper lobe boundaries}. The upper lobe boundaries 
are associated with a MI-SF transition, induced by an increase of the chemical 
potential. Reaching an upper critical value $\mu^{(u)}_c$, 
the chemical potential closes the Mott gap of the MI with $n$ particles 
per site, and it therefore adds the first particles on top of the Mott insulator.
In a bulk system these particles are essentially added in a $k=0$
state, and they undergo condensation. In a trap, the corresponding
phenomenon would be the appearance, induced by trap
squeezing, of the first particle over the 
central Mott plateau with $n$ particles per site (see Fig.~\ref{f.blueshift}(b)). 
At variance with the bulk case, these particles are not fully delocalized, 
but they still feel the overall trapping potential - hence they have a 
residual zero-point energy, due to the confinement in the trap.
To estimate this shift in the zero point energy, we can proceed in the
spirit of the extended fermionization approach \cite{Pupilloetal06}, 
(see Sec. \ref{s.protocol} and Appendix~\ref{a.fermionization}, and use the approximation
of representing the extra particle added in the trap center as moving 
on an inert background at fixed density $n$, whose only effect is to increase the hopping
amplitude of the particle from $J$ to $J(n+1)$. Hence the zero point
energy can be estimated from the solution of a harmonic oscillator
of effective mass $m^* = \hbar^2/[2J(n+1)]$ and frequency
$\hbar \omega = \sqrt{4(n+1) V_t J}$. This estimate is valid as long
as the width of the central MI plateau is larger than 
the width of the ground state of the harmonic oscillator, 
namely $\sigma = [J(n+1)/V_t]^{1/4}$. Hence the shift in 
zero-point energy due to confinement is 
$\varepsilon_{\rm zp}(n) =  \hbar \omega/2 = \sqrt{(n+1) V_t J}$.
This leads to a shift in the critical chemical potential needed to 
``dope" a particle onto the central MI plateau with respect to 
that needed in a bulk system; as a consequence, the upper critical chemical
potential $\mu_c^{(u)}$ measured by the central compressibility
is blue-shifted as
\begin{equation}
(\mu^{(u)}_c)'  =  \mu^{(u)}_c + \varepsilon_{\rm zp}(n)~.
\label{e.muushift}
\end{equation}
Similarly to what observed for the lower lobe boundary, 
one can correct \emph{a posteriori} for this shift using the experimental
parameters, and hence reconstruct 
the bulk value of  $\mu^{(u)}_c$.

\begin{figure}[h]
\begin{center}
\includegraphics[
width=94mm]{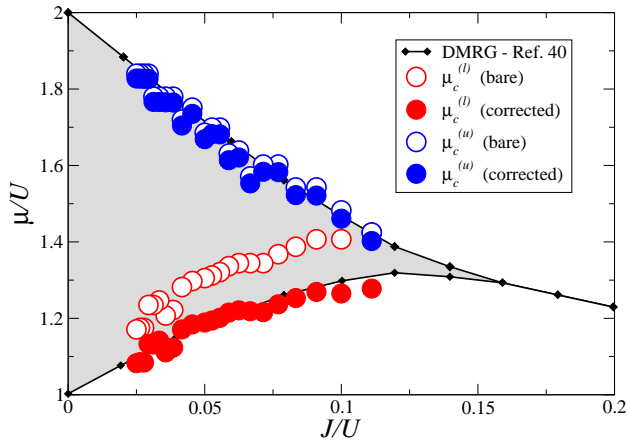}
\caption{$n=2$ Mott lobe of the 1$d$ Bose-Hubbard model as determined by trap squeezing
of $N=100$ bosons, imaged over the central 20 sites. The open circles correspond to the bare values of 
the lower and upper critical chemical potentials extracted from the 
vanishing of the central compressibility for $|C|=20$ (Fig.~\ref{f.kappa1DBH}); the 
closed circles correspond to the values corrected for blue shift, according 
to Eqs.~\eqref{e.mulshift} and \eqref{e.muushift}. The correction leads to a very
good agreement with the DMRG results of Ref.~\onlinecite{Rizzietal05}.}
\label{f.MIn2lobe}
\end{center}
\end{figure}

 Fig.~\ref{f.MIn2lobe} shows the lower and upper critical chemical potentials 
 of the $n=2$ MI lobe, as extracted from the vanishing of the central 
 compressibility of $N=100$ bosons imaged over the central 20 sites, 
 and compared with the numerically exact results of Ref.~\onlinecite{Rizzietal05}. 
 The blue shift of the lower critical chemical potential is quite sizable, but 
 it is seen to be very well corrected for by the prediction of Eq.~\eqref{e.mulshift}. 
 The blue shift of the upper critical chemical potential is instead much weaker, in 
 agreement with the prediction of Eq.~\eqref{e.muushift}. 
 This shows that trap squeezing allows to reconstruct quite accurately
 the quantum phase transition lines of the Bose-Hubbard model, even 
 in the case of a coarse imaging of a central region of $\sim 20$ sites in 
 the trap center. 
 
  Trap squeezing supplemented with imaging over
 $|C|$ sites can reliably detect MI behavior only as long as the MI plateau
 occurring in the trap center can reach the same width as that of the central  
 $C$ region: this condition guarantees that $(\mu^{(l)}_c)' < (\mu^{(u)}_c)'$, 
 namely that, upon increase the chemical potential, the (blue-shifted) lower 
 chemical potential is reached before a new superfluid core forms.
 The MI region can no longer be reliably observed if $V_t$ reaches
 the value $\bar{V}_t$ satistying the condition $(\mu^{(l)}_c)' = (\mu^{(u)}_c)'$,
 namely
 \begin{equation}
 \bar{V}_t~(|C|/2)^2 - \sqrt{(n+1)\bar{V}_t J} = \mu_c^{(u)} - \mu_c^{(l)} = \Delta_{\rm ph}
 \end{equation}
  where $\Delta_{\rm ph}$ is the particle-hole gap of the Mott insulating phase. 
  The above condition gives
  \begin{equation}
  \bar{V}_t = \frac{\Delta_{\rm ph}}{R^2} - \frac{\sqrt{(n+1) J \Delta_{\rm ph}}}{R^3} + 
  \frac{(n+1) J}{2 R^4} + {\cal O} \left( \frac{1}{R^5} \right)
  \end{equation}
  where $R = |C|/2 \gg 1$. The merging of the blue-shifted lower Mott lobe boundary with 
  the upper one is seen to occur \emph{e.g.} in Fig.~\ref{f.MIn2lobe} for $J/U \gtrsim 0.12$.
  To be able to observe the tip of the Mott lobe via trap squeezing, it would be necessary 
  to increase the total particle number so that $V_t$ can be decreased below $\bar{V}_t$,
  maintaining the same chemical potential in accordance with Eq.~\eqref{e.qTF}.  
   
  So far we have focused our discussion on the spectral properties
  associated with the compressibility. In the following we will also discuss
  the evolution of the coherence properties under trap squeezing, as an
  alternative tool to investigate the phase diagram of the system.

\begin{figure*}[ht!]
\begin{center}
\mbox{ \includegraphics[
width=0.5\textwidth]{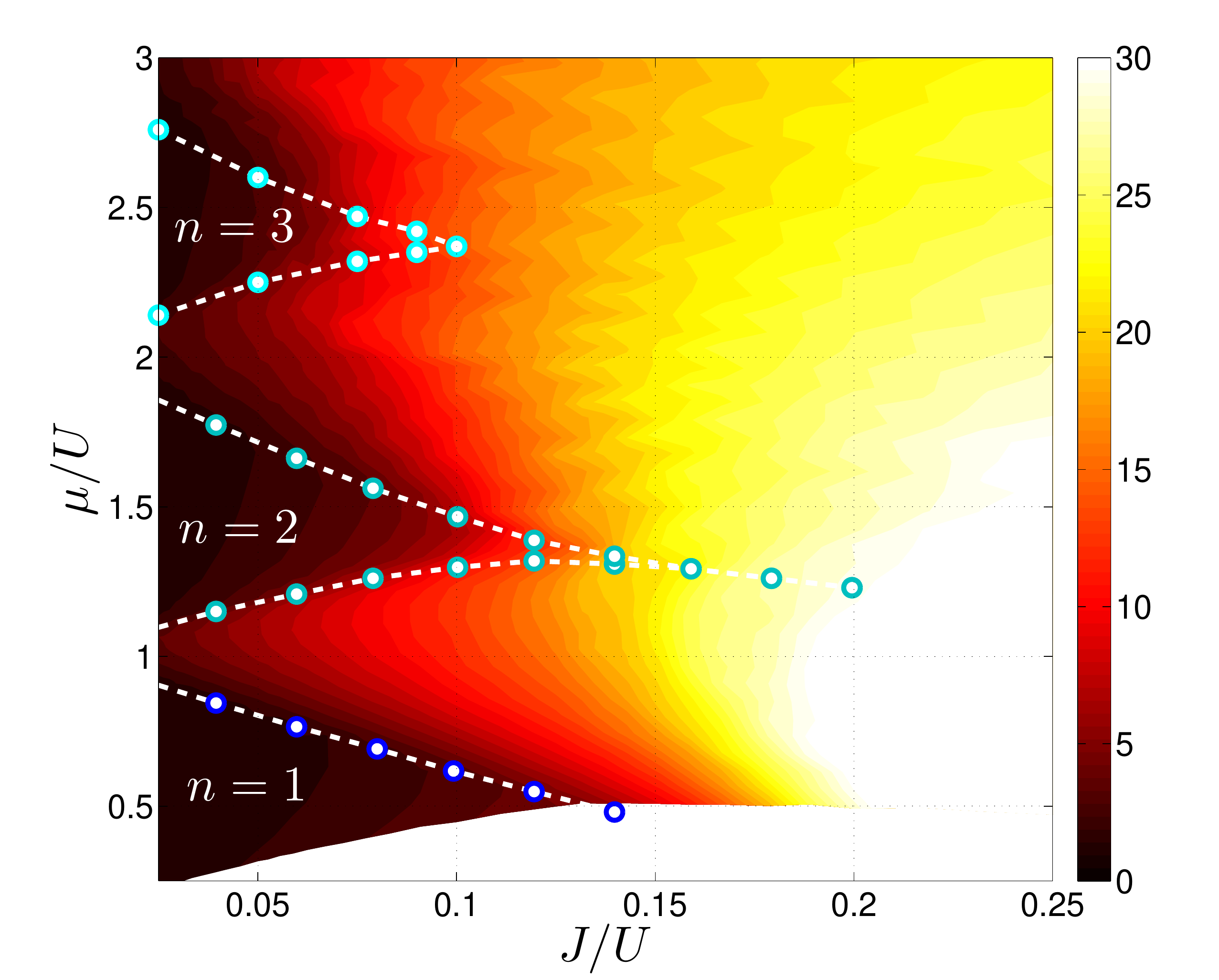} \hspace{-2.2cm} {\large $n(k=0)$} 
 \hspace{0.5cm} \includegraphics[
width=0.5\textwidth]{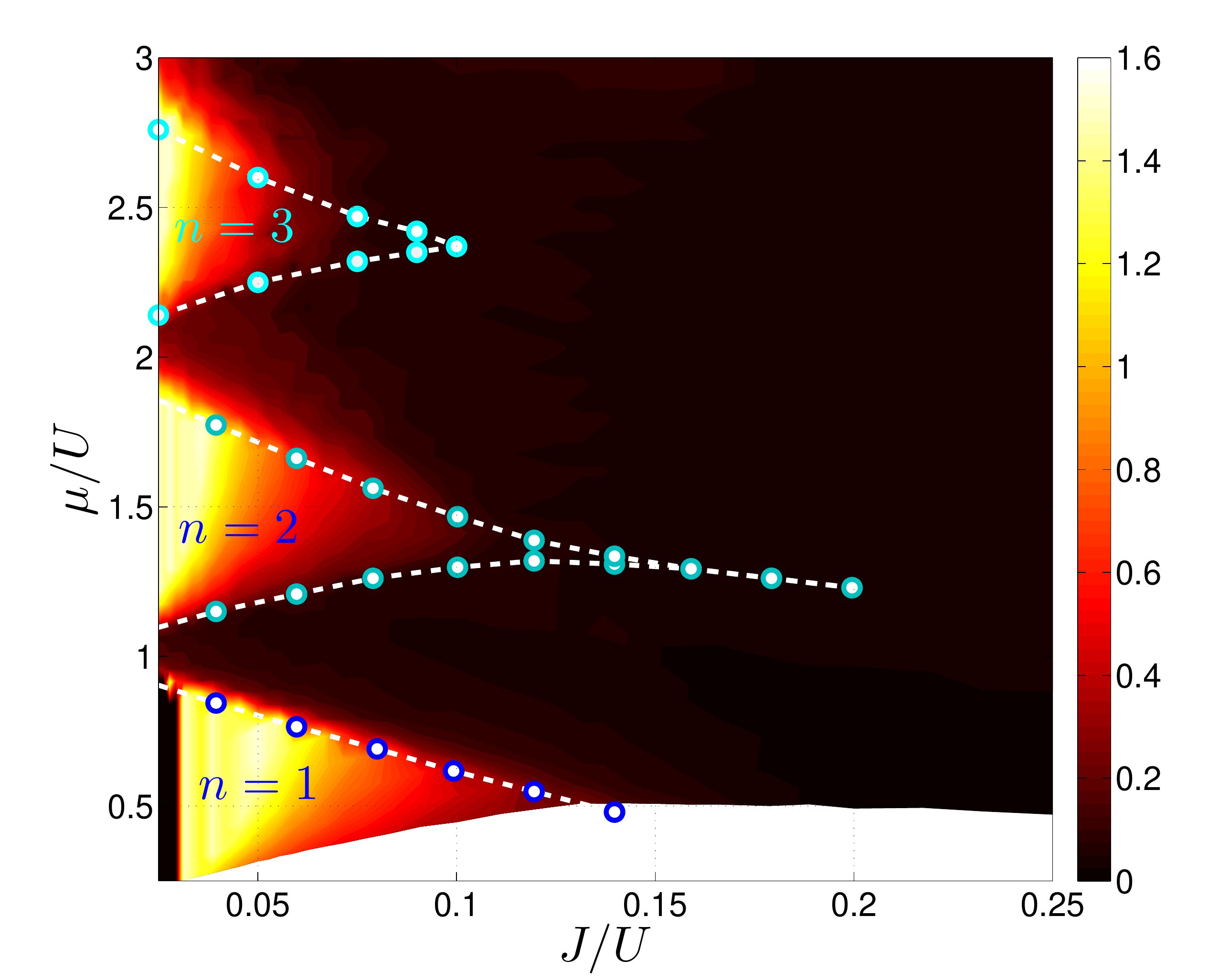} \hspace{-2.2cm} {\large FWHM ($a^{-1}$)}} 
 \caption{Evolution of time-of-flight observables for $N=100$
 bosons under variable trapping frequency. The left panel depicts the 
 condensate number $n_{_{k=0}}$, and the right panel the full width
 at half maximum (FWHM) of the momentum distribution $n_k$,
 in units of $a^{-1}$ where $a$ is the lattice spacing.} 
\label{f.TOF1DBH}
\end{center}
\end{figure*}

\subsection{Time-of-flight observables and condensate compressibility}

At variance with diagonal observables such as density and compressibility, 
which can be measured locally, 
the phase coherence properties, encoded in the momentum distribution,
are generally measured at the global level. We define the momentum distribution as 
\begin{equation}
n(k) = \frac{1}{N} \sum_{ij} e^{ik(r_i-r_j)} \langle a_i^{\dagger} a_j \rangle~.
\label{e.nk}
\end{equation}
where the normalization to the boson number $N$ is chosen so as to 
account for the actual spatial extent of the cloud  \cite{noteonbox}. 
The momentum distribution is probed via time 
of flight measurements, by collecting the interferogram of the matter wave emitted
from all sites of the optical lattice. Given the strong inhomogeneity of the 
trapped system, one can suspect that the momentum distribution
of a trapped gas is generally not a faithful portrait of the corresponding
distribution of a homogeneous system with the same global chemical 
potential. Yet some remarks are in order. For any given tolerance $\epsilon$,
one can select a central region $C$ satisfying Eq.~\eqref{e.quasih}, 
and a complementary region $\bar{C}$. 
Hence the momentum distribution can be decomposed into three 
contributions, two of which involve the central region $C$
\begin{equation}
n(k) = n_C(k) + n_{C,\bar{C}}(k) + n_{\bar{C}}(k)
\end{equation}
where
\begin{equation}
n_{C(\bar{C})} (k) = \frac{1}{N} \sum_{i,j \in C (\bar{C})}  e^{ik(r_i-r_j)} \langle a_i^{\dagger} a_j \rangle
\end{equation}
and
\begin{equation}
n_{C,\bar{C}} (k) =  \frac{1}{N} \left( \sum_{i \in C, j \in \bar{C}}  +\sum_{i \in \bar{C}, j \in C} \right)  e^{ik(r_i-r_j)} \langle a_i^{\dagger} a_j \rangle
\end{equation}
We would like to argue in the following that the behavior of the global 
momentum distribution is strongly dependent on the contribution $n_C(k)$
from the central region. Indeed the central region of the trap 
is the only \emph{simply} connected region, which implies that
the pairs of sites $(ij)$ associated with this region are the ones at the 
shortest mutual distance $|r_i - r_j|$. This latter topological aspect
has strong consequences on the coherence properties of the global 
system. Indeed one can 
envision in general two situations: 1) If the central region $C$ is 
locally in a SF state, its coherence properties are going to give 
the dominant contribution to the sum of Eq.~\eqref{e.nk}, because
$n_C(k)$ contains the largest number of strongly correlated
pairs $(ij)$ at short distance; moreover if the wings are SF too, 
the central region will also contribute with the $n_{C,\bar{C}}(k)$
term, which generally dominates over $n_{\bar{C}}(k)$
because the central region is at a higher density than 
its complement $\bar{C}$.									 
2) If the central region is in a local MI phase, extending
over a region of linear size $\sim R_C$, there might still be 
a SF halo in the region $\bar{C}$, but it will give a weak
contribution to the $n(k)$ because the pairs of sites $(ij)$ 
involved in $n_{\bar{C}}(k)$ are separated by a distance 
$\sim R_C$, and sites separated by the central MI core are 
very weakly correlated, given that particles cannot coherently propagate
across $C$. An equivalent statement to the latter one is that
in a $d$-dimensional system the SF halo surrounding
a MI region is effectively quasi-$(d-1)$-dimensional, and
consequently it has reduced coherence properties, due
to the enhanced role of quantum fluctuations in lower
dimensions. Hence it is reasonable to believe that the 
momentum distribution mainly reflects the coherence 
properties of the central $C$ region, at least for one-
and two-dimensional systems. 
 
\begin{figure}[h]
\begin{center}
\mbox{
\includegraphics[
width=0.5\textwidth]{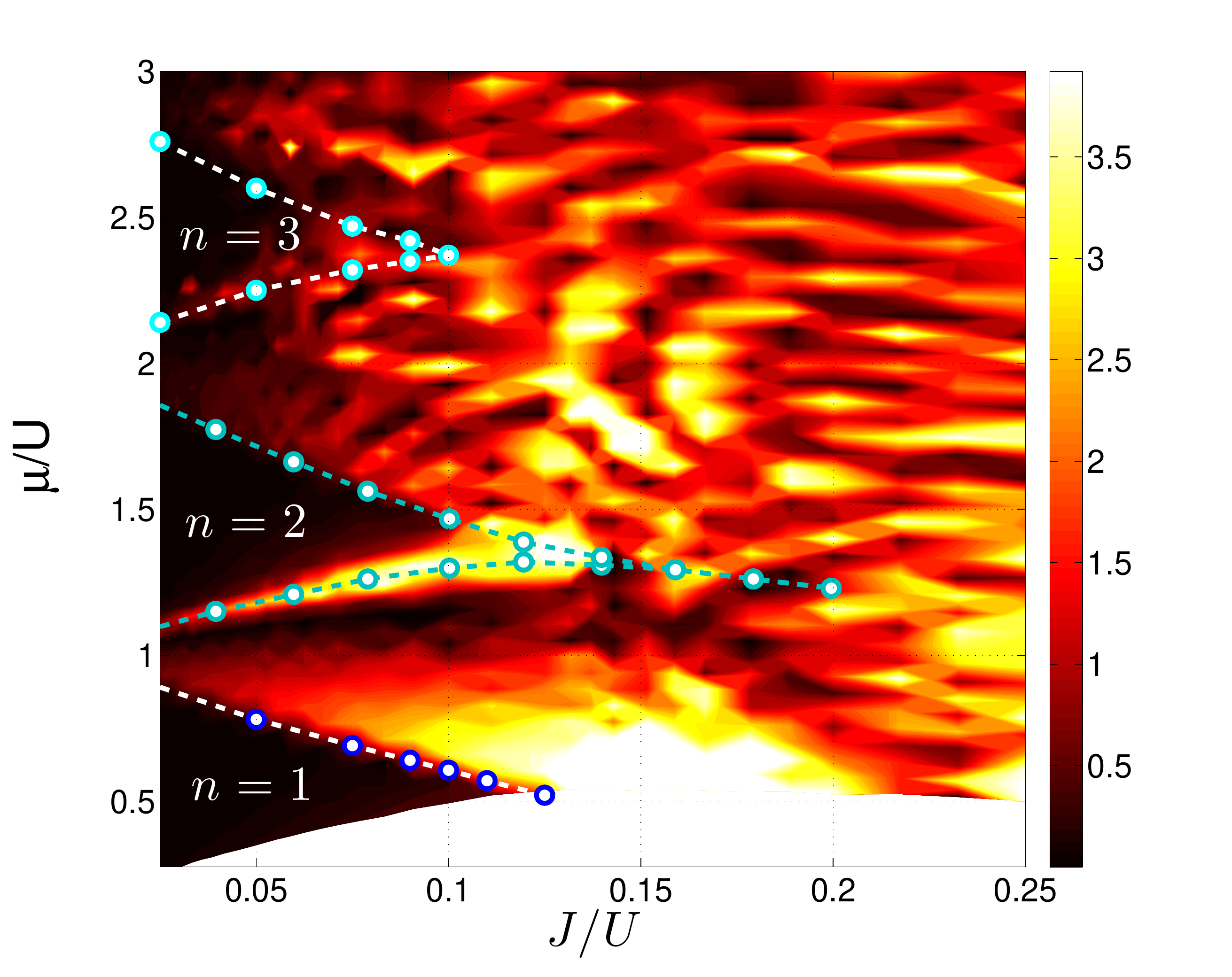} \hspace{-1.9cm} {\large $|\kappa_{_0}| U$}}
\caption{Absolute value of the condensate compressibility $\kappa_{_0} = \partial n(k=0) / \partial \mu$
(in units of $U^{-1}$) of $N=100$ bosons under variable trapping frequency.}
\label{f.denk0dmu1DBH}
\end{center}
\end{figure} 
   
Fig.~\ref{f.TOF1DBH} shows the height of the $k=0$
peak and the full width at half maximum (FWHM) of the
$n(k)$ in a system of $N=100$ bosons under trap 
squeezing control of the chemical potential, and for various 
optical lattice depths. Plotting these quantities
on the $(J/U, \mu/U)$ plane, one sees that sharp drops in the 
global coherence of the trapped system are observed 
close to the SF-MI boundary lines of the bulk phase 
diagram of the Bose-Hubbard model. As in the case of
the global compressibility, for high enough density
the sharpest features in the coherence properties
are observed for blue-shifted values of the 
chemical potential \cite{noteonRigol}. This result seems to confirm our 
expectation that the global properties of the momentum
distribution are strongly dominated by the behavior 
of the system at the trap center, which best approximates 
the bulk behavior. 

 To make this connection even more quantitative, we can 
 consider again the partition of the system into a central
 $C$ region and a complementary peripheral $\bar{C}$
 region as a partition of the trapped gas into a 
 system and a particle reservoir, respectively (Fig.~\ref{f.sys+res}). 
 Trap squeezing increases the chemical potential
 of the reservoir with respect to that of the system, and
 it hence pumps particles from the former to the 
 latter. If the system experiences (quasi-)condensation,
 a (quasi-)macroscopic fraction of the pumped particles 
 will enter the (quasi-)condensate. Alternatively, if the central
 region has a density which approaches an 
 integer value from below, $n_C < {\rm int}(n_C) + 1/2$,  
 condensation should be rather thought of in terms of holes  
 (of density ${\rm int}(n_C) + 1 - n_C$); in that case
a (quasi-)macroscopic fraction of the holes removed from the 
center under trap squeezing belongs 
to the (quasi-)condensate. In both cases, if the trap center is 
in a local (quasi-)condensate state, one expects that
the variations of the $n(k=0)$ peak in the momentum 
distribution will be proportional to (for true condensation)
or will go as a power law of (for quasi-condensation) the 
variation of the density in the trap center. 
On the other hand, if the center of the trap is in a MI
insulating phase, this prevents the particle (hole)
transfer, and it inhibits the variations of the
momentum distribution. 

From the above arguments, one expects the
behavior of the momentum distribution to be extremely
sensitive to trap squeezing; this sensitivity can be 
quantified by the \emph{condensate compressibility}
\begin{equation}
\kappa_0 =: \frac{\partial}{\partial \mu} n(k=0)~
\end{equation}
which can capture the (quasi-)macroscopic transfer
of particles from the $\bar{C}$ reservoir into a 
condensed state in the center $C$ (or viceversa for
hole condensation). The condensate compressibility
is positive for particle (quasi-)condensation in the trap 
center, and negative for hole (quasi-)condensation. Consequently
$|\kappa_0|$ is expected to be finite in presence of condensation 
in $C$ (apart from the zeros corresponding to the passage from 
particle to hole condensation), and to be essentially
vanishing corresponding to MI behavior.  Fig.~\ref{f.denk0dmu1DBH}
confirms this expectations, showing that $|\kappa_0|$ is
vanishing corresponding to the bulk MI regions (apart from the usual
blue shift of the critical chemical potentials) while it has
generally finite values corresponding to the SF regions. 
The behavior in the SF regions is highly inhomogeneous, 
due to the frequent passage from particle to hole
condensation. In any case, one can easily tell apart
the accidental vanishing of the condensate compressibility
in a SF regime from the systematic vanishing in the MI 
regime \emph{e.g.} by inspection of the FWHM, which is 
much smaller than $2\pi/a$ in the SF case, and of 
the order of $2\pi/a$ in the MI case ($a$ being the lattice
spacing).

\section{Applications: Bose glass in a three-color superlattice} 
 \label{s.3color}
 
 \subsection{Signatures of a Bose glass}
 
 In the previous section we have seen that trap squeezing
 allows to faithfully reconstruct the phase diagram of the 
 ordinary Bose-Hubbard model, either via the measurement
 of the diagonal properties in the central region of the
 trap, or via the measurement of the evolution of the 
 global off-diagonal properties (momentum distribution). 
 This latter aspect is due to the fact that spectral properties
 and coherence properties can both characterize fully
 the two phases (SF and MI) present in the phase diagram
 of the Bose-Hubbard model. On the other hand, spectral
 and coherence properties become complementary pieces
 of information in presence of disorder in the system, 
 inducing a novel phase in the phase diagram: the 
 Bose glass \cite{Fisheretal89}.  This extra phase
 is in fact characterized by a finite compressibility 
 coexisting with short-range phase coherence, and it
 needs the joint measurement of local diagonal properties
 and global momentum distribution to be detected. 
 
 \begin{figure}[h]
\begin{center}
\includegraphics[
width=60mm]{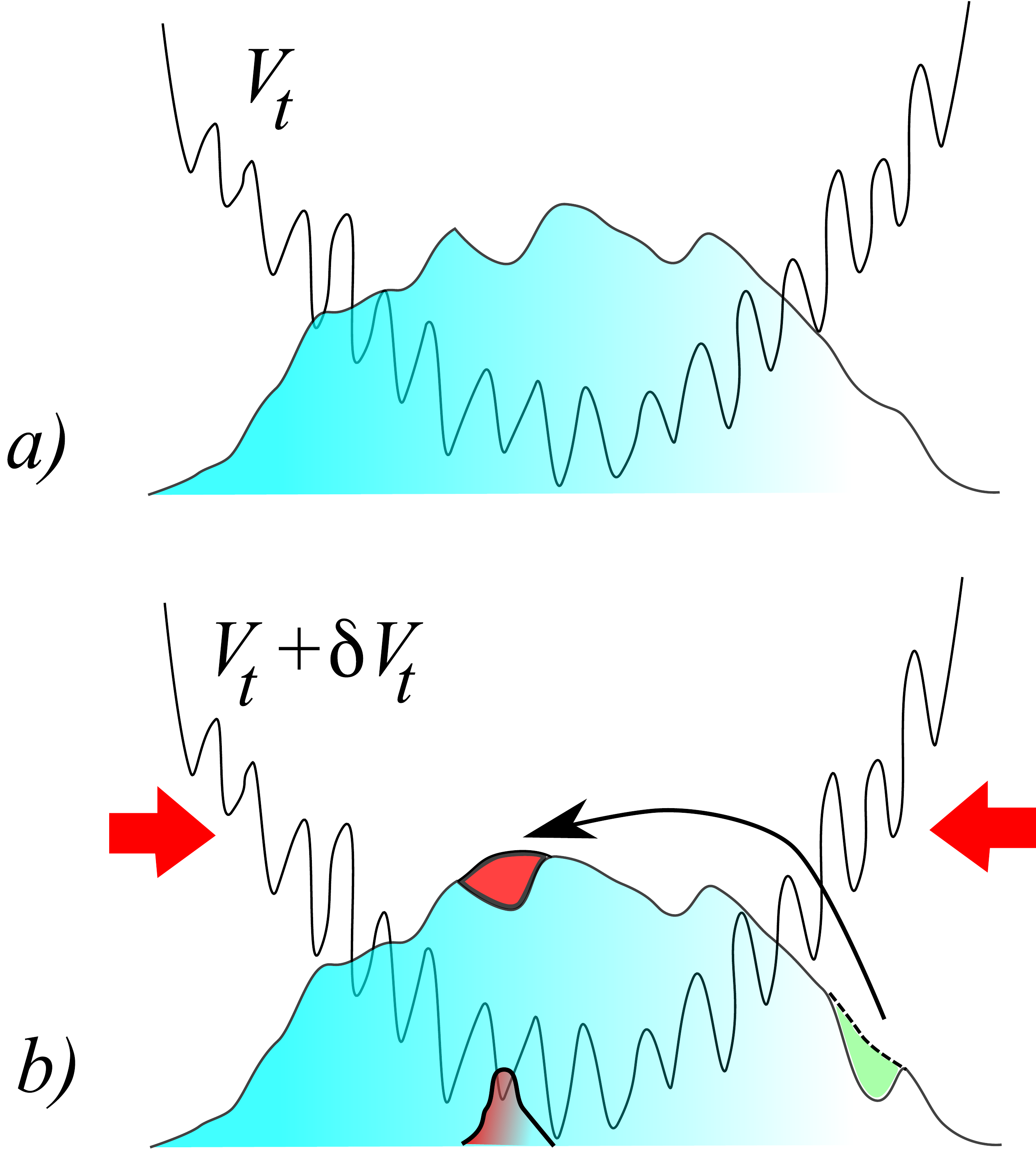} 
\caption{Trap squeezing in the Bose glass phase. \emph{a)} Schematic view
of the density profile of a boson gas in a disordered optical lattice.
\emph{b)} In presence of a Bose glass behavior in the trap center, 
upon infinitesimal trap squeezing $V_t \to V_t + \delta V_t$ a particle can be transfered from 
the trap wings to a \emph{localized} state in the center. Due to its
localized nature, this additional particle does not contribute significantly
to the coherent fraction, which remains essentially unchanged.}
\label{f.BGsqueeze}
\end{center}
\end{figure} 
 
 In Ref.~\cite{Roscilde09} we have shown that trap squeezing
 can reveal the Bose glass phase appearing in the Bose-Hubbard
 model with a one-color quasi-periodic potential, realized by a 
 two-color incommensurate optical superlattice \cite{Fallanietal07, 
 Roscilde08, Rouxetal08, Dengetal08}. The fingerprint of
 the Bose glass in a trap is the observation of a finite central
 compressibility $\kappa_C$, coexisting with a 
 weak condensate compressibility $\kappa_0$, and with
 the observation of a broad peak in the momentum distribution.
  
 Fig.~\ref{f.BGsqueeze} sketches the response to trap squeezing of a Bose glass
 occurring in the trap center. By definition, the Bose
 glass phase admits particle excitations at arbitrarily low energy,
 which means that a small increase in the trap frequency $V_t$ 
 and, consequently, in the global chemical potential $\mu$, 
 is able to transfer a particle to the center of the trap, resulting
 in a finite central compressibility $\kappa_C$. It is important
 to emphasize that this is only true \emph{on average over the disorder 
 distribution}: indeed, averaging over the disorder statistics, one
 expects always the existence of one or several disorder arrangements
 in the trap center which admit the transfer of a particle from the 
 wings to the center at an infinitesimal energy cost. Explicit disorder
 averaging is crucial, because we cannot expect in general the 
 central region of the system to be self-averaging, due to its
 relatively small size. In the following we will consider quasi-periodic
 superlattices, which admit random relative spatial phases 
 between the various Fourier components. In this case, the full
 disorder statistics in the trap is sampled by considering 
 different values of the relative phases. 
 
  If the shift in chemical potential associated with trap squeezing is 
  not bringing the system across a Bose-glass-to-superfluid transition, 
  the particles that are transfered from the wings to the center will occupy 
  \emph{localized} quasi-particle states - meaning that the density variation
  that their transfer causes will be well localized in space. 
  This implies that the new particle appearing in the center remains highly 
  spread in momentum, and that it only contributes a small amount to the 
  $k=0$ peak in the momentum distribution. As a consequence, the 
  coherence properties are only weakly altered by the particle transfer, 
  and the condensate compressibility remains small.  
  
  While a finite central compressibility $\kappa_C$ is a solid
  criterion (and a pre-requisite) for the identification of a Bose glass, 
  it is \emph{a priori} more difficult to fix a criterion on the condensate compressibility
  which would distinguish a Bose glass from a weakly
  correlated superfluid. In fact, in both cases the condensate 
  compressibility is generally finite, given that a global 
  increase of the density leads to a global change of the  
  momentum distribution, regardless of whether the 
  system exhibits condensation or not. The only rigorous approach
  to discriminate a compressible insulator from a condensate
  is to study the scaling of the condensate occupation $n(k=0)$ 
  under increasing particle number and decreasing trapping
  frequency, namely at \emph{fixed chemical potential}; 
  this will be the subject of a future publication.
  Nonetheless, a semi-quantitative criterion can be formulated
  based on the width of the momentum distribution. Indeed 
  the inverse of the FWHM of such a peak gives an estimate of the 
  correlation length $\xi$ in the system: in the following we will take
  $\xi \approx 2 {\rm (FWHM)}^{-1}$ (based on the assumption that
  the momentum distribution $n(k)$ can be approximated by
  a Lorentzian with parameter $\xi$). The system is obviously in
  an insulating state if $\xi$ is far below the width of the atomic 
  cloud. We can estimate such a width via the \emph{participation
  ratio}
  \begin{equation}
  {\rm PR} = \frac{N^2}{\sum_i \langle n_i \rangle^2}~.
  \label{e.PR}
  \end{equation}
  Hence the system can be safely considered as an insulator
  if $\xi \ll {\rm PR}$.

 \subsection{Three-color incommensurate superlattice}
 
  We now focus our attention on the 1$d$ Bose-Hubbard in a 
  two-color quasi-periodic potential, Eq.~\eqref{e.3SL}, realized via a three-color
  incommensurate superlattice. In Ref.~\cite{Roscilde08} we have shown that
  this potential, despite being quasi-random, appears to support a 
  Bose-glass phase over a continuous region of the phase diagram. 
  This is to be contrasted with the one-color quasi-periodic potential, 
  realized via two-color superlattices \cite{Fallanietal07}, which 
  features Bose glass behavior as well as incompressible 
  incommensurate-band-insulator behavior \cite{Roscilde08, Rouxetal08},
  appearing in a tight alternation upon varying the chemical potential. In other words, 
  while the one-color potential still retains features of its quasi-periodic nature, 
  the two-color potential seems to mimic very closely the behavior
  of a purely random potential.
  
  In the following we focus indeed on the regime in which the system
  immersed in a two-color potential exhibits an extended Bose-glass
  phase, namely for potential strength $V_2 = U$ which removes
  completely the Mott insulator from the phase diagram. In particular
  we consider the case of a strong potential, $V_2 = 20 J$, which
  establishes a Bose glass for all the values of the chemical potential
  we have explored. 
  The trapped version of the system is investigated in the case of $N=100$ bosons, and under
  averaging over the spatial phases $\phi$, $\phi'$ of the Fourier 
  components of the potential, Eq.~\eqref{e.3SL}. In particular
  we investigate the behavior of the phase-averaged quantities: 
  the central density $\langle n_C \rangle_{\phi}$, 
  the momentum distribution $\langle n(k) \rangle_{\phi}$ and its
  FWHM, and the participation ratio
  $\langle {\rm PR} \rangle_{\phi}$. We will plot these quantities
  as a function of the phase-averaged chemical potential, 
  which in Section \ref{s.mu} has been shown to satisfy a simple 
  scaling relationship as a function of the experimental parameters
  $V_t$ and $N$. The data on the trapped system are compared with 
  those of the bulk system, realized by a chain of $L=300$
  sites for a fixed value of the potential phases $\phi$, $\phi'$; 
  in this case disorder averaging is provided by the extended
  size of the system, which shows to be fully self-averaging
  (we have explicitly checked this aspect by comparison 
  with an even longer chain, $L=500$).
  
  \begin{figure}[h]
\begin{center}
\includegraphics[
width=85mm]{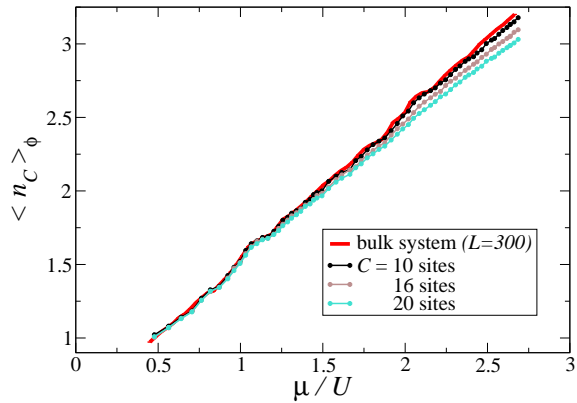} 
\caption{Phase-averaged central density of a system of $N=100$
one-dimensional bosons trapped in a parabolic potential plus a three-color 
superlattice, Eq.~\eqref{e.3SL} (see text for the parameters). The results for the trapped system
are compared with those of a bulk system with $L=300$ sites. }
\label{f.3SLdens}
\end{center}
\end{figure} 
  
   Fig.~\ref{f.3SLdens} shows the evolution of the central density
  in the trapped system for a central region $C$ of varying
  size, compared with the bulk behavior. The bulk behavior shows
  a continuous variation of the density with the chemical potential, 
  signaling a finite compressibility over the whole range
  we have considered. We observe that the phase-averaged
  central density closely follows the bulk behavior, and, in the
  case of $N=100$ bosons we consider, the dependence of the central
  density on the size of the $C$ region is very weak up to
  densities $n\lesssim 2$, signaling that a faithful measurement of 
  the $n(\mu)$ curve can be achieved at different imaging 
  resolutions. To faithfully reproduce the behavior at higher
  densities it is necessary to have a good resolution of around
  10 sites, or, alternatively, to increase the number $N$ of trapped 
  particles and reduce the trapping potential  
  (thereby keeping the product $N^{2/3} V_t^{1/3}$ constant) in order
  to increase the quasi-homogeneous region in the trap center.

\begin{figure}[h]
\begin{center}
\includegraphics[
width=85mm]{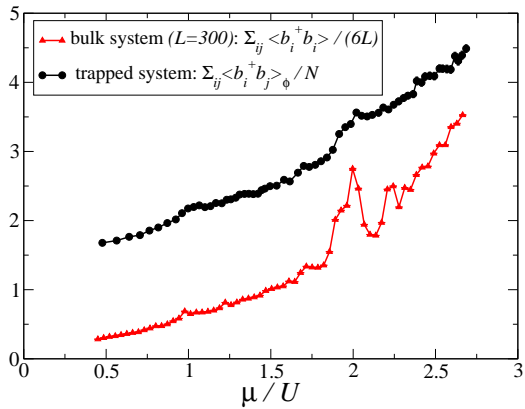} 
\caption{$k=0$ peak in the momentum distribution for a system of 
$N=100$ one-dimensional bosons trapped in a parabolic potential plus a three-color 
superlattice (same as in Fig.~\ref{f.3SLdens}). The results are compared
with those of a bulk system with $L=300$ sites.  }
\label{f.3SLenk0}
\end{center}
\end{figure} 

\begin{figure}[h]
\begin{center}
\includegraphics[
width=85mm]{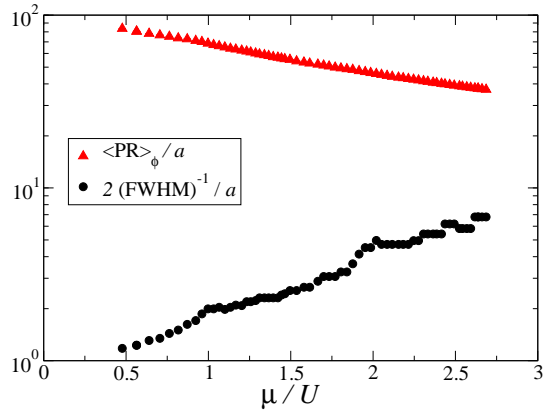} 
\caption{Correlation length of system of $N=100$ one-dimensional 
bosons trapped in a parabolic potential plus a three-color 
superlattice (same as in Fig.~\ref{f.3SLdens}), estimated
via the inverse FWHM of the momentum distribution. 
This quantity is compared with the effective cloud size, 
given by the phase-averaged participation ratio.}
\label{f.3SLxi}
\end{center}
\end{figure}

 The insulating nature of the phase stabilized in the trap center is 
revealed by the momentum distribution. In Fig.~\ref{f.3SLenk0} we show
the evolution of the peak in the momentum distribution under
trap squeezing. The comparison with the data of the bulk system
shows that the trapped system reproduces the main features 
of the bulk behavior, including some subtle ones (for instance,
the anomaly for $\mu/U \approx 2$); this further confirms 
that the momentum distribution is dominated by the behavior
in the trap center, which best approximates the bulk behavior. 
In the case of the bulk system, the inspection of the superfluid 
density \cite{Roscilde08} allows to conclude that the system is in 
an insulating state for all explored chemical potentials, as the superfluid
density turns out to be identically vanishing everywhere.  
This conclusion cannot be easily drawn on the basis of the 
observation of the momentum distribution. In fact, at variance
with the Mott insulating state, the condensate compressibility 
$\kappa_0$ appears to be finite everywhere, although large
excursions of the chemical potential (\emph{e.g.} from 0.5 $U$
to $U$) only slightly change the condensate population 
(by about $10\%$ in the case mentioned). As already mentioned, 
the most dramatic signature of the insulating behavior would be 
revealed upon scaling the system size. Yet, in this case the
insulating behavior is so pronounced that it is already revealed 
at the level of the single system size we considered. In fact, 
as shown in Fig.~\ref{f.3SLxi}, the correlation length estimated
via the inverse of the full width at half maximum is considerably
smaller than the system size, estimated via the participation ratio
(it is more than an order of magnitude smaller for $\mu/U \lesssim 2$).
Hence the observation of the central density and its
evolution under trap squeezing, together with the measurement of
the momentum distribution, would experimentally allow to 
conclude unambiguously about the Bose-glass nature of the
phase realized in this system.

\section{Central vs. global compressibility} 
 \label{s.centralvsglobal}
 
  Recent experiments have probed for the first time the 
  compressibility of a fermionic gas in an optical lattice
  via trap squeezing and \emph{in situ} imaging of the cloud 
  width \cite{Schneideretal08}. The same technique is obviously 
  applicable to bosons as well, and it has been proposed as a
  practical tool for the detection of the Bose glass \cite{DelandeK08}. 
  Probing the response of the global width of the atomic cloud to 
  trap squeezing has the obvious advantage that the imaging
  resolution required in this case is much less than that
  demanded by the technique we propose. Here we would like to 
  point out a few drawbacks, particularly associated with the
  detection of the Bose-glass phase. A discussion of the
  drawbacks for the case of repulsive fermions can be found
  in Ref.~\onlinecite{Scarolaetal09}.  
  
\begin{figure}[h]
\begin{center}
\includegraphics[
width=85mm]{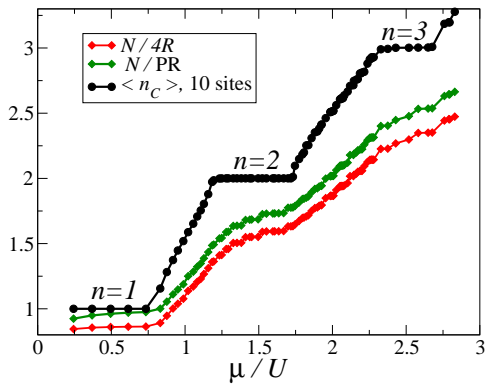} 
\caption{Central density (averaged over 10 sites) of a system of $N=100$ trapped bosons, 
described by the one-dimensional Bose-Hubbard model with $U=20 J$ in a parabolic trap. This quantity
is compared with the effective global density of the atomic cloud, estimated
via the inverse of the cloud radius $R$ and via the inverse of the participation ratio.}
\label{f.R_1DBH}
\end{center}
\end{figure} 
  
  In Ref.~\onlinecite{Schneideretal08} the cloud width is 
  estimated via the \emph{cloud radius}, namely the second moment of the density distribution
  $\langle n_i \rangle$ \cite{noteonn} (centered around the origin)
  \begin{equation}
  R = \sqrt{\frac{1}{N} \sum_i |{\bm r}_i|^2 \langle n_i \rangle} ~.
  \end{equation}
  The response of the atomic cloud to the variation of the trap
  frequency is then quantified in terms of the variation in the 
  cloud radius $R$; given that $N/R^d$ gives an estimate of the 
  effective average density of the cloud, one can define a \emph{global cloud
  compressibility} $\kappa_g$ as the derivative of $R^{-d}$
  with respect to the trapping potential
 \begin{equation}
  \kappa_g = - \frac{1}{R^{d+1}} \frac{\partial R}{\partial V_t}~.
  \end{equation}
  This is a measure of the
  compressibility of an inhomogeneous trapped system, 
  which collects contributions from all its various parts. In particular, 
  no matter what state the system core is in, the dilute
  halo around the core is always in a compressible state, so that
  the global compressibility is necessarily always finite, although it
  will be significantly higher or smaller depending on whether the
  core is in a locally compressible state or not. Hence the realization
  of a specific bulk phase in the system core does not result in 
  a \emph{qualitatively} different behavior of the global compressibility, 
  but only a \emph{quantitatively} different one. This aspect makes the 
  discrimination of the phases realized in the system core very hard, 
  and even more so the location of the phase boundaries. 
  In particular, as we will discuss below, the identification of some 
  phases might even be completely missed. 
  
   Fig.~\ref{f.R_1DBH} shows the global density $N/R$ for 100
   bosons described by the
  1$d$ Bose-Hubbard model with $U=20 J$ and for a variable
  trapping potential, controlling the chemical potential of the system. 
  An alternative definition of the global average density, which 
  turns out to be much closer to the true average density, is
  $N /$PR, based on the participation ratio PR, Eq.~\eqref{e.PR}, 
  which gives the effective volume occupied by the system. 
  The PR is very close to size of the region over which 
  the atomic density is non-zero, while the cloud radius $R$   
  is closer to the FWHM of the density profile. Despite the apparent
  variety in the definitions, Fig.~\ref{f.R_1DBH}  shows that  
  $N/R$ and $N/$PR exhibit precisely the same behavior, 
  and they essentially differ by a multiplicative factor. Hence we will
  use them interchangeably in the following.  
 
  In Fig.~\ref{f.R_1DBH} the global densities $N/R$
  and  $N/$PR are compared with the central density $\langle n_C \rangle$
  for a central region of 10 sites - as shown in Ref.~\onlinecite{Roscilde09}
  for the same parameter set, this latter quantity reproduces rather 
  faithfully the density of the bulk system. For the chemical potential
  excursion considered in this study, we observe three Mott plateaus,
  which are clearly manifested in the central density, and which
  also show up in the global densities via a significant change
  of slope in the dependence on the chemical potential 
  \cite{noteonslope}. In this case
  one could argue that the global compressibility contains essentially
  the same information as the central one. At the same time, it is
  evident that important pieces of information are missing in the global 
  compressibility: under the assumption that a loss in global compressibility
  is always equivalent to the appearance of an incompressible phase
  in the system core, it is quite hard to locate the phase boundaries
  between core-compressible and core-incompressible phases, as
  the changes in slope are quite smooth (despite the commensurate-incommensurate
  transitions in the core being very sharp for what concerns the central
  compressibility, as we have already discussed in Sec.~\ref{s.1DBH}). 
  Moreover, the global densities contain very little information on
  the central density at which the core-incompressible phases are 
  realized. As a general remark, one could say that the global density
  and global compressibility can give information on the alternation of phases and their
  nature provided that one has a \emph{prior} knowledge of the topology of the phase
  diagram of the system, and of the particular phases that are realized
  there. 
  
\begin{figure}[h]
\begin{center}
\includegraphics[
width=90mm]{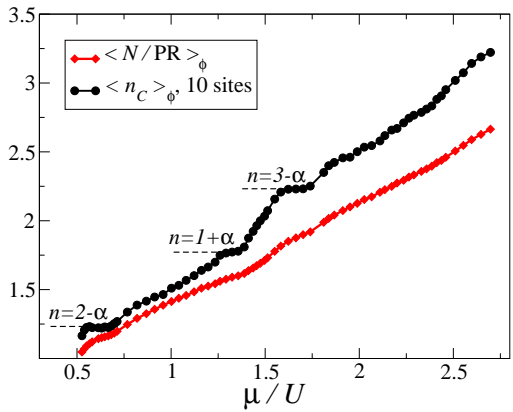} 
\caption{Central density (averaged over 10 sites) of a system of $N=100$ bosons 
trapped in a parabolic potential plus a two-color incommensurate 
superlattice (see text). This quantity is averaged over 
different realizations of the relative spatial phase of the superlattice.    
Comparison is made with the global density, estimated
via the phase-averaged inverse participation ratio.}
\label{f.R_1D2SL}
\end{center}
\end{figure} 
   
   A much harder (yet more general) situation is the one in which one does 
   not know the succession of phases realized in the system - although
   the set of the possible phases might still be known. This is a generic
   situation in which one would like an analog quantum simulator to work. 
   As an example of a complex phase diagram, we consider the case of the 
   Bose-Hubbard model in an incommensurate one-color potential 
   \cite{Roscilde08, Rouxetal08, Dengetal08}.
  As already mentioned elsewhere in this paper, this system exhibits
  four bosonic phases: Mott insulator, superfluid, Bose glass, and incommensurate
  band insulator. Following Ref.~\cite{Roscilde08}, we consider a system of $N=100$ bosons
  in a strong incommensurate potential $V_2 = U = 20 J$ with incommensuration
  parameter $\alpha = 830/1076$, and under variable trapping potential. 
  Fig.~\ref{f.R_1D2SL} shows the phase-averaged global density of such a system
  $\langle N/ {\rm PR} \rangle_{\phi}$, compared with the phase-averaged 
  central density $\langle n_C \rangle_{\phi}$ for a region of 10 sites.
  As seen in Ref.~\onlinecite{Roscilde08} this latter quantity reproduces most of the
  features of the density of a bulk system for low enough values, $n \lesssim 2.5$.
 In particular a distinctive feature is represented by plateaus at incommensurate
 densities, related to the potential parameter $\alpha$, and corresponding to 
 incommensurate band insulating phases. The plateau regions are 
 separated by compressible regions manifesting either Bose glass or
 superfluid behavior (for such strong external potential the Mott insulator
 is completely washed out, and no plateau appears at integer densities). 
 
 When inspecting the global density, one can still observe a weak 
 decrease of its slope as a function of the chemical potential 
 corresponding to the incommensurate plateaus. Nonetheless, 
 without prior knowledge of the bulk phase diagram, it is very hard
 to understand the nature of this less compressible behavior. 
 First of all, it is a priori unclear whether a local minimum in the compressibility
 should always be associated with the appearance of a strictly 
 incompressible behavior in the core, and, even working
 under this assumption, the estimate of the phase boundaries is 
 somewhat arbitrary. 
 Moreover, it is not possible in principle to decide whether the less
 compressible phases correspond to ordinary Mott insulators or to some other form
 of incompressible insulating state,  because
 one lacks definite information on the filling at the trap center. 
  Hence we can conclude that the global density and compressibility
 do not allow in general to reconstruct the phase diagram of the Bose-Hubbard 
 model in a quasi-periodic potential.

 \section{Conclusions}
 \label{s.conclusions}
 
  In this paper we have shown how the control on the trapping
  potential for bosons in optical lattices provides a very practical
  experimental knob on the total chemical potential of the system.
  Indeed, even in the strongly correlated case of bosons
  described by the Bose-Hubbard model with or without an 
  external superlattice potential, the global chemical potential 
  at zero-temperature turns out to be related to the trapping
  potential (and to the other Hamiltonian 
  parameters) via a simple scaling relation. This
  relation takes the form of a modified Thomas-Fermi scaling relation, corrected 
  by quantum effects, and valid in the case of large enough
  densities or weak enough interactions, in which
  lattice commensuration effects are negligible. Corrections
  beyond this simple expression are quantitatively captured
  at the level of a simple lattice mean-field theory 
  (Gutzwiller Ansatz), which agrees surprinsingly well with
  quantum Monte Carlo results. The possibility of tuning 
  the chemical potential via the trap gives access to the exploration 
  of the phase diagram of Hamiltonian models in the grand-canonical ensemble, 
  even if cold-atom experiments are generally performed
  at a fixed number of particles.    
    
   If supplemented with an imaging technique able to 
  measure selectively the average density $n_C$ in a finite
  region $C$ around the trap center, the \emph{a priori} knowledge of 
  the global chemical potential $\mu$ of the system in an optical lattice allows 
  to reconstruct the density-\emph{vs.}-$\mu$ curve of the model
 implemented in the experiment \emph{without} the trap. 
 In particular this gives access to the bulk compressibility $\kappa$, 
 estimated via the central compressibility $\kappa_C = \partial n_C / \partial \mu$. 
 We show that central compressibility measurements tell apart
 very clearly Mott insulator from superfluid phases, and hence they
 allow to fully reconstruct the phase diagram of the Bose-Hubbard model 
 at zero temperature in the grand-canonical ensemble. The addition
 of an incommensurate superlattice \cite{Fallanietal07} introduces
 Bose glass regions in the phase diagram, whose unambiguous 
 detection can be achieved via the joint measurement of 
 the central compressibility and of the global coherence of the 
 system, as well as its response to trap squeezing.    
 
 The broader perspective of a truly useful quantum simulation
 with cold atoms requires to implement computationally
 more challenging models than lattice bosons - \emph{e.g.,} to simulate 
 lattice fermions. To do so with minimal ``classical input", on needs
 a strategy to know in advance all the parameters of the system 
 without the necessity an \emph{ab-initio} classical simulation of the model
 of interest. While it is still unclear how to gain \emph{a priori} knowledge on 
 the temperature without extensive classical simulations \cite{Trotzkyetal09, Jordensetal09},
 the case study of the Bose-Hubbard model, presented in this work, suggests 
 that mean-field variational Ansatzes might be successful to reconstruct \emph{a priori}
 the chemical potential of strongly correlated system in their ground state,
 even when the same variational Ansatzes do not provide an accurate
 description of the microscopic behavior of the system.  

 \section{Acknowledgements}
We thank L. Fallani for suggesting the imaging technique 
described in Section \ref{s.imaging}, and M. Rigol for stimulating suggestions.

\appendix
\section{Jordan-Wigner fermionization and extended fermionization}\label{a.EF}
\label{a.fermionization}

Hardcore boson operators $[b_i,b_j^{\dagger}] = 0$
 ($i\neq j$), $\{b_i,b_i^{\dagger}\} = 1$ can be mapped
 onto spinless fermions $f_i$, $f_i^{\dagger}$ via the Jordan-Wigner 
 transformation \cite{Liebetal61}.  
 \begin{equation}
\label{Jordan-Wigner transformation}
b_i^\dagger=f_i^\dagger\prod_{k=1}^{i-1}e^{-i\pi f_k^\dagger f_k}, 
\hspace{0.5 cm} b_i=\prod_{k=1}^{i-1}e^{i\pi f_k^\dagger f_k}f_i, 
\end{equation}
 so that the Hamiltonian of infinitely repulsive bosons in an 
 external potential $V_i$ becomes that of non-interacting spinless fermions:
 \begin{equation}
 {\cal H} =  -J \sum_i \left(f_i f_{i+1}^{\dagger} + 
 {\rm h.c.}\right) + \sum_i V_i f_i^{\dagger} f_i
 \label{e.fermiham}
 \end{equation} 
 for which we assume open boundary conditions.
 
 This approach has been recently extended to the case of strongly
 repulsive 1$d$ softcore bosons in a trap \cite{Pupilloetal06}
 for $U/J > (U/J)_{c,n}$, where $(U/J)_{c,n}$ 
 is the critical
 value for the superfluid-to-Mott-insulator transition at the 
 largest \emph{integer} density $n$ reproduced locally in 
 the trap. Under this assumption, the density profile in the trap
 takes a characteristic ``wedding-cake" structure, namely regions
 at densities in the interval $[m-1,m]$ (with integer $m$)
 are separated from regions with densities in the interval $[m,m+1]$
 by a sizable Mott plateau at integer density $m$. If $U/J$
 is sufficiently large, density fluctuations in the plateau can 
 be neglected, which means that density fluctuations in the 
 $[m-1,m]$ region are essentially decoupled from those in the
 $[m,m+1]$ region, and that the lattice sites in each of these
 regions only experience density fluctuations between the two
 extremal values of the average density. This fundamental 
 restriction of the local Hilbert space to a set of two-level systems
 implies the possibility of fermionizing each layer of the cake
 individually (extended fermionization); the fermionization has
 to be supplemented with the information that particles in the $m$-th layer have an effectively
 larger hopping due to bosonic enhancement $J \to m J$, 
 and that they feel an overall potential energy $U(m-1)$ per site.
 The extended-fermionization Hamiltonian reads then
 \begin{eqnarray}
 {\cal H} &=&  \sum_{i,m} - m J \left(f_{i,m} f_{i+1,m}^{\dagger} + 
 {\rm h.c.}\right)\nonumber \\
 &+& \sum_{i,m} \left[ U(m-1) + V_i \right] f_{i,m}^{\dagger} f_{i,m}
 \label{e.fermihamext}
 \end{eqnarray} 
 where $f_{i,m}$, $f_{i,m}^{\dagger}$ are the spinless fermions
 associated with the $m$-th layer.

\section{Thomas-Fermi results for the atom number distribution in 
low-dimensional systems}\label{a.TF}

 We consider a gas of weakly interacting bosons in a
 three-dimensional trap of frequencies $(\omega_x, \omega_y, \omega_z)$ 
 and in a standing wave potential along $3-d$ spatial 
 dimensions, defining an array of $d$-dimensional trapping 
 elements (tubes for $d=1$, pancakes for $d=2$). In the
 weakly interacting case, we describe the system within 
 Gross-Pitaevskii (GP) theory in continuum space for $d$ dimensions,
 and on a lattice for the remaining $3-d$ ones. To calculate the
 density profile of the atoms, we make use of the Thomas-Fermi
 approximation, which amounts to minimizing the potential-energy
 part of the GP energy functional. 
 
 \subsection{$d=1$ (tubes)}
 In this case the system is immersed in two standing waves along the $x$
 and $y$ directions, with wavelength $\lambda_x$ and $\lambda_y$
respectively, defining a 2$d$ array of tubes. Being $\bm d_x = (d_x, 0)$, $ \bm d_y = (0,d_y)$
the vectors of the 2$d$ array, with $d_\alpha=\lambda_\alpha/2$ 
($\alpha=x,y$), we identify the position of the tubes as
$\bm R(i_x,i_y) = i_x ~\bm d_x + i_y ~\bm d_y$. Imagining that the atoms
can only occupy the lowest Wannier states of each tube in the $x$ and $y$
directions ($w_x$ and $w_y$, respectively), we can discretize the space in those directions, so that the GP wavefunction 
can be written as $\Psi = \Psi(i_x, i_y; z)$. The potential energy part of the associated 
GP functional reads:
\begin{eqnarray}
{E}^{\rm (pot)}_{\rm GP}[\Psi, \Psi^*] = 
  \sum_{i_x, i_y} 
\int dz \Big[ ~ \frac{\tilde g^{(1d)}}{2} |\Psi(i_x,i_y;z)|^4  + ~~~~\nonumber\\
 \left( \epsilon_x i_x^2 +  \epsilon_y i_y^2  + \frac{1}{2} m \omega_z^2 z^2 - \mu^{(1d)} \right ) 
|\Psi(i_x,i_y;z)|^2 \Big]  ,~~~~~~
\end{eqnarray}
where $\epsilon_\alpha = m\omega_\alpha^2 d_\alpha^2/2$, and  
\begin{equation}
\tilde g^{(1d)}= g \int ~dx ~dy~ |w_x(x)|^4 ~|w_y(y)|^4
\end{equation}
is the effective coupling constant for particles trapped in the lowest Wannier
states in the $x$ and $y$ direction, related to the bare coupling constant
$4\pi \hbar^2 a_s/m$ ($a_s$ = $s$-wave scattering length). Taking a Gaussian approximation for 
the Wannier states \cite{Zwerger01} one obtains
\begin{equation}
\tilde g^{(1d)}\approx g \frac{k_x k_y}{2\pi} \left( \frac{V_{0x} V_{0y}}{E_{rx} E_{ry}} \right)^{1/4}
\end{equation}
where $k_\alpha = 2\pi/\lambda_\alpha$, $V_{0\alpha}$ is the lattice depth
in the $\alpha$ direction, and $E_{r\alpha} = \hbar^2 k_{\alpha}^2/(2m)$
the associated recoil energy. 

The Thomas-Fermi approximation gives for the number of atoms in 
each tube the expression
\begin{equation}
N^{(1d)}(i_x,i_y) = \int dz ~\frac{ \mu^{(1d)} -  \left( \epsilon_x i_x^2 +  \epsilon_y i_y^2  + \frac{1}{2} m \omega_z^2 z^2\right)}{\tilde g^{(1d)}} 
\label{e.N2dTF}
\end{equation}
where the chemical potential $\mu^{(1d)}$ is consistently given by:
\begin{equation}
 \mu^{(1d)} = \left[ \frac{15 ~\tilde g^{(1d)} d_x d_y  N}{8 \pi} \left( \frac{m \bar\omega^2}{2} \right)^{3/2} \right]^{2/5}
\end{equation}
with $\bar\omega = (\omega_x \omega_y \omega_z)^{1/3}$. 
Integrating in Eq.~\eqref{e.N2dTF} on the region in which the integrand is positive, 
one finally obtains the expression \cite{noteonParedes}:
\begin{equation}
N^{(1d)}(i_x,i_y) = \frac{4 \sqrt{2}}{3 \tilde g^{(1d)} \sqrt{m\omega_z^2}} \left( \mu^{(1d)} - \epsilon_x i_x^2 - \epsilon_y i_y^2\right)^{3/2}~.
\label{e.N1d}
\end{equation}

\subsection{$d=2$ (pancakes)}
In this case the system is immersed in a standing wave along the $z$
direction with wavelength $\lambda_z$, defining a 1$d$ array of pancakes.
In analogy with what done in the previous section, we define a discretized
GP wavefunction $\Psi = \Psi(x, y; i_z)$, whose associated potential energy 
is described by the GP functional
\begin{eqnarray}
{E}^{\rm (pot)}_{\rm GP}[\Psi, \Psi^*] = 
  \sum_{i_z} 
\int dx ~dy \Big[ ~ \frac{\tilde g^{(2d)}}{2} |\Psi(x,y,i_z)|^4  + ~~~~\nonumber\\
 \left(\frac{1}{2} m \omega_x^2 x^2 + \frac{1}{2} m \omega_x^2 y^2 +  \epsilon_z i_z^2 - \mu^{(2d)} \right ) 
|\Psi(x,y;i_z)|^2 \Big]  ,~~~~~~
\end{eqnarray}
where we have introduced the effective coupling constant
\begin{equation}
\tilde g^{(2d)}= g \int ~dz~|w_z(z)|^4 \approx g \frac{k_z}{\sqrt{2\pi}} \left( \frac{V_{0z}}{E_{rz}} \right)^{1/4}
\end{equation}
with evident meaning of the symbols. 

Integrating the Thomas-Fermi expression for the density within each pancake, 
we obtain the atom number distribution 
\begin{equation}
N^{(2d)}(i_z) = \frac{\pi}{m \tilde g^{(1d)} \omega_x \omega_y} \left( \mu^{(2d)} - \epsilon_z i_z^2 \right)^{2}
\label{e.N2d}
\end{equation}
with chemical potential
\begin{equation}
 \mu^{(2d)} = \left[ \frac{15 ~\tilde g^{(2d)} d_z  N}{8 \pi} \left( \frac{m \bar\omega^2}{2} \right)^{3/2} \right]^{2/5}~.
\end{equation}


\begin{thebibliography}{99}
\bibitem{note.on.3DSL} We notice that the $\mu$ data 
for the $d=3$ system in a superlattice are much
more scattered than for all the other cases, as shown
on the right panel of Fig.~\ref{f.3Dmu}. This is 
due to strong size effects on the system considered,
which are inevitable in $d=3$ due to the short linear
size of each spatial dimension (we have considered
cubes with up to $L=18$ sites on each side, but the
particles are actually confined on a region with
linear extent which is even smaller). In presence of
a superlattice potential, a system on such a finite
size experiences only a few wells of the potential 
in each spatial dimension, and hence the nonlinearities due to
such potential become relevant and cannot be simply 
averaged out as in the TF/AL approach.
\bibitem{noteongap} We estimate gaps as small as 
$\sim 10^{-4} J$, and whose smallness might be limited
by the resolution of the grid with which we sample
the trapping strength.
\bibitem{noteonbox} The standard definition
of the lattice momentum distribution would involve a normalization
to the volume $L^d$ of the quantization box, but in the case of a trapped system
the quantization box is arbitrary. As an alternative, here we imagine that the quantization
box is fixed to the number of bosons $N$ - which is a convenient convention, given that
this number is kept fixed in our study of a trapped system. In this case 
the condensate occupation $n(k=0)$ takes the meaning of the number 
of particles with momenta in the box $|k| \leq \pi/(a N)$, 
where $a$ is the lattice spacing.
\bibitem{noteonn} In the case of Ref.~\onlinecite{Schneideretal08} 
 $\langle n_i \rangle$ is actually the column-integrated
 density of a three-dimensional system, but this aspect 
 is actually irrelevant for our discussion).
\bibitem{noteonslope} We avoid extracting the central 
and global compressibilities by differentiating the related
densities, because of the significant numerical noise associated
with the derivative. Nonetheless the changes in slope in
Fig.~\ref{f.R_1DBH} are evident.
\end{thebibliography}
\end{document}